\documentclass[preprint,12pt]{elsarticle}
\pdfoutput=1
\usepackage{caption}
\usepackage{amsmath}
\usepackage{amssymb}
\usepackage{amsthm}
\usepackage{mathrsfs}

\usepackage{subcaption}
\usepackage{booktabs} %The package enhances the quality of tables in LATEX, providing extra commands as well as behind-the-scenes optimisation
\usepackage{multirow} %Create tabular cells spanning multiple rows

\usepackage{url}
\usepackage{hyperref}
\usepackage{xcolor}
\usepackage{cleveref} %Intelligent cross-referencing

\usepackage{lineno}

\usepackage{color}
\usepackage{epstopdf}
\usepackage{float}
\usepackage[left=2cm, right=2cm]{geometry}
\renewcommand{\vec}[1]{\boldsymbol{#1}}

\biboptions{numbers,sort&compress}

\journal{Journal of Computational Physics}

\begin{document}

% \begin{CJK*}{UTF8}{gbsn}y

\begin{frontmatter}



\title{Unified gas-kinetic wave-particle methods VII: diatomic gas with rotational and vibrational nonequilibrium}

\author[a]{Yufeng Wei}
\author[a]{Yajun Zhu}
\author[a,b,c]{Kun Xu\corref{cor1}}

\cortext[cor1] {Corresponding author.}
\ead{makxu@ust.hk}

\address[a]{Department of Mathematics, Hong Kong University of Science and Technology, Hong Kong, China}
\address[b]{Department of Mechanical and Aerospace Engineering, Hong Kong University of Science and Technology, Hong Kong, China}
\address[c]{HKUST Shenzhen Research Institute, Shenzhen, 518057, China}

\begin{abstract}
Hypersonic flow around a vehicle in near space flight is associated with multiscale non-equilibrium physics at a large variation of local Knudsen number from the leading edge highly compressible flow to the trailing edge particle free transport. To accurately capture the solution in all flow regimes from the continuum Navier--Stokes solution to the rarefied gas dynamics in a single computation requires genuinely multiscale method. The unified gas-kinetic wave-particle (UGKWP) method targets on the simulation of such a multicale transport. Due to the wave-particle decomposition, the dynamics in the Navier--Stokes wave and kinetic particle transport has been unified systematically and efficiently under the unified gas-kinetic scheme (UGKS) framework. In this study, the UGKWP method with the non-equilibrium among translation, rotation and vibration modes, is developed based on a multiple temperature relaxation model. The real gas effect for high speed flow in different flow regimes has been properly captured. Numerical tests, including Sod tube, normal shock structure, hypersonic flow around two-dimensional cylinder and three-dimensional flow around a sphere and space vehicle, have been conducted to validate the UGKWP method. In comparison with the discrete velocity method (DVM)-based Boltzmann solver and particle-based direct simulation Monte Carlo (DSMC) method, the UGKWP method shows remarkable advantages in terms of computational efficiency, memory reduction, and automatic recovering of multiscale solution.
\end{abstract}


\end{frontmatter}



\section{Introduction}\label{sec:intro}

For high-speed flying vehicle in near space, the highly compressed gas at the leading edge and the strong expansion wave in the trailing edge can cover the whole flow regimes with several orders of magnitude on the differences of particle mean free path \cite{bird1994molecular}.
Multiscale flow with a large variation of local Knudsen number is involved in the computation of the flow field around the vehicle.
For high-speed and high-temperature flow, both rotational and vibrational modes of diatomic gas will be activated with significant impact
on aerodynamic heating and forcing \cite{boyd2017nonequilibrium}. In the aerospace engineering practice, an accurate and efficient multiscale method being capable of simulating both continuum and rarefied flow with the inclusion of molecular translation, rotation, and vibration nonequilibrium is of great importance.

The Boltzmann equation is the fundamental governing equation in rarefied gas dynamics. Theoretically, it can capture multiscale flow physics in all Knudsen regimes, with the enforcement of resolving the flow physics in the particle mean free path and mean collision time scale.
For highly non-equilibrium flow, there are mainly two kinds of numerical methods to solve the Boltzmann equation, i.e., the stochastic particle method and the deterministic discrete velocity method.
The stochastic methods employ discrete particles to simulate the statistical behavior of molecular gas dynamics \cite{bird1994molecular,fan2001statistical,shen2006rarefied,sun2002direct,baker2005variance,homolle2007low,degond2011moment,
pareschi2000asymptotic,ren2014asymptotic,dimarco2011exponential}. This kind of Lagrangian-type scheme achieves high computational efficiency and robustness in rarefied flow simulation, especially for hypersonic flow.
However, it suffers from statistical noise in the low-speed simulation due to its intrinsic stochastic nature.
Meanwhile, in the near continuum flow regime, the treatment of intensive particle collisions makes the computational cost very high.
The deterministic approaches
apply a discrete distribution function to solve the kinetic equations and naturally obtain accurate solutions without statistical noise \cite{chu1965kinetic,JCHuang1995,Mieussens2000,tcheremissine2005direct,Kolobov2007,LiZhiHui2009,ugks2010,wu2015fast,aristov2012direct,
li2004study,li2019gas,ugks2010,guo2013discrete,chen2017unified,chen2015comparative}.
At the same time, the deterministic method can achieve high efficiency by using numerical acceleration techniques, such as implicit algorithms \cite{yang1995rarefied,Mieussens2000,zhu2016implicit,zhu2017implicit,zhu2018implicit,jiang2019implicit},
memory reduction techniques \cite{chen2017unified}, and adaptive refinement method \cite{chen2012unified}, fast evaluation of the Boltzmann collision term \cite{mouhot2006fast,wu2013deterministic}.
Asymptotic preserving (AP) schemes \cite{filbet2010class,dimarco2013asymptotic} can be developed to release the stiffness of the collision term
at the small Knudsen number case. However, for most AP schemes only the Euler solution in the hydrodynamic limit is recovered.
Additionally, for hypersonic and rarefied flow, the deterministic methods have to discretize the particle velocity space
with a high resolution to capture nonequilibrium distribution, which brings huge memory consumption and computational cost, especially for the three-dimensional calculation. Moreover, for both stochastic and deterministic methods,
once the gas evolution process is split into collisionless free transport and instant collision, a numerical dissipation being proportional to the time step is usually unavoidable. Therefore, the mesh size and the time step in these schemes have to be less than the mean free path and the particle mean collision time, respectively, to avoid the physical dissipation being overwhelmingly taken over by the numerical one in the continuum regime, such as the laminar boundary layer computation at high Reynolds number. In order to remove the constraints on the mesh size and time step
in the continuum flow regime, the unified gas-kinetic scheme (UGKS) and discrete UGKS (DUGKS) with the coupled particle transport and collision in the flux evaluation has been constructed successfully \cite{ugks2010,jiang2019implicit,guo2021progress}.
At the same time, the multiscale particle methods have been constructed as well \cite{fei2020unified,fei2021efficient}.

Combining the advantages of the deterministic and the stochastic methods,
a unified gas-kinetic wave-particle (UGKWP) method \cite{liu2020unified, zhu2019unified} was proposed under the UGKS framework \cite{xu-book}, as well as the simplified versions \cite{PhysRevE.102.013304,DUGKWP-yang}.
The coupled multiscale transport and collision in UGKWP is modeled according to cell's Knudsen number and is used in the flux evaluation across the cell interface and inner cell relaxation.
The UGKWP method releases the restriction on the mesh size and time step being less than the particle mean free path and particle collision time.
Also, the wave-particle decomposition in UGKWP makes the scheme adaptively become a particle method in highly rarefied flow regime and a hydrodynamic flow solver in the continuum flow regime.
In the continuum flow limit at a small cell's Knudsen number, the UGKWP gets back to the gas kinetic scheme (GKS) for the Navier--Stokes solution \cite{xu2001}.
Thus, the UGKWP method could achieve high efficiency both in the continuum and rarefied regimes. In the intermediate transition regime, the distributions of wave and particle are fully controlled by the time accurate integral solution of the kinetic model equation.
Different from the hybrid methods with domain decompositions for different solvers with interfaces to separate them,
the UGKWP method employs an adaptive wave-particle decomposition in each cell with a unified treatment in the whole computational domain.
The contributions of wave and particle are weighted by the local Knudsen number (${\rm Kn}_c = \tau /\Delta t$) defined by the ratio of particle collision time $\tau$ over the numerical time step $\Delta t$, with the weights $\exp(-{\rm Kn}_c)$ and $(1 - \exp(-{\rm Kn}_c))$.
As a result, the UGKWP method becomes a physically consistent and numerically efficient solver for multiscale flow.
The methodology of UGKWP has been extended to other multiscale transport processes, such as
radiation, plasma, and multiphase flow \cite{li2020unified,liu2020plasma,yang2022unified}.

In the previous works \cite{liu2020unified, zhu2019unified}, the Bhatnagar--Gross--Krook (BGK) \cite{BGK1954} model and Shakhov model \cite{shakhov1968generalization} were employed to describe the multiscale evolution of monatomic gas flow.
For diatomic gases, the internal degrees of freedom, such as rotation and vibration, should be considered \cite{wujunlin,zhang2015vib,liu2014unified,xu2021rot,wu2021derivation,li2022kinetic,fei2022unified}, especially for the high-speed and high-temperature flow.
The BGK-type model was extended to diatomic gas by introducing additional internal energy variables
in the distribution function \cite{morse1964kinetic,rykov1978macroscopic,andries2000gaussian,zhang2015vib,bernard2019bgk}. The Rykov model \cite{rykov1978macroscopic} was also incorporated in the UGKWP method to include the diatomic effect with molecular translational and rotational nonequilibrium only \cite{xu2021rot}.
In this study, we present the UGKWP method with the inclusion of vibrational mode for diatomic gas.
The vibrational model is used to describe the relaxation process from non-equilibrium
to the equilibrium state \cite{bird1994molecular}, where three equilibrium states are employed to take into account of the elastic and inelastic collisions and the detailed energy exchange between the translational, rotational and vibrational degrees of freedom.
With the inclusion of molecular vibrational mode,
the UGKWP method has to take into account several groups of particles with different temperature.
But, it provides more accurate solution for multiscale transport for high-speed and high-temperature flow.
In this paper, in order to clearly present the algorithm development, the scheme with the BGK-type relaxation model will be constructed and validated in many cases from one dimensional to three dimensional flow simulations. The scheme with the inclusion of additional heat flux modification through the Shakhov and Rykov models can be done easily under the current framework.

The paper is organized as follows. Section \ref{sec:vib} presents the kinetic model of diatomic gas with molecular vibration. Since the UGKWP method is an enhanced unified gas-kinetic particle (UGKP) method employing the adaptive wave-particle decomposition, the UGKP method will be introduced first in Section \ref{sec:kp}. Then the UGKWP method with molecular vibration will be presented in Section \ref{sec:wp}. Numerical validation of the current method will be carried out in Section \ref{sec:test}
and a conclusion will be drawn in Section \ref{sec:conclusion}.

\section{Kinetic model equation for diatomic gas}\label{sec:vib}

\subsection{Kinetic model with molecular translation, rotation and vibration}\label{subsubsec:vibBGK}
Considering molecular rotation and vibration, the kinetic model equation for diatomic gases can be written as
\begin{equation}\label{eq:BGK-vib}
\frac{\partial f}{\partial t}+\vec{u} \cdot \frac{\partial f}{\partial \vec{r}}
=\frac{g_{t}-f}{\tau}
+\frac{g_{t r}-g_{t}}{Z_{r} \tau}
+\frac{g_{M  }-g_{tr}}{Z_{v} \tau},
\end{equation}
where $f = f\left( {{\vec{r}},{\vec{u}}, {\vec{\xi}}, {\varepsilon_v}, t} \right)$ is the distribution function for gas molecules at physical space location $\vec{r}$ with microscopic translational velocity $\vec{u}$, rotational motion ${\vec{\xi}}$, and vibrational energy $\varepsilon_{v}$ at time $t$.
$\tau$ is the mean collision time or relaxation time to represent the mean time interval of two successive collisions.
The rotational and vibrational relaxation times are defined as
\begin{equation*}
	{\tau _{rot}} = {Z_r}\tau,
\end{equation*}
\begin{equation*}
	{\tau _{vib}} = {Z_v}\tau,
\end{equation*}
where $Z_r$ and  $Z_v$ are the rotational and vibration collision numbers, respectively.

The elastic collision process of molecules' translational motions and the inelastic collision process of internal energy exchange are described by the right-hand side of Eq.~\eqref{eq:BGK-vib} with three equilibrium states. The equilibrium state $g_t$ with three different temperatures for molecular translation, rotation and vibration gives
\begin{equation*}
	g_{t}= \rho \left(\frac{\lambda_{t}}{\pi}\right)^{\frac{3}{2}} e^{-\lambda_{t} {\vec{c}}^2}
	\left(\frac{\lambda_{r}}{\pi}\right) e^{-\lambda_{r} {\vec{\xi}}^{2}} \frac{4 \lambda_{v}}{K_v(\lambda_v)} e^{-\frac{4 \lambda_{v}}{K_v(\lambda_v)} \varepsilon_{v}},
\end{equation*}
where $\vec{c} = \vec{u} - \vec{U}$ denotes the peculiar velocity, and ${\vec{c}}^2 = (u - U)^2 + (v-V)^2 + (w-W)^2$ and ${\vec{\xi}}^2 = \xi_{1}^2 + \xi_{2}^2$. The intermediate equilibrium state $g_{tr}$ has the same temperature $\lambda_{tr}$ of molecular translation and rotation, but a different temperature $\lambda_v$ for vibration, which indicates complete energy exchange between translational and rotational degrees of freedom, and a frozen process of vibrational energy
\begin{equation*}
	g_{tr}=\rho\left(\frac{\lambda_{tr}}{\pi}\right)^{\frac{3}{2}} e^{-\lambda_{t r} {\vec{c}}^2}
	\left(\frac{\lambda_{t r}}{\pi}\right) e^{-\lambda_{t r} {\vec{\xi}}^2} \frac{4 \lambda_{v}}{K_v(\lambda_v)} e^{-\frac{4 \lambda_{v}}{K_v(\lambda_v)} \varepsilon_{v}}.
\end{equation*}
After sufficient collisions, the equilibrium state with equal-partitioned energy for each degree of freedom
\begin{equation*}\label{eq:g-M}
	g_{M}=\rho\left(\frac{\lambda_M}{\pi}\right)^{\frac{3}{2}} e^{-\lambda_M {\vec{c}}^2}
	\left(\frac{\lambda_M}{\pi}\right) e^{-\lambda_M {\vec{\xi}}^2} \frac{4 \lambda_M}{K_v(\lambda_M)} e^{-\frac{4 \lambda_M}{K_v(\lambda_M)} \varepsilon_{v}},
\end{equation*}
will be reached.

In these equilibrium states, $\lambda$ is computed from the corresponding internal energy.
Specifically, we have
\begin{eqnarray*}
&\lambda_{t}=&\frac{3 \rho}{4} / (\rho E_t), \\
&\lambda_{r}=&\frac{K_r \rho}{4} / (\rho E_r), \\
&\lambda_{v}=&\frac{K_v(\lambda_v)\rho}{4} / (\rho E_v), \\
&\lambda_{t r}=&\frac{(3+K_r) \rho}{4} / (\rho E_{tr}), \\
&\lambda_M=&\frac{[3+K_r+K_v(\lambda_M)] \rho}{4} / (\rho E_M),
\end{eqnarray*}
and
\begin{eqnarray*}
&\rho E_{t}&= \frac{1}{2} \int {{\vec{c}}^2 f {\rm d}{\vec{\Xi}}}, \\
&\rho E_{r}&= \frac{1}{2} \int{{\vec{\xi}}^{2} f {\rm d} \vec{\Xi}}, \\
&\rho E_{v}&= \int {\varepsilon_{v} f {\rm d}\vec{\Xi}}, \\
&\rho E_{t r} &=\frac{1}{2} \int{({\vec{c}}^2 + {\vec{\xi}}^{2}) f	{\rm d}{\vec{\Xi}}}, \\
&\rho E_M     &= \int {\left[\frac{1}{2}({\vec{c}}^{2} + \vec{\xi}^{2}) + \varepsilon_{v}\right] f {\rm d}\vec{\Xi}},
\end{eqnarray*}
where
\begin{equation*}
\int{ (\cdot) {\rm d}\vec{\Xi}} = \int_{-\infty}^{\infty} {\rm d}\vec{u} \int_{-\infty}^{\infty} {\rm d}{\vec{\xi}} \int_{0}^{\infty} {(\cdot) {\rm d}{\varepsilon_v}},
\end{equation*}
and $K_r$ and $K_v(\lambda)$ denote the number of rotational and vibrational degrees of freedom, respectively. $\lambda_t$, $\lambda_r$, $\lambda_v$, $\lambda_{tr}$, $\lambda_M$ are associated with the
translational temperature $T_t$, rotational temperature $T_r$, vibrational temperature $T_v$, the translation-rotation average temperature $T_{tr}$ and the fully relaxed temperature $T_M$, respectively by $\lambda = m / (2 k_B T)$, where $m$ is molecular mass, $k_B$ is the Boltzmann constant. It should be noted that the number of vibrational degrees of freedom $K_v(\lambda)$ is determined by the vibrational temperature in each equilibrium state, i.e.,
\begin{equation*}\label{eq:Kv}
	K_{v}(\lambda)=\frac{4\Theta_{v}k_B\lambda/m}{e^{2\Theta_{v}k_B\lambda/m}-1},
\end{equation*}
where $\Theta_v$ is the characteristic temperature of vibration for diatomic gases, e.g.,  $3371$ K for nitrogen and $2256$ K for oxygen \cite{shen2006rarefied}.

\begin{figure}[H]
	\centering
	\includegraphics[width=10cm]{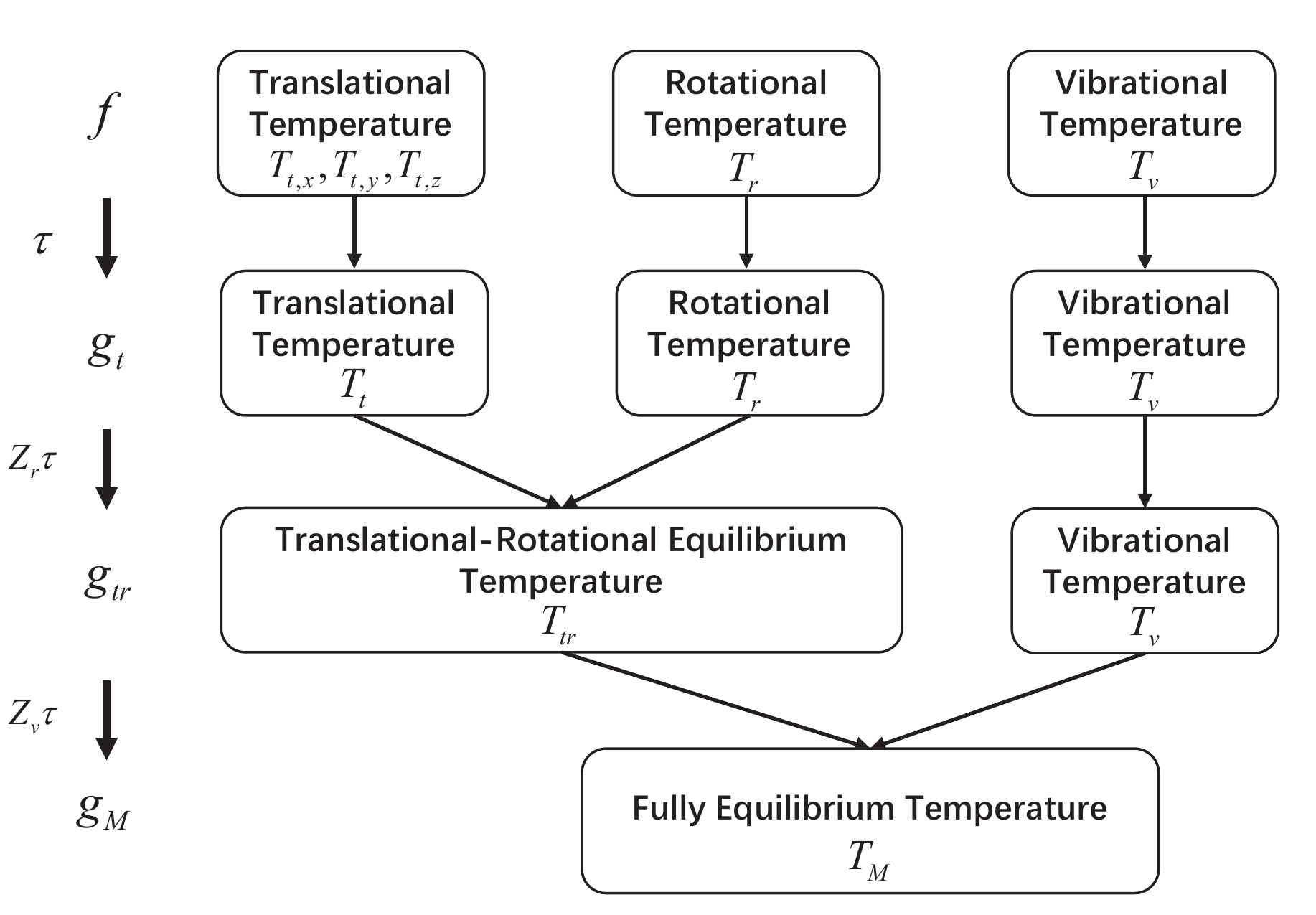}
	\caption{Relaxation process for the vibrational model.}
	\label{fig:BGK-vib}
\end{figure}

With the above three equilibrium states, the energy exchange between molecular translation, rotation, and vibration can be well described by adjusting the collision numbers $Z_r$ and $Z_v$. Experimental observation shows that the rotational relaxation is faster than the vibrational one, i.e., $1 < Z_r < Z_v$. From the relaxation terms on the right-hand side of Eq.~\eqref{eq:BGK-vib}, the relaxation process
can be divided into three stages as shown in Fig.~\ref{fig:BGK-vib}. Firstly, the non-equilibrium distribution function $f$ has different translational, rotational, and vibrational temperatures. After time $\tau$, the elastic collisions drive the distribution function $f$ approaching the translational equilibrium state $g_t$. In the second stage, the inelastic collisions happen within time $Z_r \tau$ to exchange the translational and rotational energy, which drives the distribution function approaching the rotational equilibrium state $g_{tr}$ with the same translational and rotational temperature $T_{tr}$. In the last stage, gas molecules encounter sufficient elastic and inelastic collisions within time $Z_v \tau$, and the internal energy is fully exchanged between each degree of freedom. At this time, the full equilibrium state $g_M$ with the same temperature $T_M$ for translation, rotation, and vibration is achieved.

\newpage
\section{Unified gas-kinetic particle method}\label{sec:kp}

\subsection{General framework}

The unified gas-kinetic particle (UGKP) method is a particle implementation of the UGKS under the finite volume framework, where the discrete particles are employed to describe the non-equilibrium gas distribution function,
and the evolution of particles recovers the multiscale nature in different flow regimes.

Here, we re-write the kinetic model equation in a BGK-type as
\begin{equation}\label{eq:BGK-vib-2}
\frac{{\partial {f}}}{{\partial t}} + {\vec{u}} \cdot \frac{{\partial {f}}}{{\partial {\vec{r}}}} = \frac{{g}^\ast - {f}}{\tau },
\end{equation}
where ${g}^\ast$ is the effective equilibrium state, defined as the convex combination of three modified equilibrium distribution function
\begin{equation}\label{eq:g*}
{g}^\ast = \left(1 - \frac{1}{Z_r}\right){{g}_{t}} + \left(\frac{1}{Z_r} - \frac{1}{Z_v}\right){{g}_{tr}} + \frac{1}{Z_v} {{g}_M}.
\end{equation}
Along the characteristic line, the integral solution of the kinetic model equation gives
\begin{equation}\label{eq:integral-solution1}
f(\vec{r},t)
=
\frac{1}{\tau}\int_{0}^t e^{-(t-t')/\tau}
g^{*}(\vec{r}',t')
{\rm{d}} t'
+ e^{-t/\tau}f_0(\vec{r}-\vec{u}t),
\end{equation}
where $ f_0(\vec{r}) $ is the initial distribution function at the beginning of each step $t_n$, and
$g^\ast(\vec{r}, t)$ is the effective equilibrium state distributed in space and time around $\vec{r}$ and $t$. The integral solution describes
an evolution process from non-equilibrium to equilibrium state through particle collision.

In the UGKS, with the expansion of initial distribution function and equilibrium state
\begin{equation}\label{eq:2nd-expansion}
\begin{aligned}
f_0(\vec{r}) &= f_0 + \vec{r} \cdot \frac{\partial f}{\partial \vec{r}}, \\
g^\ast(\vec{r}, t) &= g_0^\ast + \vec{r} \cdot \frac{\partial g^\ast}{\partial \vec{r}} + \frac{\partial g^\ast}{\partial t} t,
\end{aligned}	
\end{equation}
the second-order accurate flux for macroscopic flow variables across cell interface $ij$ can be constructed from the integral solution
\begin{equation} \label{eq:flux}
\begin{aligned}
\vec{F}_{ij} &= \frac{1}{\Delta t} \int_0^{\Delta t} \int \vec{u} \cdot \vec{n}_{ij} f_{ij}(t) {\vec{\psi}} {\rm d}\vec{\Xi} {\rm d}t \\
&= \int \vec{u} \cdot \vec{n}_{ij} \left[C_1 g_0^\ast
+ C_2 \vec{u} \cdot \frac{\partial g^\ast}{\partial \vec{r}} + C_3 \frac{\partial g^\ast}{\partial t} \right]{\vec{\psi}} {\rm d}\vec{\Xi} +
  \int \vec{u} \cdot \vec{n}_{ij} \left[C_4 f_0 + C_5 \vec{u} \cdot \frac{\partial f}{\partial \vec{r}} \right]{\vec{\psi}} {\rm d}\vec{\Xi} \\
&= {\vec{F}}_{ij}^{eq} + {\vec{F}}_{ij}^{fr}
\end{aligned}
\end{equation}
where ${\vec{n}_{ij}}$ is the normal vector of the cell interface, and
\begin{equation*}
\vec{\psi} = \left( 1, \vec{u}, \frac{1}{2} {\vec{u}}^2 + \frac{1}{2} {\vec{\xi}}^2 + {\varepsilon_v},
\frac{1}{2} {\vec{\xi}}^2, {\varepsilon_v} \right)^{T}.
\end{equation*}
$\vec{F}^{fr}_{ij}$ and $\vec{F}^{eq}_{ij}$ are the macroscopic fluxes in the free transport and collision processes, respectively. The integrated time coefficients are
\begin{equation*}
\begin{aligned}
C_1 &= 1 - \frac{\tau}{\Delta t} \left( 1 - e^{-\Delta t / \tau} \right) , \\
C_2 &= -\tau + \frac{2\tau^2}{\Delta t} - e^{-\Delta t / \tau} \left( \frac{2\tau^2}{\Delta t} + \tau\right) ,\\
C_3 &=  \frac12 \Delta t - \tau + \frac{\tau^2}{\Delta t} \left( 1 - e^{-\Delta t / \tau} \right) , \\
C_4 &= \frac{\tau}{\Delta t} \left(1 - e^{-\Delta t / \tau}\right), \\
C_5 &= \tau  e^{-\Delta t / \tau} - \frac{\tau^2}{\Delta t}(1 -  e^{-\Delta t / \tau}).
\end{aligned}
\end{equation*}

The UGKS updates both the gas distribution function and macroscopic flow variables under a finite volume framework.
In the UGKP method, the particle will be used to follow the evolution of gas distribution function directly and keep the finite volume version for the updates of macroscopic flow variables.
On the microscopic scale, the particle evolution follows the evolution solution in Eq.~\eqref{eq:integral-solution1}, where the particle free transport and collision will be taken into account.
On the macroscopic scale, the fluxes across the cell interface for the updates of macroscopic flow variables inside each control volume are evaluated by Eq.~\eqref{eq:flux}.

Denote a simulation particle as $P_k(m_k,\vec{r}_k,\vec{u}_k, e_{r,k}, e_{v,k})$, which represents a package of real gas molecules at location $\vec{r}_k$ with particle mass $m_k$, microscopic velocity $\vec{u}_k$, rotational energy $e_{r,k}$ and vibrational energy $e_{v,k}$. According to the integral solution, the cumulative distribution function of particle's collision is
\begin{equation*}\label{tc-distribution}
\mathcal{G}(t)=1-\exp(-t/\tau),
\end{equation*}
then the free transport time of a particle within one time step $\Delta t$ would be
\begin{equation}\label{freetime}
t_{f} = \min(-\tau \ln \eta, \Delta t),
\end{equation}
where $\eta$ is a random number uniformly distributed in $(0,1)$. In a numerical time step from $t^{n}$ to $t^{n+1}$, according to the free transport time $t_f$, the simulation particles can be categorized into collisionless particles ($t_f = \Delta t$) and collisional particles ($t_f < \Delta t$).

In the free transport process, i.e., $t < t_f$, no collisions would happen, and the particles move freely and carry the initial information. The trajectory of particle $P_k$ could be fully tracked by
\begin{equation}\label{stream}
\vec{x}_{k} = \vec{x}_{k}^{n} + \vec{u}_k t_{f,k}.
\end{equation}
During the free transport process, the effective net flux across interfaces of cell $i$ can be evaluated by
\begin{equation}\label{particleevo}
\vec{W}_{i}^{fr}=\frac{1}{\Delta t}\left( \sum_{\vec{x}_k \in \Omega_i}\vec{\phi}_k - \sum_{\vec{x}_k^n \in \Omega_i} \vec{\phi}_{k}\right),
\end{equation}
where $\vec{\phi}_{k} = (m_k, m_k \vec{u}_k, \frac{1}{2} m_k \vec{u}_k^2 + {m_k}{e_r} + {m_k}{e_v}, {m_k}{e_r}, {m_k}{e_v})^T$.
The free transport flux $\vec{F}_{ij}^{fr}$ in Eq.~\eqref{eq:flux} has been recovered by the particles' movement.

In the free transport process, the particle during the time interval $(0, t_f)$ is fully tracked.
The collisionless particles with $t_f = \Delta t$ are kept at the end of the time step.
The collisional particles with $t_f < \Delta t$ would encounter collision at $t_f$ and they are only tracked up to this moment.
Then, all collisional particles are removed, but their accumulating mass, momentum, and energy inside each cell can be still updated through the evolution of macroscopic variables. These collisional particles can be re-sampled from the updated macroscopic variables at the beginning of next time step from equilibrium state if needed.

The equilibrium flux $\vec{F}_{ij}^{eq}$ in Eq.~\eqref{eq:flux} contains three terms, i.e., $g^\ast$, $\partial_{\vec{r}} g^\ast$ and $\partial_t g^\ast$, which are only related to the equilibrium states and can be fully determined by the macroscopic flow variables.
Here we re-write Eq.~\eqref{eq:g*} as
\begin{equation*} \label{eq:g*-2}
g^\ast  = g_t
+ \frac{ g_{tr} - g_t }{Z_r}
+ \frac{ g_M - g_{tr} }{Z_v}.
\end{equation*}
The previous study\cite{xu2021rot} shows that
\begin{equation*}
	\frac{ g_{tr} - g_t }{Z_r}  = \mathcal{O}(\tau),\quad
	\frac{ g_M - g_{tr} }{Z_v} = \mathcal{O}(\tau),\quad
	\tau  \to 0.
\end{equation*}
Therefore, the terms with coefficients $C_2$ and $C_3$ in the equilibrium flux related to the spatial and temporal gradients $\partial_{\vec{r}} g^\ast$ and $\partial_t g^\ast$ would be on the order of $\mathcal{O}(\tau^2)$ or $\mathcal{O}(\tau \Delta t)$, which can be ignored in the continuum regime $\Delta t > \tau$ for recovering NS limit.

Once the Maxwellian distribution and its derivatives around the cell interface are determined, the equilibrium flux $\vec{F}^{eq}_{ij}$ can be obtained by
\begin{equation}\label{eq:Feq}
\vec{F}^{eq}_{ij} =
\int \vec{u} \cdot \vec{n}_{ij} \left[C_1 g_0^\ast
+ C_2 \vec{u} \cdot \frac{\partial g_t}{\partial \vec{r}}
+ C_3 \frac{\partial g_t}{\partial t} \right]{\vec{\psi}} {\rm d}\vec{\Xi}.
\end{equation}
The macroscopic variables for the determination of equilibrium state $g_0^\ast$ at cell interface $ij$ are coming from the
colliding particles from both sides of the cell interface
\begin{equation*}
\vec{W}_{ij}=\int  \left[ g_{t}^l H[\bar{u}_{ij}] + g_{t}^r (1 - H[\bar{u}_{ij}])\right] \vec{\psi} {\rm d} \vec{\Xi},
\end{equation*}
where $\bar{u}_{ij}=\vec{u}\cdot\vec{n}_{ij}$, and $H[x]$ is the Heaviside function.
The gradient of the equilibrium state is obtained from the gradient of macroscopic flow variables ${\partial \vec{W}_{ij}}/{\partial {\vec{r}}}$,
see \ref{sec:appendix-1}.
In this study, the spatial reconstruction of macroscopic flow variables is carried out by the least-square method with Venkatakrishnan limiter \cite{venkatakrishnan1995convergence}. As to the temporal gradient, the compatibility condition on Eq.~\eqref{eq:BGK-vib-2}
\begin{equation*}
	\int {(g_t - f) \vec{\psi} {\rm d}\Xi} = 0
\end{equation*}
is employed to give
\begin{equation*}
\frac{\partial\vec{W}_{ij}}{\partial t}
= - \int \vec{u} \cdot \frac{\partial g_t}{\partial \vec{r}} \vec{\psi} {\rm d}\vec{\Xi}.
\end{equation*}
Correspondingly, the temporal gradient of equilibrium state $\partial_t g_t$ can be evaluated from the above ${\partial\vec{W}_{ij}}/{\partial t}$. With $g_0^\ast$, $\partial_{\vec{r}} g_t$ and $\partial_t g_t$, the equilibrium flux $\vec{F}_{ij}^{eq}$ can be fully determined.

During the collision process, inelastic collisions will happen, which lead to energy exchange between the degrees of freedom of molecular translation, rotation and vibration.
As a result, source terms appear in the macroscopic governing equations, i.e.,
\begin{equation} \label{eq:source}
\vec{S} = \int_{t^n}^{t^{n+1}} \frac{g^{*}-f}{\tau} \vec{\psi}
{\rm{d}} \vec{\Xi} {\rm{d}} t
= \int_{t^{n}}^{t^{n+1}} \vec{s} {\rm{d}} t,
\end{equation}
where $\vec{s}$ can be expressed as
\begin{equation*}
{\vec{s}} = \left(0,\vec{0},0,
		\frac{\rho E_{r}^{tr} - \rho E_r}{Z_r \tau} + \frac{\rho E_r^M - \rho E_{r}^{tr}}{Z_v \tau},
		\frac{\rho E_v^M - \rho E_v}{Z_v \tau} \right)^T.
\end{equation*}
The intermediate equilibrium energy $\rho E_r^{tr}$ is determined under the assumption $\lambda_r = \lambda_t = \lambda_{tr}$, and thus
\begin{equation}\label{eq:rhoErTR-rhoEvTR}
	\rho E_r^{tr} = \frac{K_r\rho}{4 \lambda_{tr}},\quad
	\quad \lambda_{tr} = 	
	\frac{( K_r + 3 )\rho }{ 4 ( \rho E - \frac{1}{2} \rho {\vec{U}}^2 - \rho E_v )}.
\end{equation}
The rotational and vibrational energy at the full equilibrium state $\rho E_r^M$ and $\rho E_v^M$ are determined under the assumption $\lambda_v = \lambda_r = \lambda_t = \lambda_{M}$, and thus
\begin{equation}\label{eq:rhoErM-rhoEvM}
	\rho E_r^M = \frac{K_r \rho}{4 \lambda_M},\quad
	\rho E_v^M = \frac{K_v(\lambda_M) \rho}{4 \lambda_M}\quad
	{\text{ and }}
	\quad {\lambda_{M}} = 	
	\frac{\left(K_v(\lambda_M) + K_r + 3\right)\rho}{4\left(\rho E - \frac{1}{2} \rho {\vec{U}}^2\right)}.
\end{equation}

With consideration of numerical stability, the source term is usually treated in an implicit way, such as the trapezoidal rule for rotational and vibrational energies
\begin{equation*}
\begin{aligned}
S_r &= \frac{\Delta t}{2} \left(s_r^n + s_r^{n+1}\right) \\
 	&= \frac{\Delta t}{2} \left[\frac{(\rho E_r^{tr})^n - (\rho {E_r})^n}{Z_r \tau} + \frac{(\rho E_r^M)^n - (\rho E_r^{tr})^n}{Z_v \tau}\right] \\
    &+\frac{\Delta t}{2} \left[\frac{(\rho E_r^{tr})^{n+1} - (\rho {E_r})^{n+1}}{Z_r \tau} + \frac{(\rho E_r^M)^{n+1} - (\rho E_r^{tr})^{n+1}}{Z_v \tau}\right],\\
S_v &=\frac{\Delta t}{2} \left(s_v^n + s_v^{n+1}\right) = \frac{\Delta t}{2} \left[\frac{(\rho E_v^M)^n - (\rho E_v)^n}{Z_v \tau} + \frac{(\rho E_v^M)^{n+1} - (\rho E_v)^{n+1}}{Z_v \tau}\right].
\end{aligned}	
\end{equation*}

\subsection{Updates of macroscopic variables and discrete particles}\label{subsec:update}

Under the finite volume framework, according to the conservation law, the updates of macroscopic variables can be written as
\begin{equation}\label{eq:update}
{\vec {W}}_i^{n + 1}
=
{\vec {W}}_i^n
- \frac{\Delta t}{\Omega_i}
\sum\limits_{j \in N(i)} {{\vec{F}}^{eq}_{ij}{\cal A}_{ij}}
+ \frac{\Delta t}{\Omega_i} \vec{W}^{fr}_i
+ {\vec{S}}_i,
\end{equation}
where ${\vec{W}}^{fr}_{i}$ is the net free streaming flow of cell $i$ calculated by particle tracking in the free transport process in Eq.~\eqref{particleevo}, the equilibrium flux ${\vec{F}}^{eq}_{ij}$ is evaluated from macroscopic flow variables and their gradients in Eq.~\eqref{eq:Feq}, and the source term ${\vec{S}}_i$ in Eq.~\eqref{eq:source} has values only for the last two components of macroscopic flow variables ${\vec{W}}_i$, indicating energy exchange between molecular translation, rotation and vibration.

Based on the fluxes, the conservative flow variables $\rho^{n+1}, (\rho \vec{U})^{n+1}, (\rho E)^{n+1}$ can be updated directly. Then $\lambda_M^{n+1}$, $(\rho E_r^M)^{n+1}$ and $(\rho E_v^M)^{n+1}$ can be obtained from the updated conservative flow variables by Eq.~\eqref{eq:rhoErM-rhoEvM}, and the vibrational energy $(\rho E_v)^{n+1}$ with implicit source term can be solved in an explicit way without iterations
\begin{equation} \label{eq:update-ErEv2}
(\rho E_v)^{n+1} = \left(1 + \frac{\Delta t}{2 Z_v \tau}\right)^{-1} \left[(\rho {E_v})^\dagger +
\frac{\Delta t}{2} \left(s_v^n + \frac{(\rho E_v^M)^{n+1}}{Z_v \tau} \right) \right].
\end{equation}
Similarly, $\lambda_{tr}^{n+1}$ and $(\rho E_r^{tr})^{n+1}$ can be obtained by Eq.~\eqref{eq:rhoErTR-rhoEvTR} with the updated $(\rho E_v)^{n+1}$, then the rotational energy $(\rho E_r)^{n+1}$ can be renewed by
\begin{equation} \label{eq:update-ErEv1}
(\rho {E_r})^{n+1} = \left(1 + \frac{\Delta t}{2 Z_r \tau}\right)^{-1}
\left[ (\rho {E_r})^\dagger + \frac{\Delta t}{2} \left(s_r^n  + \frac{(\rho E_r^{tr})^{n+1}}{Z_r \tau} + \frac{(\rho E_r^M)^{n+1} - (\rho E_r^{tr})^{n+1}}{Z_v \tau}\right) \right].
\end{equation}
Here $(\rho E_v)^\dagger$ and $(\rho E_r)^\dagger$ are the  updated intermediate vibrational and rotational energies with inclusion of the fluxes only. It would be noticed that in Eq.~\eqref{eq:rhoErM-rhoEvM} the vibrational degrees of freedom $K_v(\lambda_M)$ rely on the full equilibrium temperature $\lambda_M$. The explicit expression of $\lambda_M$ cannot be given due to the complexity of function $K_v(\lambda_M)$. In the current study, $\lambda_M$ is computed by iterations
\begin{equation*}
\lambda_M^{i + 1} = \frac{\left[ K_v(\lambda_M^i) + K_r + 3\right] \rho}{4 \left(\rho E - \frac{1}{2} \rho \vec{U}^2 \right)} \quad \text{with} \quad
\lambda_M^0 = \frac{\left(K_r + 3\right)\rho }{4\left(\rho E - \frac{1}{2} \rho \vec{U}^2\right)}.
\end{equation*}
Numerical tests show that the relative error can approach to $\mathcal{O}(10^{-16})$ after $5 \sim 6$ iterations.

Substitute Eq.~\eqref{eq:2nd-expansion} into the integral solution Eq.~\eqref{eq:integral-solution1} of kinetic model equation, the time evolution of distribution function along the characteristic line can be given as
\begin{equation*}
	f(\vec{r},t) = (1-e^{-t/\tau}) g^\ast(\vec{r}^\prime, t^\prime) + e^{-t/\tau} f_0(\vec{r} - \vec{u} t),
\end{equation*}
where
\begin{equation*}
	\vec{r}^\prime = \vec{u} \left(\frac{t e^{-t/\tau}}{1-e^{-t/\tau}} -\tau\right), \quad t^\prime = \left( \frac{t e^{-t/\tau}}{1-e^{-t/\tau}}-\tau \right) + t.
\end{equation*}
It indicates that the collisional particles will follow the near-equilibrium state $g^\ast(\vec{r}^\prime,t^\prime)$ after collision within the time step $t_f < \Delta t$. With the updated macroscopic flow variables, these untracked collisional particles within the time  $t \in (t_f, \Delta t)$ can be re-sampled from the hydro-particle macroscopic  quantities
\begin{equation}\label{eq:hydro-kp}
	  \vec{W}_i^{h, n+1}
	= \vec{W}_i^{n+1} - \vec{W}_i^{p, n+1}
	= \vec{W}_i^{n+1} - \frac{1}{\Omega_i} \sum_{x_k^{n+1} \in\Omega_i} \vec{\phi}_k,
\end{equation}
where $\vec{W}_i^{p, n+1}$ is from the the collisionless particles remaining in cell $i$.
With the macroscopic quantities and the form of effective equilibrium state $g^\ast$, the corresponding particles can be generated.
Details of sampling from a given distribution function are provided in \ref{sec:sampling}.

The free transport and collision processes for both microscopic discrete particles and macroscopic flow variables have been described above.
Here, we give a summary of the procedures of the UGKP method. Following the illustration in \cite{zhu2019unified}, the algorithm of UGKP method for diatomic gases with molecular translation, rotation and vibration can be
summarized as follows:

\begin{figure}[H]
	\centering
	\begin{subfigure}[b]{0.24\textwidth}
		\includegraphics[width=\textwidth]{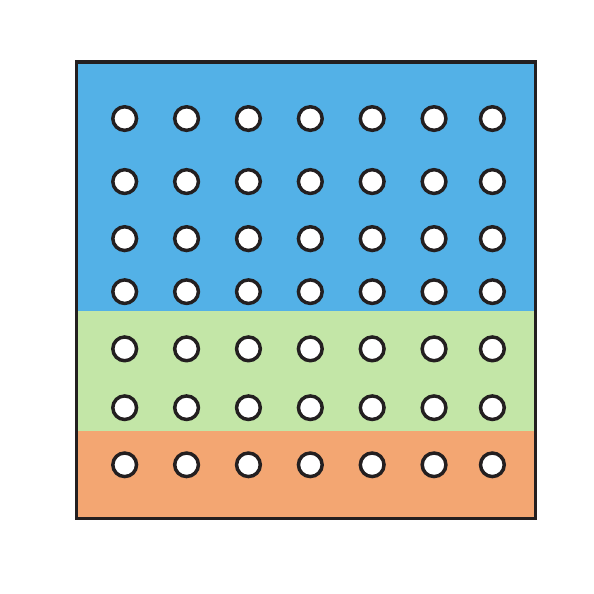}
		\caption{}
		\label{ugkp1}
	\end{subfigure}
	\begin{subfigure}[b]{0.24\textwidth}
		\includegraphics[width=\textwidth]{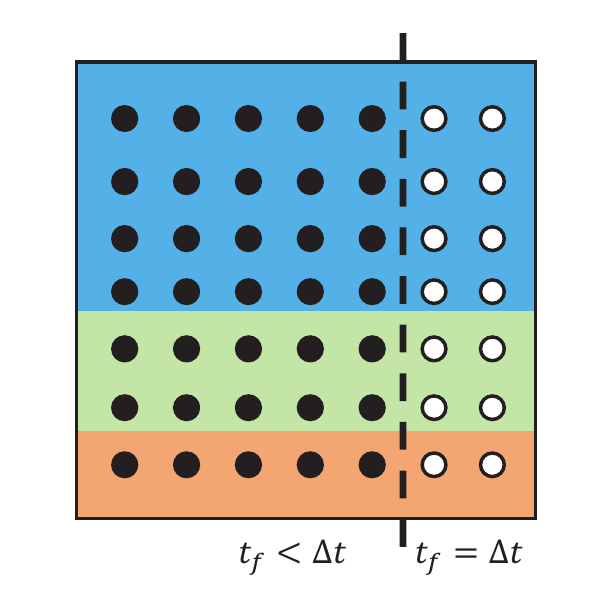}
		\caption{}
		\label{ugkp2}
	\end{subfigure}
	\begin{subfigure}[b]{0.24\textwidth}
		\includegraphics[width=\textwidth]{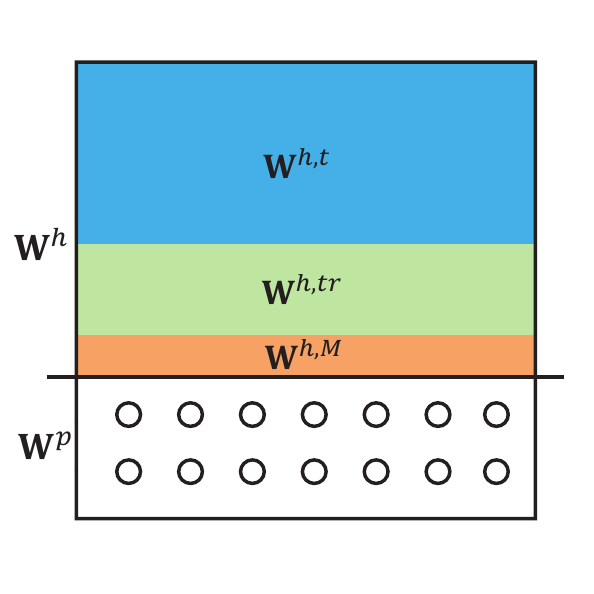}
		\caption{}
		\label{ugkp3}
	\end{subfigure}
	\begin{subfigure}[b]{0.24\textwidth}
		\includegraphics[width=\textwidth]{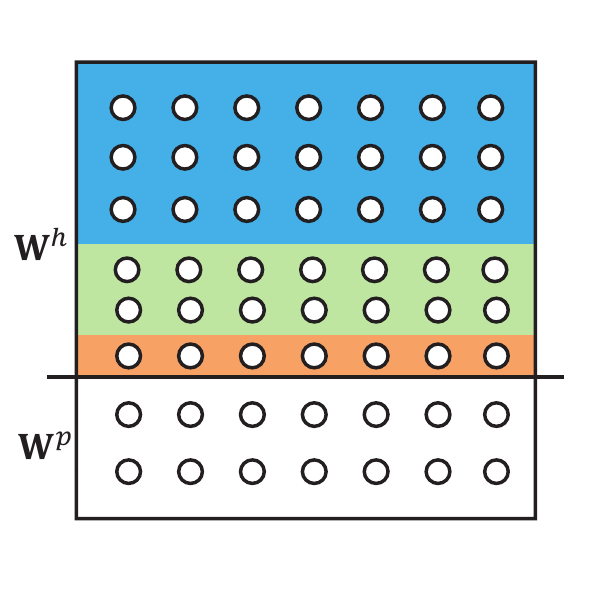}
		\caption{}
		\label{ugkp4}
	\end{subfigure}
	\caption{Diagram to illustrate the composition of the particles during time evolution in the UGKP method. (a) Initial field, (b) classification of the collisionless particles (white circle) and collisional particles (solid circle) according to the free transport time $t_f$, (c) update solution at the macroscopic level, and (d) update solution at the microscopic level.}
	\label{fig:ugkp}
\end{figure}

\begin{description}
	\item[Step 1]
	For the initialization, sample particles from the given initial conditions \ref{sec:sampling} as shown in Fig.~\ref{fig:ugkp}(a).
	\item[Step 2]
		Generate the free transport time $t_f$ for each particle by Eq.~\eqref{freetime}, and classify the particles into collisionless particles (white circles in Fig.~\ref{fig:ugkp}(b)) and collisional ones (solid circles in Fig.~\ref{fig:ugkp}(b)).
		Stream the particles for free transport time by Eq.~\eqref{stream}, and evaluate the net free streaming flow $\vec{W}^{fr}_i$ by Eq.~\eqref{particleevo}.
	\item[Step 3]
	    Reconstruct macroscopic flow variables and compute the equilibrium flux $\vec{F}^{eq}_{ij}$ by Eq.~\eqref{eq:Feq}.		
	\item[Step 4]
		Update the macroscopic flow variables $\vec{W}_i$ by Eqs~\eqref{eq:update}, \eqref{eq:update-ErEv2} and \eqref{eq:update-ErEv1}.
		Obtain the updated hydro-particle macroscopic  quantities of collisional particles $\vec{W}^h_i$ by extracting the macro-quantities of collisionless particles
		$\vec{W}^p_i$ from the total flow variables $\vec{W}_i$ in Eq.~\eqref{eq:hydro-kp} as shown in Fig.~\ref{fig:ugkp}(c).		
	\item[Step 5]
		Delete the collisional particles at $t_f$ and re-sample these particles from the updated hydro-particle macroscopic  variables $\vec{W}^h_i$ as shown in Fig.~\ref{fig:ugkp}(d), as the initial state in Fig.~\ref{fig:ugkp}(a) at the beginning of next time step.
	\item[Step 6] Go to Step 2. Continue time step evolution or stop the calculation at finishing time.
\end{description}

\section{Unified gas-kinetic wave-particle method}\label{sec:wp}

In UGKP method, based on the updated hydro-particle macroscopic variables $\vec{W}^h_i$  of collisional particles, these particles
will be re-sampled from equilibrium state at the beginning of next time step. However, some of these re-sampled particles will get collision in the next time step and get eliminated again.
Therefore, in the unified gas-kinetic wave-particle (UGKWP) method, only free transport particles in the next time step will be re-sampled from $\vec{W}^h_i$. In the continuum regime at very small Knudsen number, it is possible that no free particles will get re-sampled.

\begin{figure}[H]
	\centering
	\includegraphics[width=0.9\textwidth]{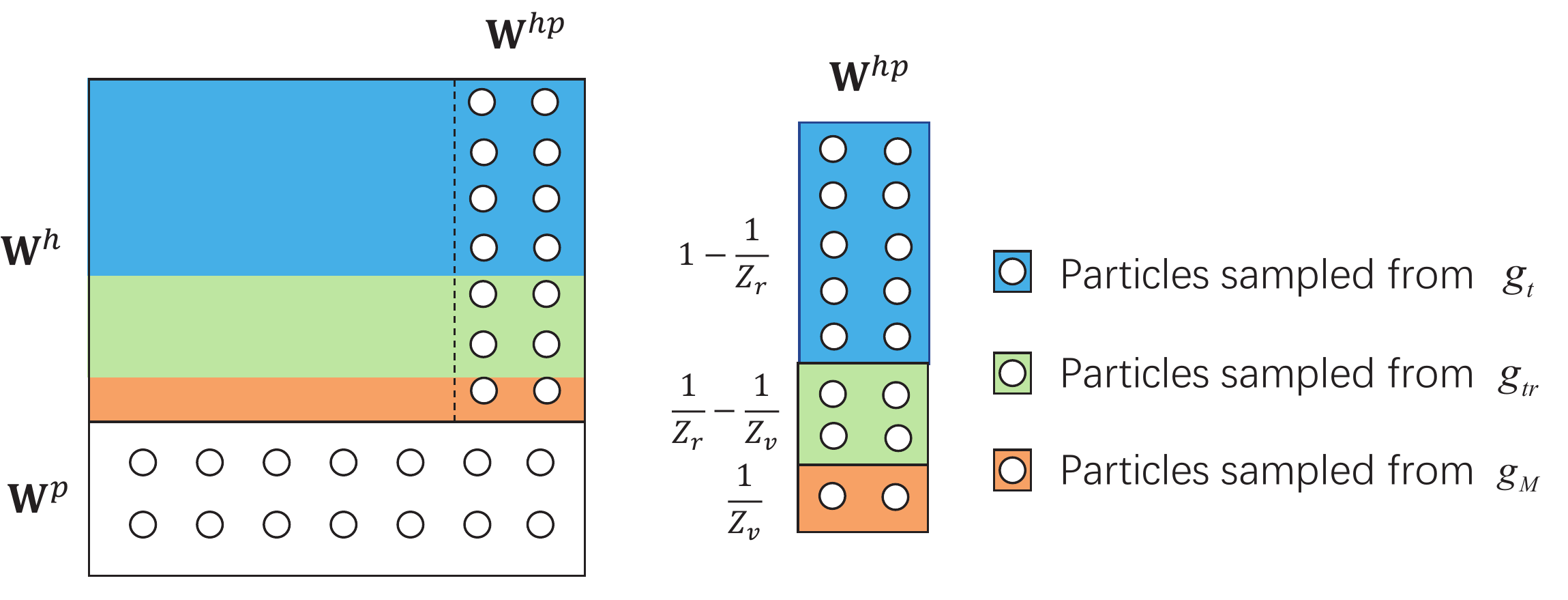}
	\caption{Sampling particles for the UGKWP method with vibrational model.}
	\label{fig:sample-Whp}
\end{figure}

The collisionless particles with $t_f = \Delta t$ will be sampled from  $\vec{W}^h_i$.
According to the integral solution, the collisionless particles will take a fraction of $\vec{W}^h_i$ by the amount
\begin{equation*}
{\vec{W}}^{hp}_i = e^{\frac{-\Delta t}{\tau}}{\vec{W}}^{h}_{i}.
\end{equation*}
As shown in Fig.~\ref{fig:sample-Whp}, there is no need to sample particles from the hydrodynamic part $(\vec{W}^h_i -{\vec{W}}^{hp}_i)$.
The free transport flux from these un-sampled collisional particles can be evaluated analytically
\begin{equation*}\label{Ff1}
\vec{F}^{fr,h}_{ij}=\int
\vec{u}\cdot\vec{n}_{ij}
\left[
C_4^\prime g^\ast_0
+ C_5^\prime \vec{u} \cdot \frac{\partial g_t}{\partial \vec {r}}\right]
\vec{\psi} {\rm{d}} \vec{\Xi} ,
\end{equation*}
where
\begin{equation*}
\begin{aligned}
C_4^\prime &= \frac{\tau}{\Delta t} \left(1 - e^{-\Delta t / \tau}\right)
            -  e^{-\Delta t / \tau}, \\
C_5^\prime &= \tau  e^{-\Delta t / \tau}
- \frac{\tau^2}{\Delta t}(1 -  e^{-\Delta t / \tau})
+ \frac12\Delta t e^{-\Delta t / \tau}.
\end{aligned}
\end{equation*}
Then, the update of macroscopic flow variables in the UGKWP method becomes
\begin{equation}\label{eq:update-Wp}
	{\vec{W}}_i^{n + 1}
	= {\vec{W}}_i^n
		- \frac{\Delta t}{\Omega_i}
		  \sum\limits_{j \in N(i)}  {\vec{F}}^{eq}_{ij}{\cal A}_{ij}
		- \frac{\Delta t}{\Omega_i}
		  \sum\limits_{j \in N(i) } {\vec{F}}^{fr,h}_{ij} {\cal A}_{ij}
		+ \frac{\Delta t}{\Omega _i} \vec{W}^{fr,p}_{i}
		+ \vec{S}_i.
\end{equation}

\begin{figure}[H]
	\centering
	\begin{subfigure}[b]{0.24\textwidth}
		\includegraphics[width=\textwidth]{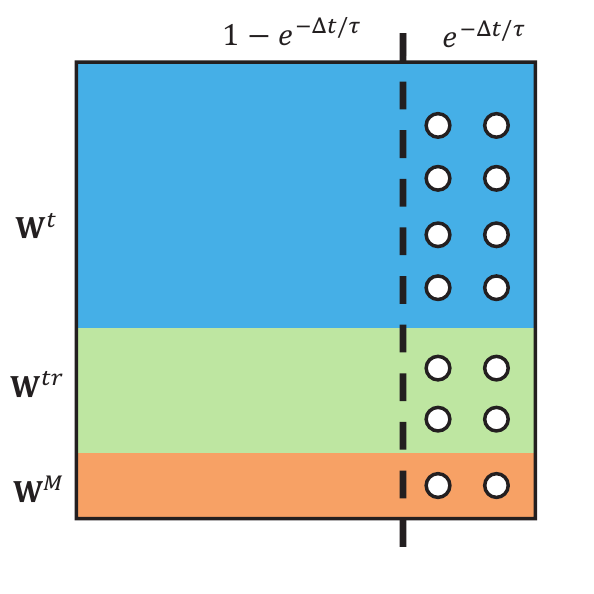}
		\caption{}
		\label{ugkwp1}
	\end{subfigure}
	\begin{subfigure}[b]{0.24\textwidth}
		\includegraphics[width=\textwidth]{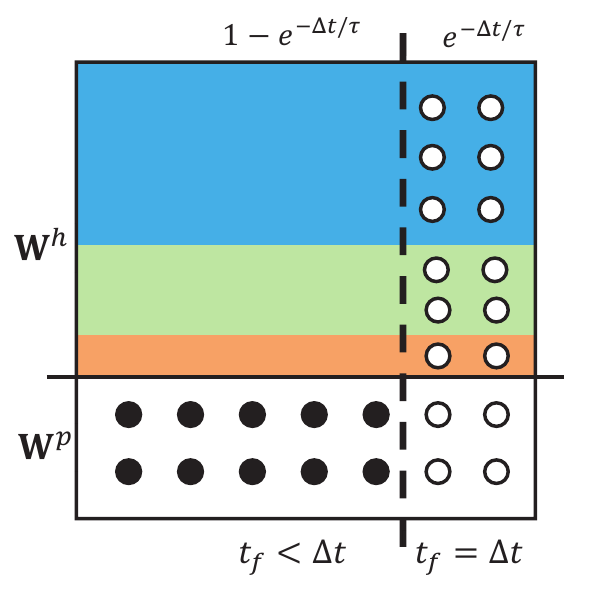}
		\caption{}
		\label{ugkwp2}
	\end{subfigure}
	\begin{subfigure}[b]{0.24\textwidth}
		\includegraphics[width=\textwidth]{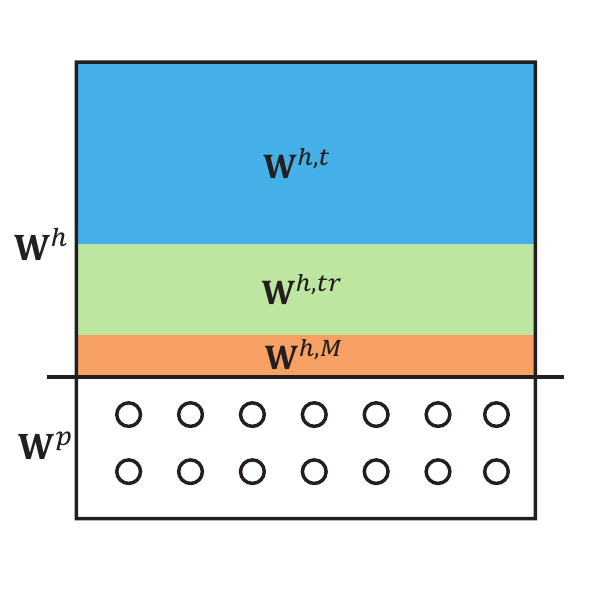}
		\caption{}
		\label{ugkwp3}
	\end{subfigure}
	\begin{subfigure}[b]{0.24\textwidth}
		\includegraphics[width=\textwidth]{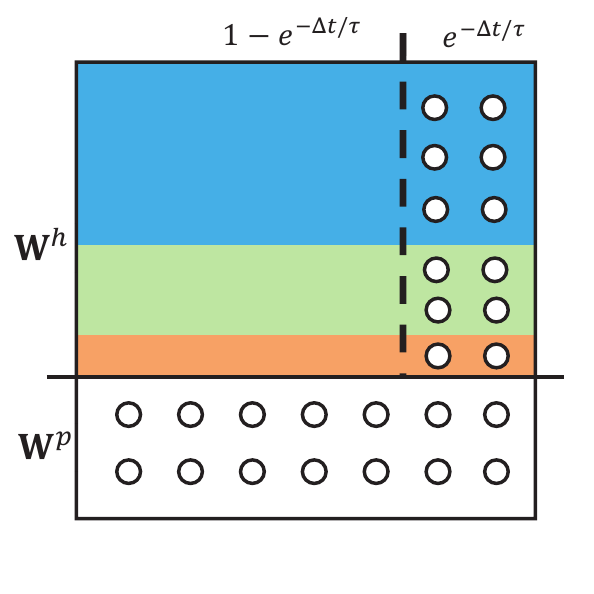}
		\caption{}
		\label{ugkwp4}
	\end{subfigure}
	\caption{Diagram to illustrate the composition of the particles during time evolution in the UGKWP method. (a) Initial field, (b) classification of the collisionless and collisional particles for $\vec{W}^p_i$, (c) update on the macroscopic level, and (d) update on the microscopic level.}
	\label{fig:ugkwp}
\end{figure}

The algorithm of the UGKWP method for diatomic gases can be summarized as follows.
\begin{description}
	\item[Step 1]
		For the initialization, sample collisionless particles from $\vec{W}_i^{hp}$ with $t_f = \Delta t$ by \ref{sec:sampling}
		as shown in Fig.~\ref{fig:ugkwp}(a).
		For the first step, $\vec{W}^h_i = \vec{W}_i^{n = 0}$.		
	\item[Step 2]
		Generate the free transport time $t_f$ by Eq.~\eqref{freetime} for the remaining particles from previous step evolution with total amount $\vec{W}_i^p$, and classify the particles into collisionless particles (white circles in Fig.~\ref{fig:ugkwp}(b)) and collisional ones (solid circles in Fig.~\ref{fig:ugkwp}(b)).
		Stream the particles for free transport time by Eq.~\eqref{stream}, and evaluate the net free streaming flow $\vec{W}^{fr}_i$ by Eq.~\eqref{particleevo}.
	\item[Step 3]
		Reconstruct macroscopic flow variables and compute the free transport flux of collisional particles $\vec{F}_{ij}^{fr,h}$ by Eq.~\eqref{Ff1} and the equilibrium flux $\vec{F}^{eq}_{ij}$ by Eq.~\eqref{eq:Feq}.			
	\item[Step 4]
		Update the macroscopic flow variables $\vec{W}_i$ by Eqs~\eqref{eq:update-Wp}, \eqref{eq:update-ErEv2}, and \eqref{eq:update-ErEv1}.
		Obtain the updated macroscopic quantities for collisional particles $\vec{W}^h_i$ by extracting the macro-quantities of collisionless particles
		$\vec{W}^p_i$ from the total flow variables $\vec{W}_i$ in Eq.~\eqref{eq:hydro-kp} as shown in Fig.~\ref{fig:ugkwp}(c).	
	\item[Step 5]
	Delete the collisional particles at $t_f$ ($t_f < \Delta t $).
Re-sample the collisionless particles from $\vec{W}^{hp}_i$ with $t_f = \Delta t$ at the beginning of next time step, as shown in Fig.~\ref{fig:ugkwp}(d).
	\item[Step 6]
	Go to Step 2, continue time evolution or stop the calculation.
\end{description}

The UGKP method uses particles to represent the gas distribution function. However, the UGKWP method adopts a hybrid formulation of wave and particles to recover the gas distribution function. The wave representation of the equilibrium part can be described by the corresponding macroscopic flow variables without sampling these particles explicitly. The non-equilibrium part is represented by surviving particles. It is realized that in the rarefied regime, the UGKWP method is dominated by particle evolution, which results in a particle method.
While in the continuum regime, the UGKWP method is  mainly about the evolution of macroscopic variables, and the scheme becomes a hydrodynamic NS solver, the so-called gas-kinetic scheme (GKS) \cite{xu2001}. Therefore, the UGKWP achieves much better computational efficiency and lower memory consumption than the purely particle methods in the transition and continuum flow regimes, and gives more accurate physical solutions than the NS solvers in the non-equilibrium regime.

\section{Numerical Validation}\label{sec:test}

In this section, the UGKWP method with molecular vibration (WP-vib) will be used in many test cases.
Since most of the cases are external flow, the determination of the initial condition of free stream at different Knudsen number will be provided here first.
For a specific gas, the density in the free stream corresponding to a given Knudsen number is
\begin{equation*}
	\rho = \frac{4\alpha(5-2\omega)(7-2\omega)}{5(\alpha+1)(\alpha+2)}\sqrt{\frac{m}{2\pi k_B T}}\frac{\mu}{L_{ref}{\rm Kn}},
\end{equation*}
where $m$ is the molecular mass and $L_{ref}$ is the reference length to define the Knudsen number. The dynamic viscosity is calculated from the translational temperature by the power law
\begin{equation*}\label{eq:pow-law}
	\mu  = \mu_{ref} \left( \frac{T}{T_{ref}} \right)^{{\omega}},
\end{equation*}
where $\mu_{ref}$ is the reference dynamic viscosity at the temperature $T_{ref}$.

In the tests, diatomic gas of nitrogen gas is employed with molecular mass $m=4.65\times 10^{-26}$ kg, $\alpha=1.0$, $\omega=0.74$, and the reference dynamic viscosity $\mu_{ref} = 1.65\times 10^{-5}$ ${\rm Nsm^{-2}}$ at the temperature $T_{ref} = 273$ K. For non-dimensional cases, the freestream or upstream values are used to non-dimensionalize the flow variables, i.e.,
\begin{equation*}
	\begin{aligned}
	\rho_0 &= \rho_\infty, \quad U_0 = \sqrt{2 k_B T_{\infty} / m}, \\
	T_0 &= T_\infty, \quad \text{or} \quad p_0 = p_\infty.
	\end{aligned}
\end{equation*}
In addition, according to the reference \cite{tumuklu2016particle}, the vibrational collision number can be evaluated by
\begin{equation}\label{eq:Zv-tumuklu}
	Z_v
	=
	\frac{5}{ 5 + K_v(\lambda_v) }
	\frac{C_1}{ T_t^\omega }
	\exp\left( C_2 T_t^{-1/3} \right), 	
\end{equation}
and the rotational collision number is computed by
\begin{equation} \label{eq:Zr-tumuklu}
	Z_r = \frac{Z_v}{Z_v + Z_r^{DSMC}},
\end{equation}
with
\begin{equation*}
	Z_r^{DSMC}
	= \frac{3}{5}
	\frac{ Z_r^\infty }
	     { 1 + ( \sqrt{\pi} / 2 ) \left( \sqrt{ T^\ast / T_t } \right) + ( T^\ast / T_t ) ( { \pi^2/4 + \pi } )},
\end{equation*}
where $C_1 = 6.5$ and $C_2 = 220.0$ are adopted.

\subsection{Sod Tube}
The Sod shock tube problem is computed at different Knudsen numbers to verify the capability of the UGKWP method for simulating the continuum and rarefied flows. The non-dimensional initial condition is
\begin{equation*}
	(\rho, U, V, W, p)= \begin{cases}(1,0,0,0,1), & 0<x<0.5, \\ (0.125,0,0,0.1), & 0.5<x<1.\end{cases}
\end{equation*}
The spatial discretization is carried out by a three-dimensional structured mesh with $100 \times 5 \times 5$ uniform cells. The inlet and outlet of the tube are treated as far field,
and the side walls are set as symmetric planes. The Courant--Friedrichs--Lewy (CFL) number is taken as 0.5. Constant values of $Z_r=3.5$ and $Z_v = 10$ are used for all cases. The results at the time $t = 0.12$ are investigated.

\begin{figure}[H]
	\centering
	\subfloat[]{\includegraphics[width=0.33\textwidth]{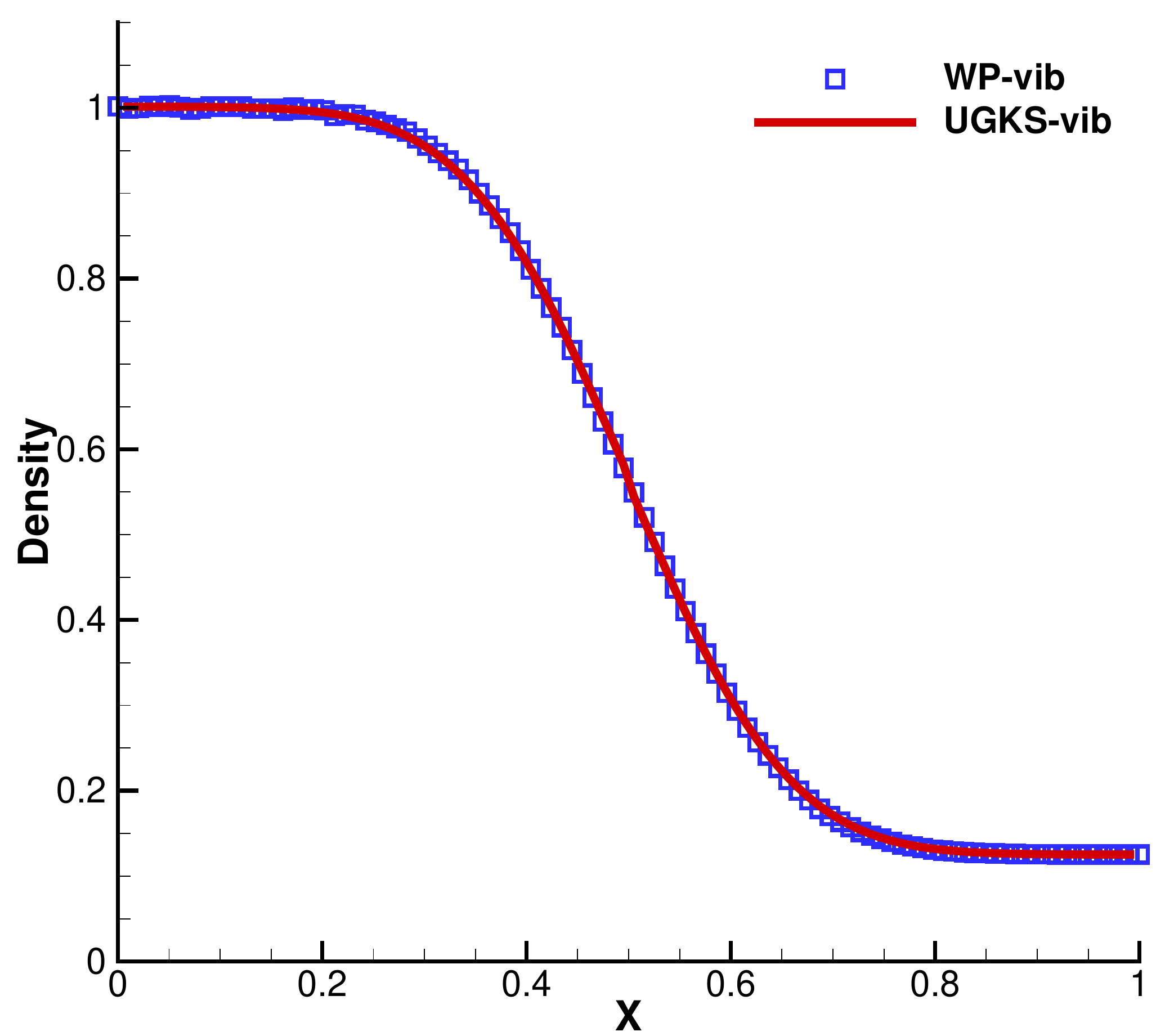}}
	\subfloat[]{\includegraphics[width=0.33\textwidth]{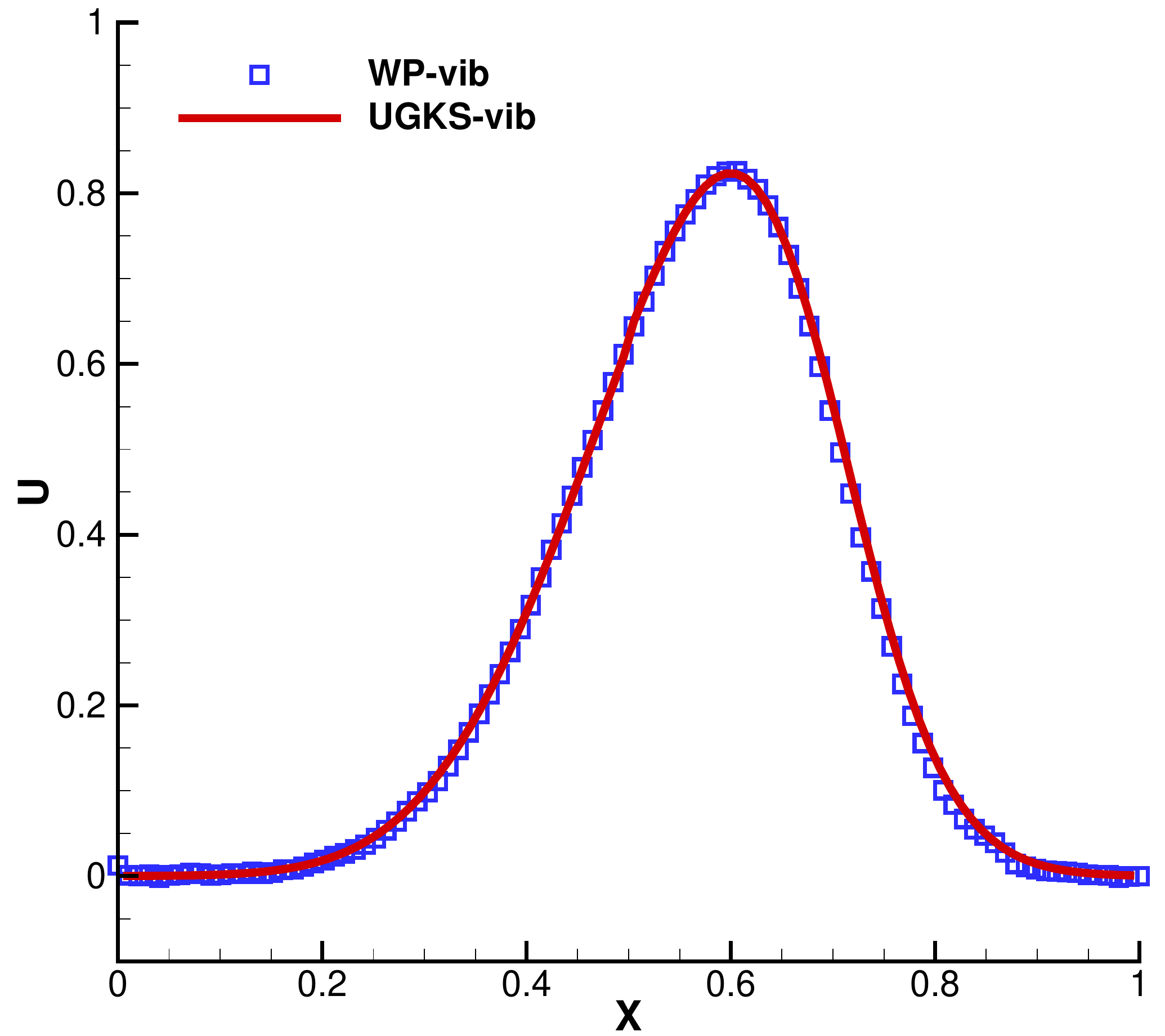}}
	\subfloat[]{\includegraphics[width=0.33\textwidth]{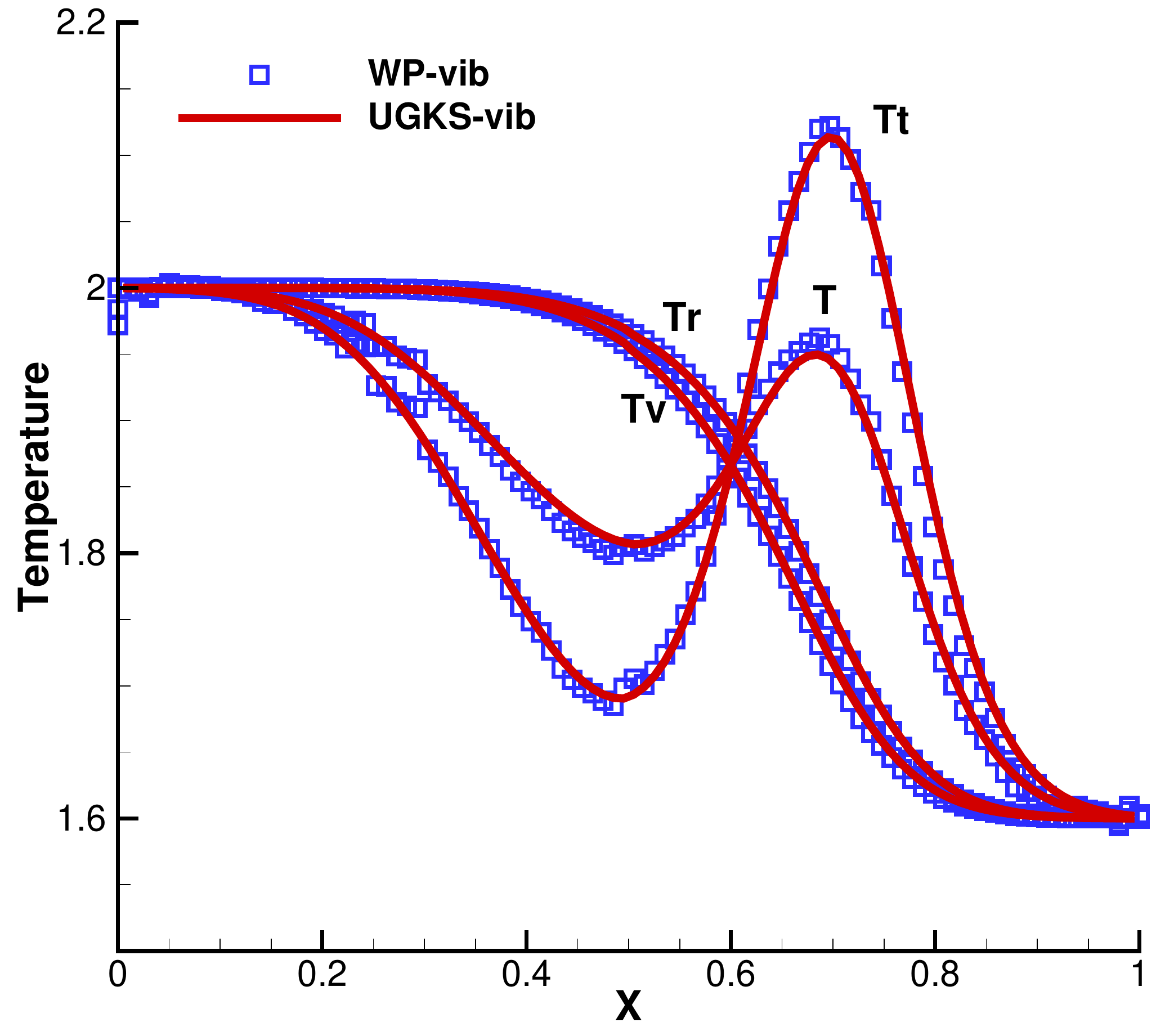}}
	\caption{Sod tube at ${\rm Kn} = 10$. (a) Density, (b) velocity, and (c) temperatures.}
	\label{fig:sod-kn10}
\end{figure}

\begin{figure}[H]
	\centering
	\subfloat[]{\includegraphics[width=0.33\textwidth]{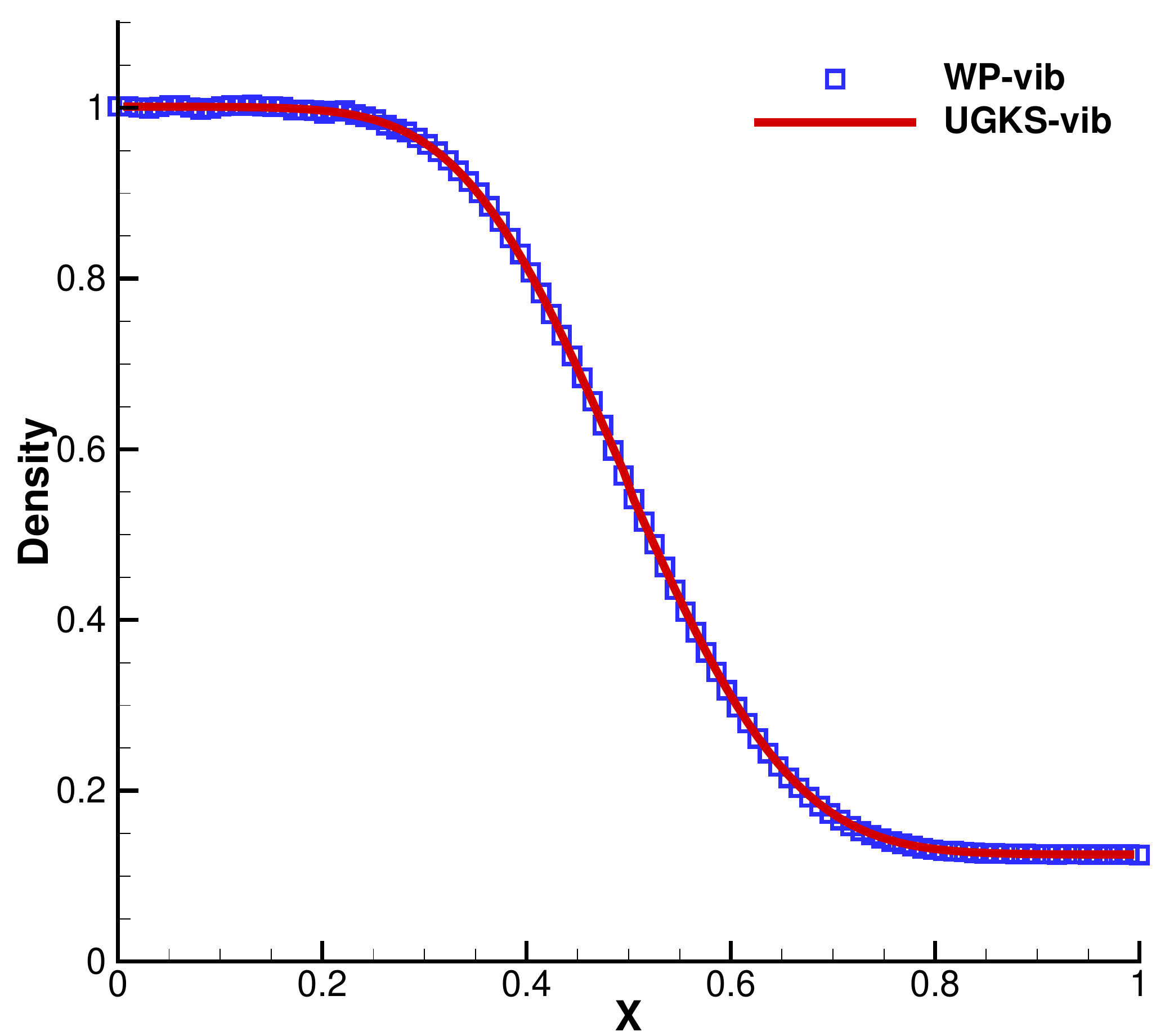}}
	\subfloat[]{\includegraphics[width=0.33\textwidth]{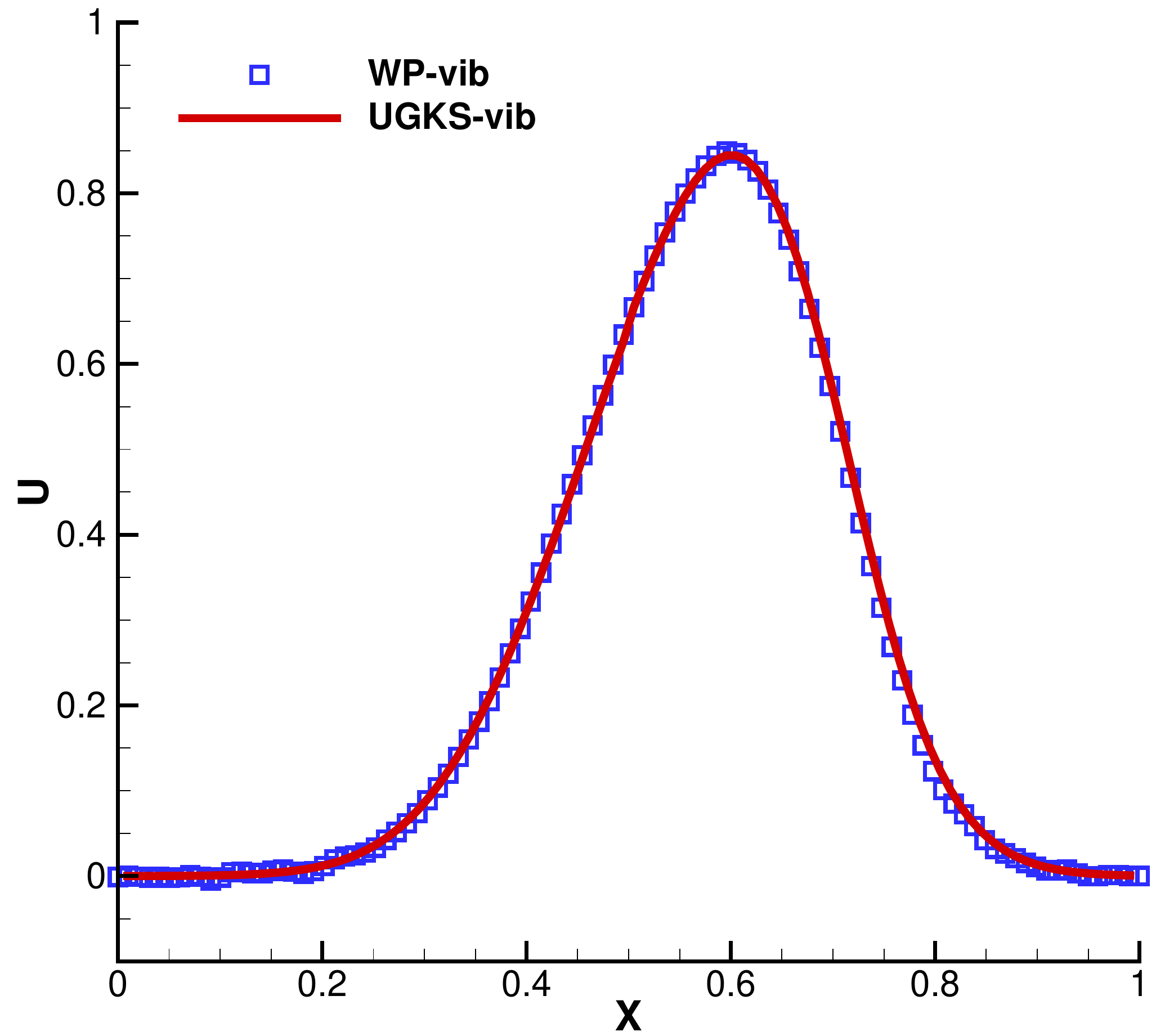}}
	\subfloat[]{\includegraphics[width=0.33\textwidth]{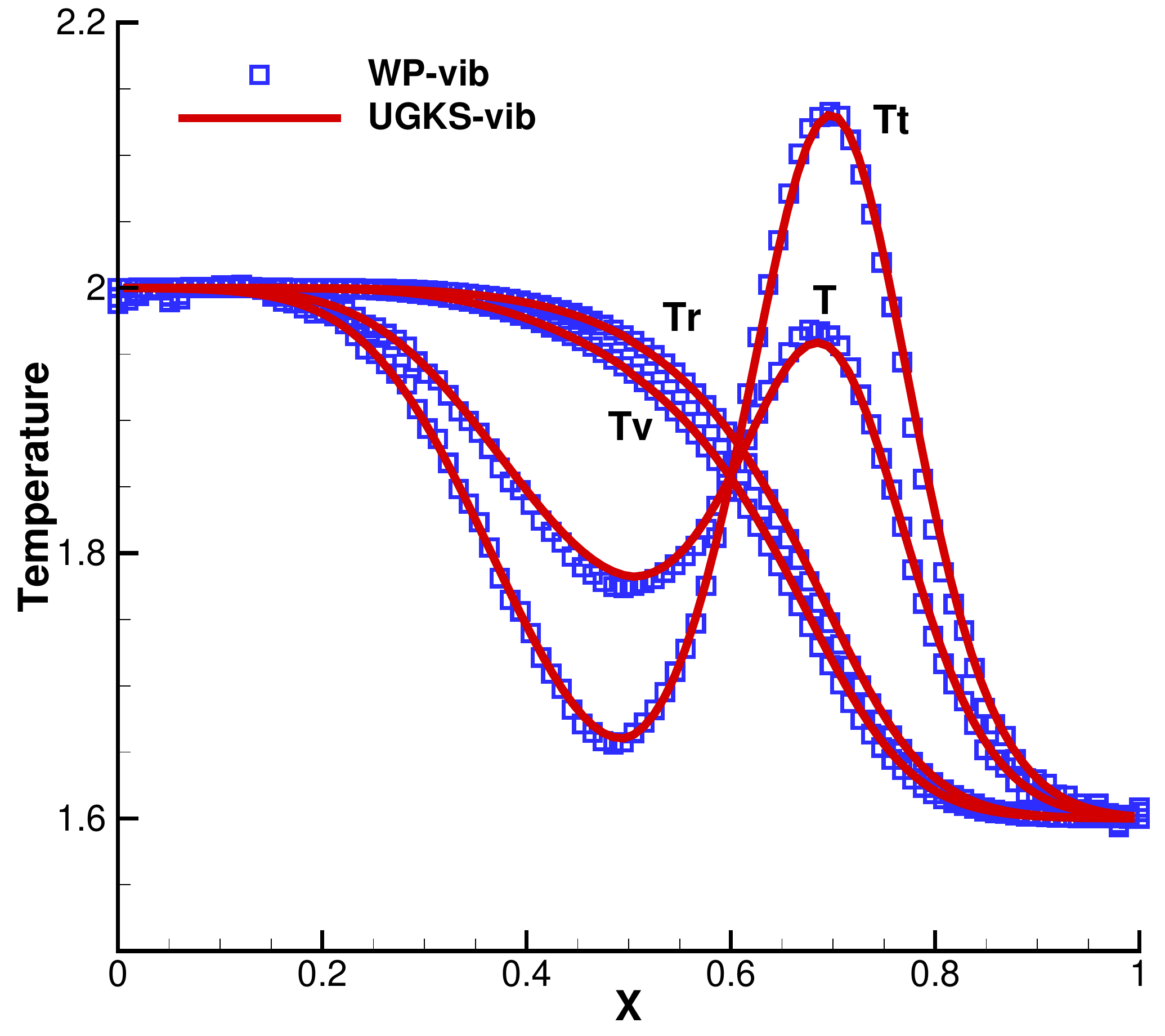}}
	\caption{Sod tube at ${\rm Kn} = 0.1$. (a) Density, (b) velocity, and (c) temperatures.}
	\label{fig:sod-kn1e-1}
\end{figure}

\begin{figure}[H]
	\centering
	\subfloat[]{\includegraphics[width=0.33\textwidth]{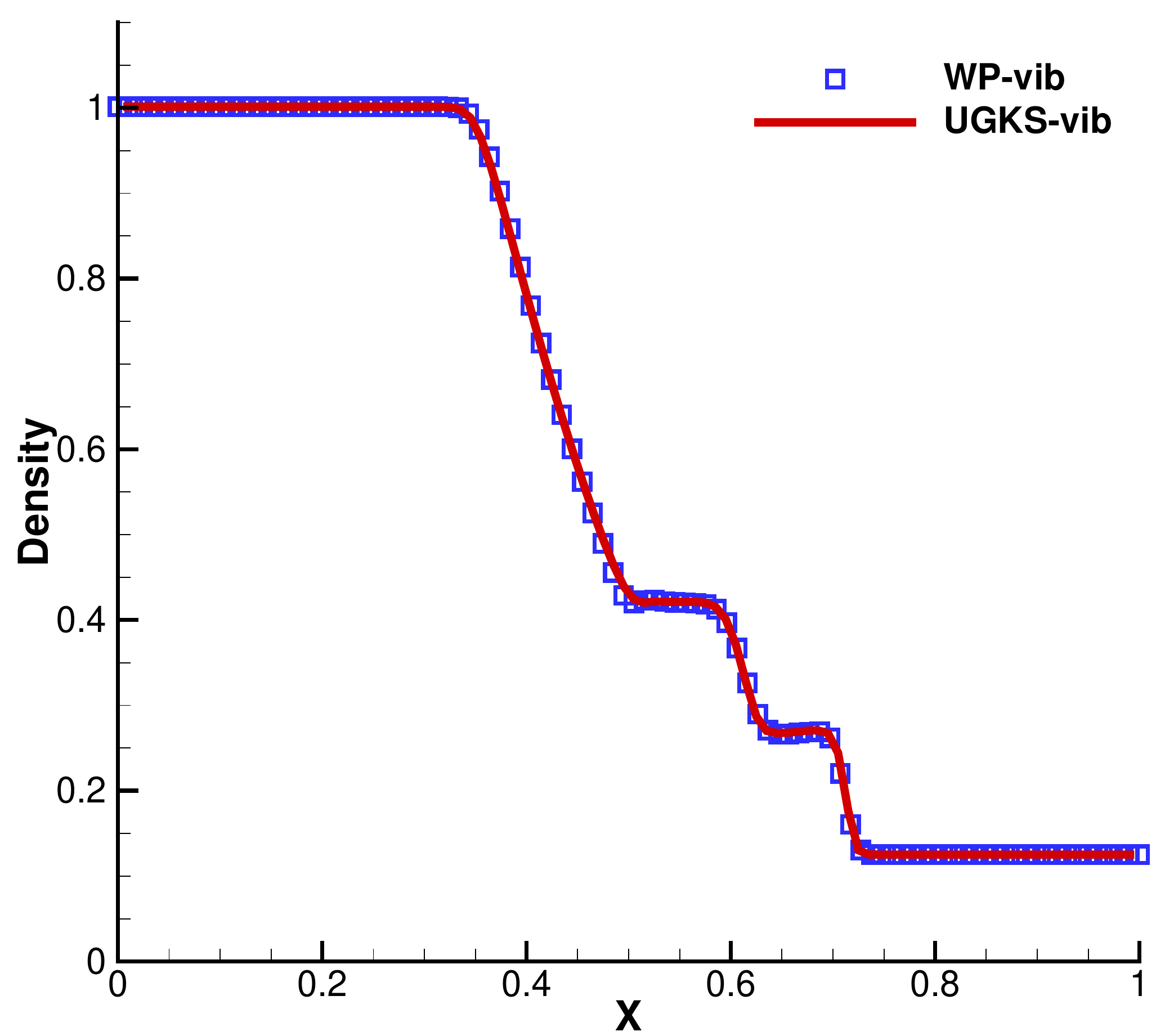}}
	\subfloat[]{\includegraphics[width=0.33\textwidth]{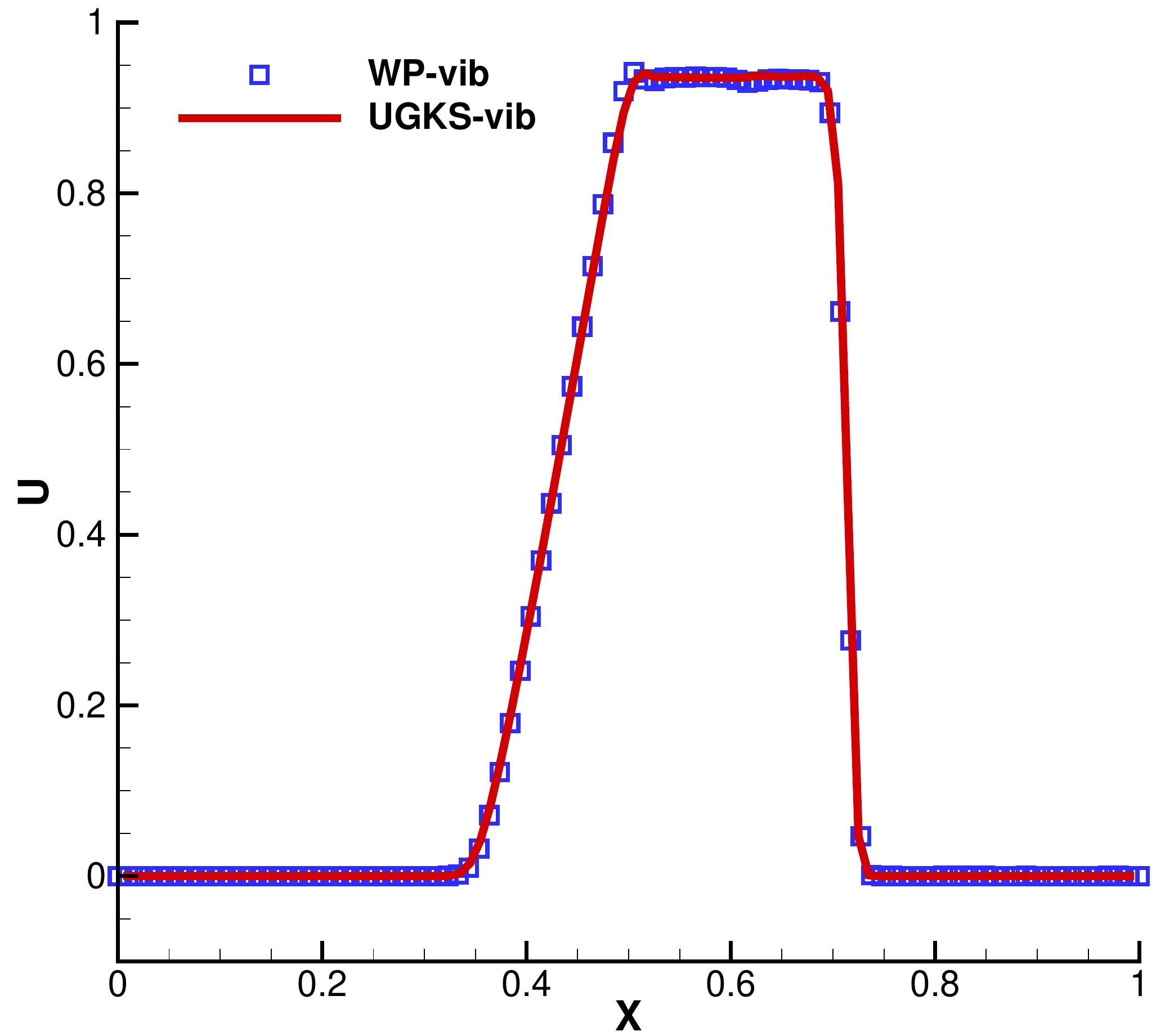}}
	\subfloat[]{\includegraphics[width=0.33\textwidth]{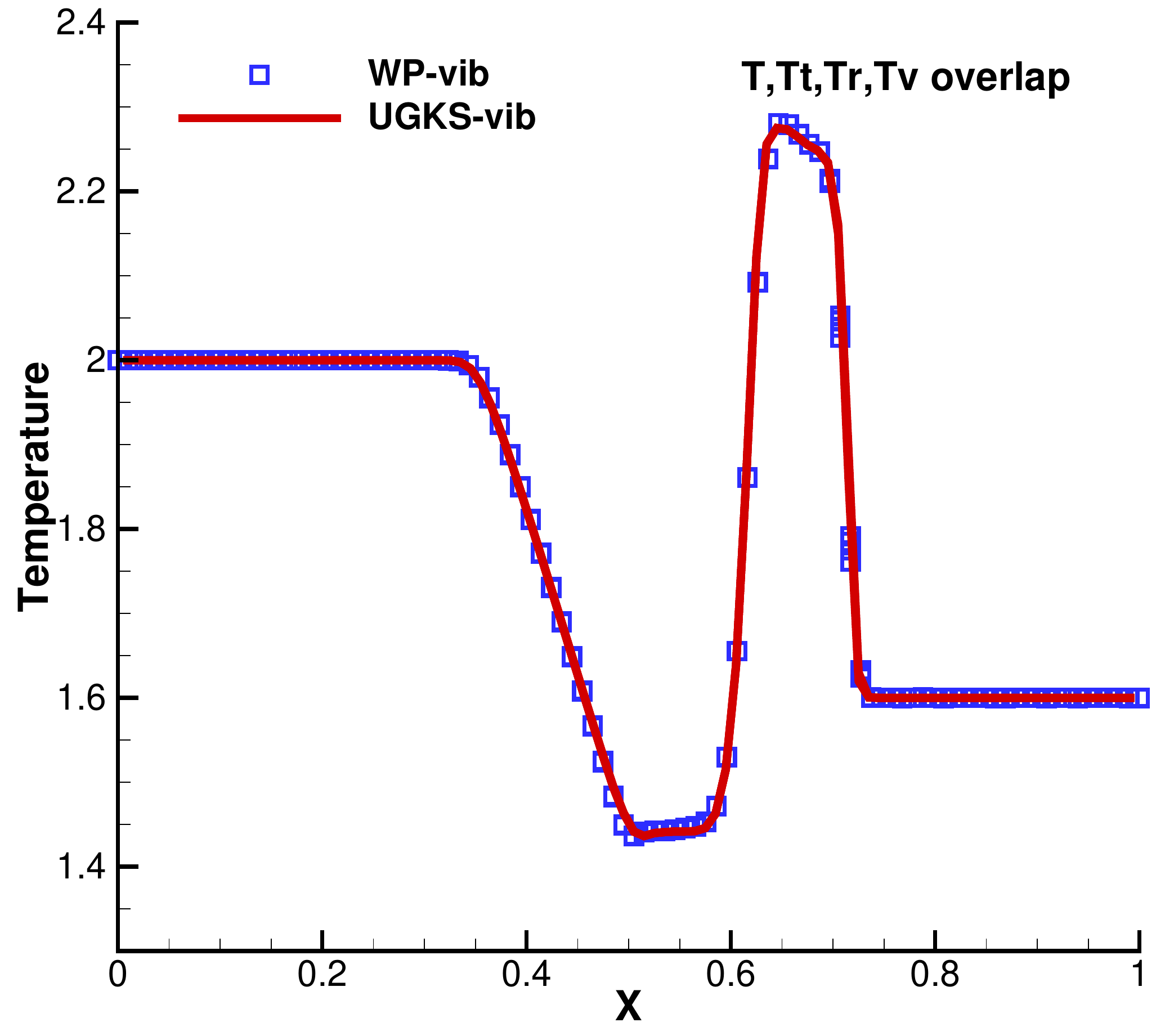}}
	\caption{Sod tube at ${\rm Kn} = 10^{-4}$. (a) Density, (b) velocity, and (c) temperatures.}
	\label{fig:sod-kn1e-4}
\end{figure}

The density, velocity as well as the temperatures including the translational, rotational, vibrational and the average temperatures obtained by WP-vib and UGKS at different Knudsen numbers are plotted in Fig.~\ref{fig:sod-kn10}--\ref{fig:sod-kn1e-4}. In the calculation, the preset reference number of particles are $N_{r} = 2\times 10^4$, $2\times 10^4$ and $400$ for the cases at ${\rm Kn} = 10$, $0.1$, and $10^{-4}$, respectively. Sufficient simulation particles are employed so that satisfactory solutions are obtained with no need of time-averaging treatment for the unsteady flow. The three-dimensional flow field obtained by the WP-vib is projected to the $x$ direction by taking average over the cells on $y$-$z$ plane to further reduce the statistical noises. For all these cases, the WP-vib results agree well with the UGKS solutions with the same vibrational relaxation model. It has been shown that the WP-vib is capable of numerical simulations in both continuum and rarefied regimes.

\subsection{Shock Structure}
For diatomic gas with vibrational degrees of freedom, the initial conditions of the normal shock wave in the upstream and downstream with different specific heat ratios are determined by the conservation, which is given in \ref{sec:app-shock}.
The computational domain $[-25, 25]$ has a length of $50$ times of the particle mean free path and is divided by $200$ cells uniformly.
The left and right boundaries are treated as far field condition. The CFL number is taken as $0.5$.

In this study, a strong shock wave at upstream Mach number ${\rm Ma}_1 = 10$ is investigated,
and the upstream temperature is $T_1 = 226.149$ K. The rest parameters could be obtained from the non-dimensional initial condition
\begin{equation*}
	\begin{cases}\rho_1 = 1,  \qquad U_1=8.3666, \lambda_1=1, & x<0, \\
		         \rho_2 = 6.9294, U_2=1.2074, \lambda_2=0.05736, & x \geq 0.\end{cases}
\end{equation*}
The rotational and vibrational collision numbers keep constant as $Z_r = 5$ and $Z_v = 28$.

In kinetic theory, the particle collision time depends on the particle velocity. In order to cope with this physical reality,
the relaxation time of the high-speed particles is amended by \cite{xu2021modeling}
\begin{equation*}\label{eq:tauStar}
	\tau^*= \begin{cases}\tau, & \text { if }|\vec{u}-\vec{U}| \leq b \sqrt{RT}, \\
		    \frac{1}{ 1 + a^*|\vec{u}-\vec{U}| / \sqrt{RT} } \tau, & \text { if }|\vec{u}-\vec{U}|>b \sqrt{RT},\end{cases}
\end{equation*}
with two parameters $a= 0.1$ and $b=5$.

\begin{figure}[H]
	\centering
	\subfloat[]{\includegraphics[width=0.5\textwidth]{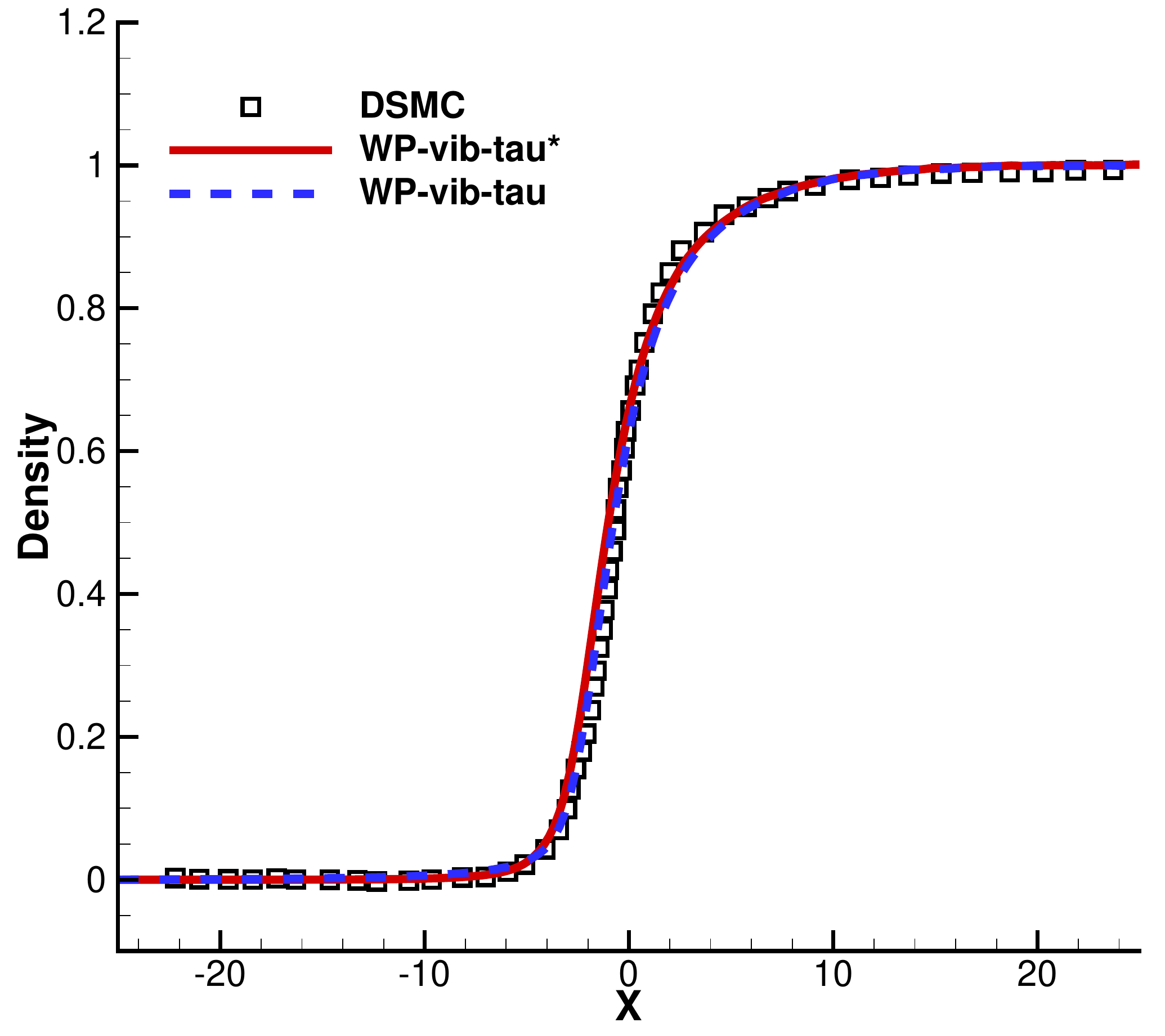}}
	\subfloat[]{\includegraphics[width=0.5\textwidth]{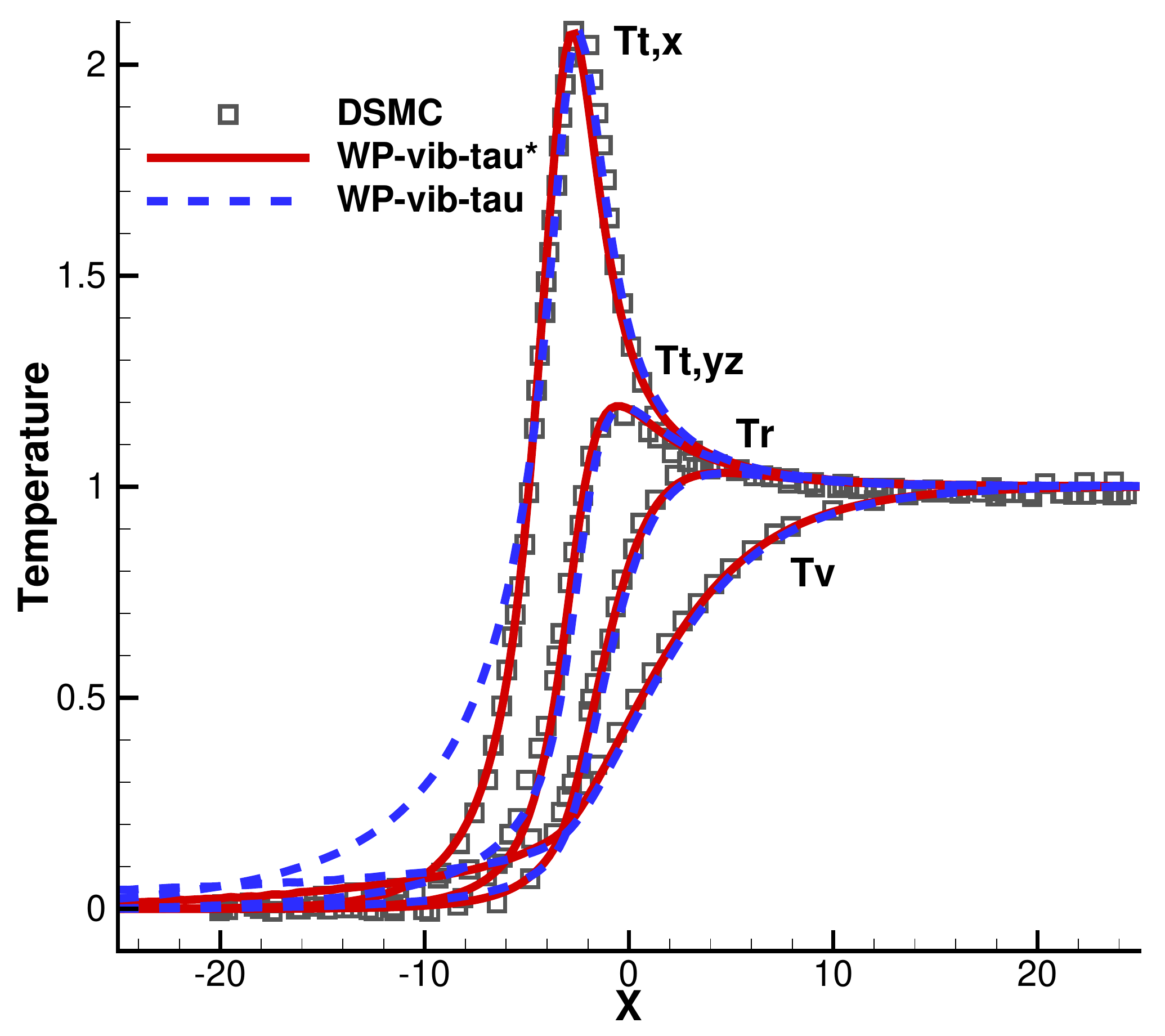}}
	\caption{Shock structure at ${\rm Ma} = 10$. (a) Density and (b) temperatures compared
	with the DSMC.}
	\label{fig:shock}
\end{figure}

To reduce the statistical noise, $5 \times 10^3$ simulation particles are used in each cell. The time-averaging is taken from 2500th step over 12500 steps. The normalized density and temperature from the original WP-vib, the modified WP-vib with $\tau^*$, and the DSMC \cite{cai2008one} simulation are plotted in Fig.~\ref{fig:shock}, which shows the good agreement between WP-vib and DSMC data.

In Fig.~\ref{fig:shock}(b), $T_r$ and $T_v$ denote rotational and vibrational temperature respectively. $T_{t,x}$ denotes the translational temperature in $x$ direction, and $T_{t,yz}$ is the average translational temperature in $y$ and $z$ directions, which are obtained from
\begin{equation*}\label{eq:Tt-x}
	T_{t,x} = \frac{1}{ \rho R } \int { ( u-U )^2 } f {\rm d}{\vec \Xi},
\end{equation*}
and
\begin{equation*}
	T_{t,yz} = \frac{1}{ 2 \rho R } \int { \left[  ( v-V )^2 + ( w-W )^2  \right] f {\rm d}{\vec \Xi}}.
\end{equation*}

\begin{figure}[H]
	\centering
	\subfloat[]{\includegraphics[width=0.5\textwidth]{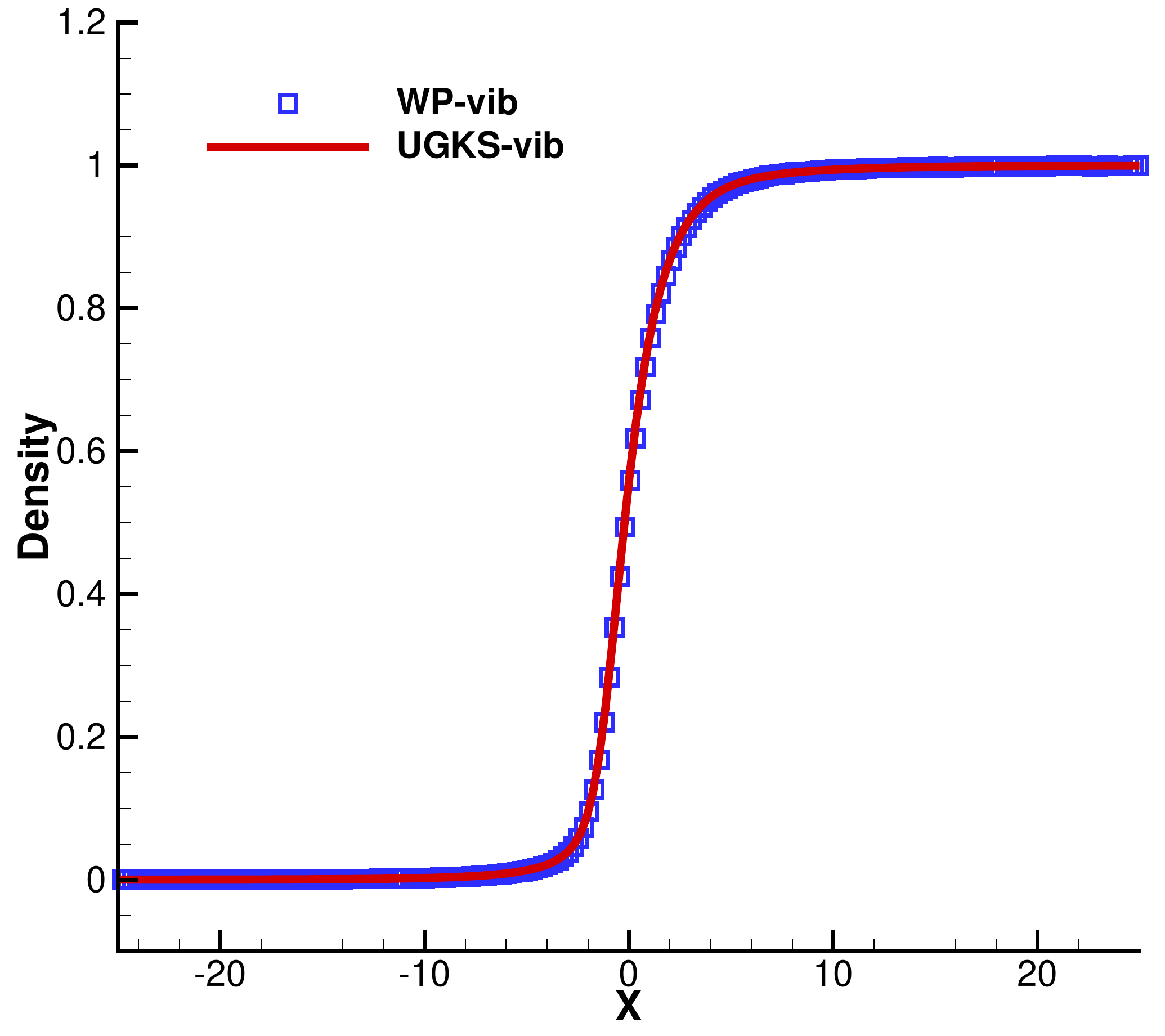}}
	\subfloat[]{\includegraphics[width=0.5\textwidth]{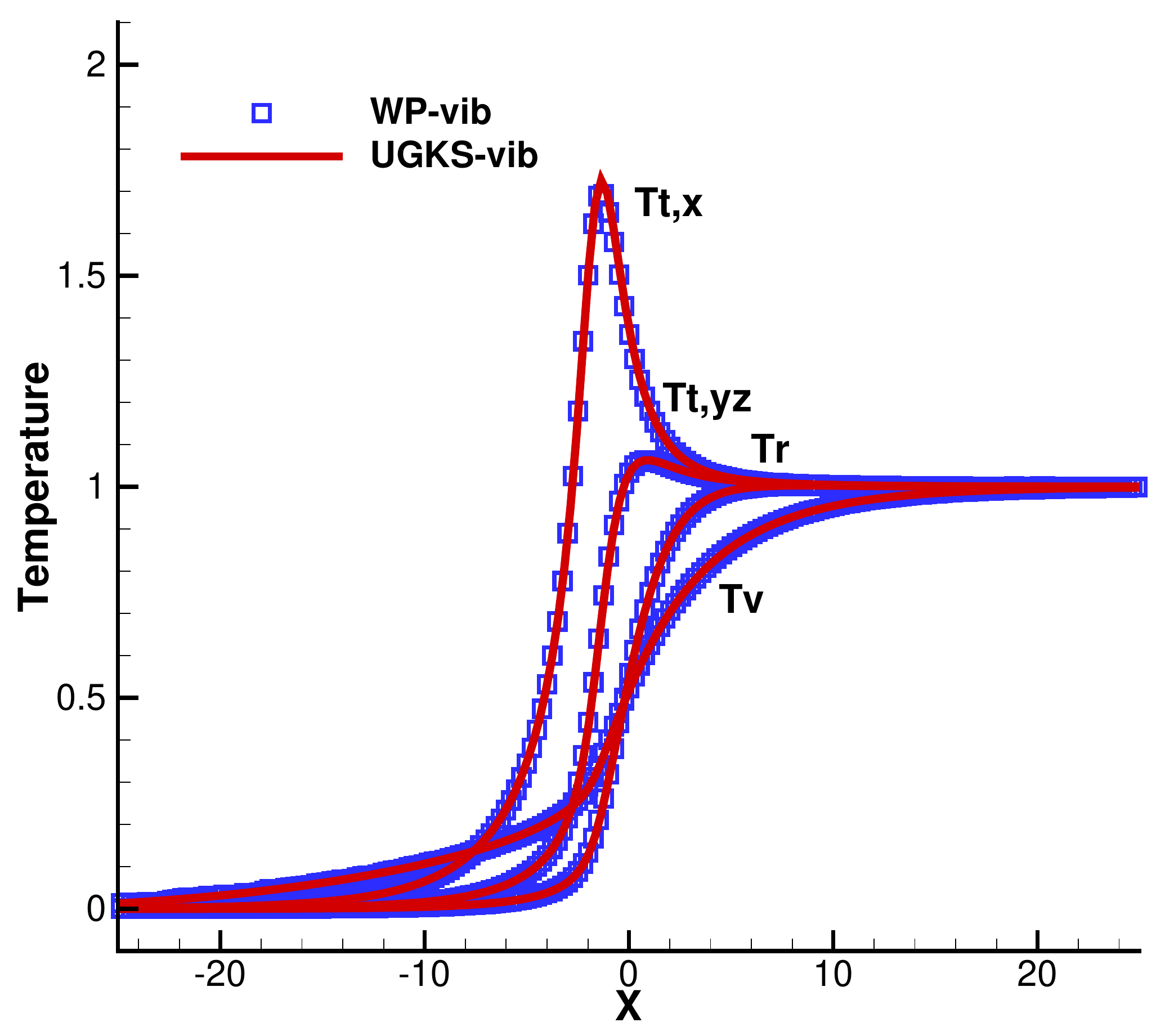}}
	\caption{Shock structure at ${\rm Ma} = 4$. (a) Density and (b) temperatures compared
	with the UGKS.}
	\label{fig:shock-Ma4}
\end{figure}

\begin{figure}[H]
	\centering
	\subfloat[]{\includegraphics[width=0.5\textwidth]{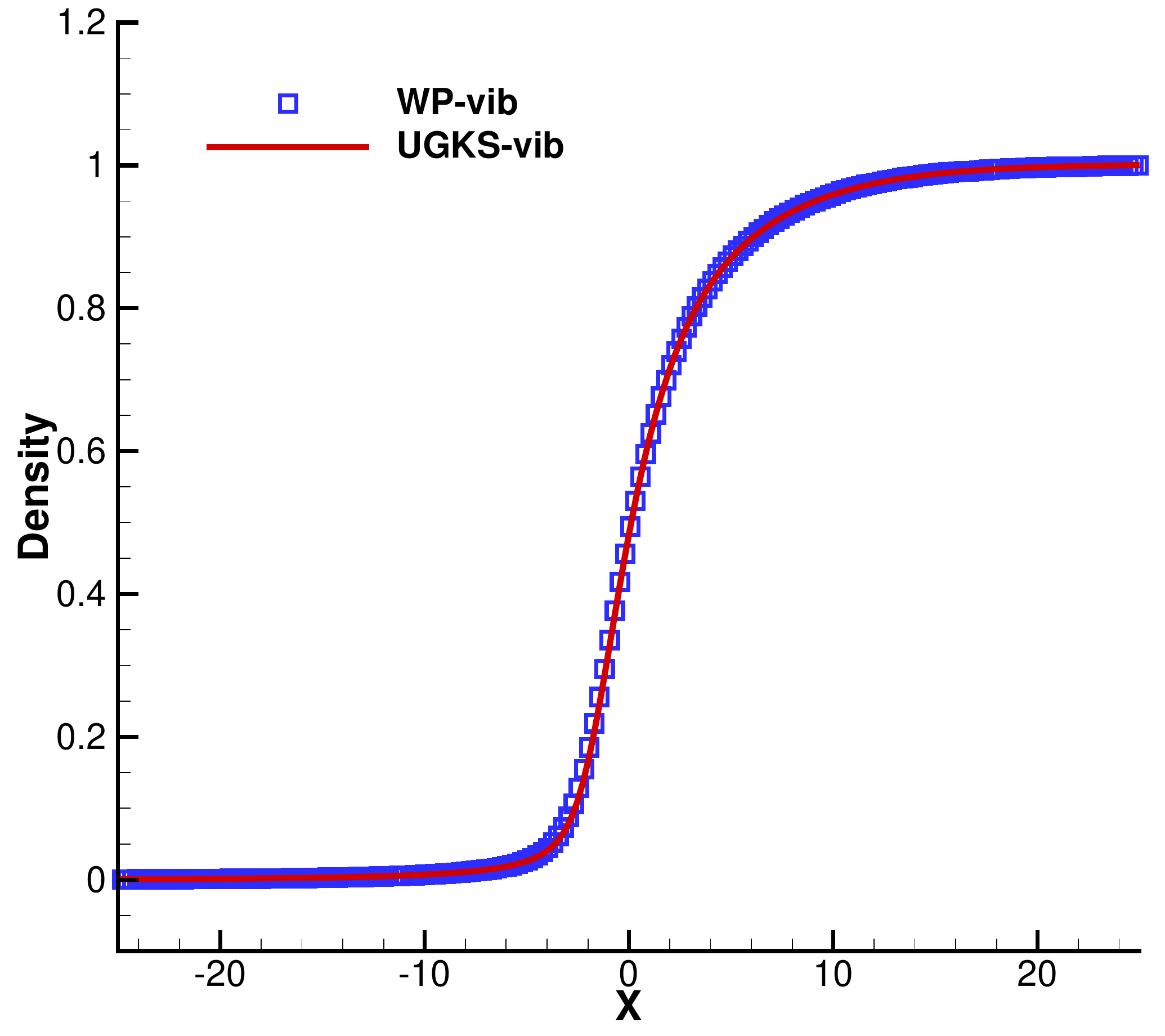}}
	\subfloat[]{\includegraphics[width=0.5\textwidth]{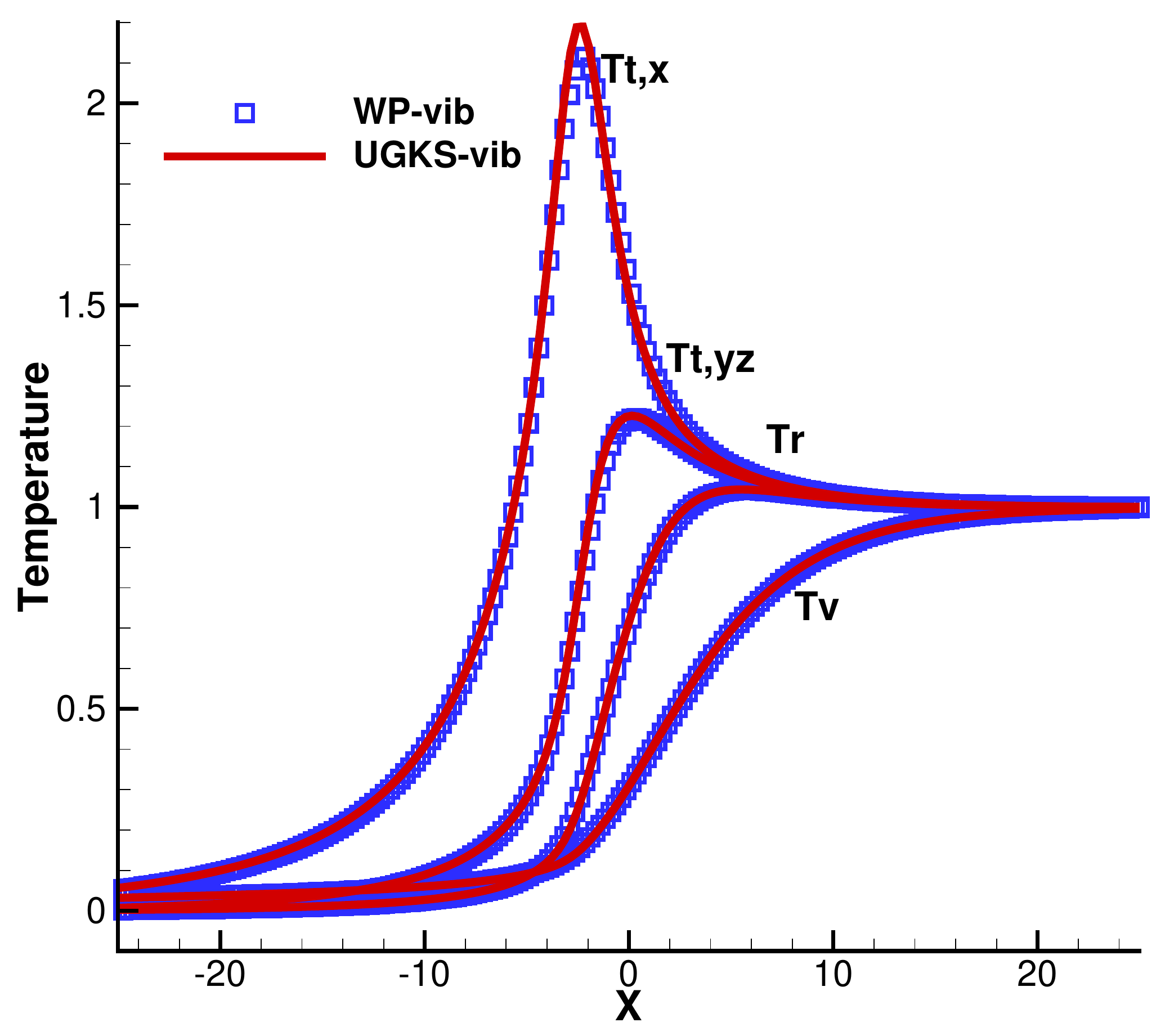}}
	\caption{Shock structure at ${\rm Ma} = 15$. (a) Density and (b) temperatures compared
	with the UGKS.}
	\label{fig:shock-Ma15}
\end{figure}

To further validate the WP-vib, the shock structure at ${\rm Ma} = 4$ and ${\rm Ma} = 15$ with the same parameters set at ${\rm Ma} = 10$ and using the unmodified relaxation time $\tau$ are simulated. Fig.~\ref{fig:shock-Ma4} and Fig.~\ref{fig:shock-Ma15} show that the agreement in the results from UGKS and UGKWP methods.

\subsection{Flow around a circular cylinder}

\begin{figure}[H]
	\centering
	\subfloat[]{\includegraphics[width=0.3\textwidth]
	{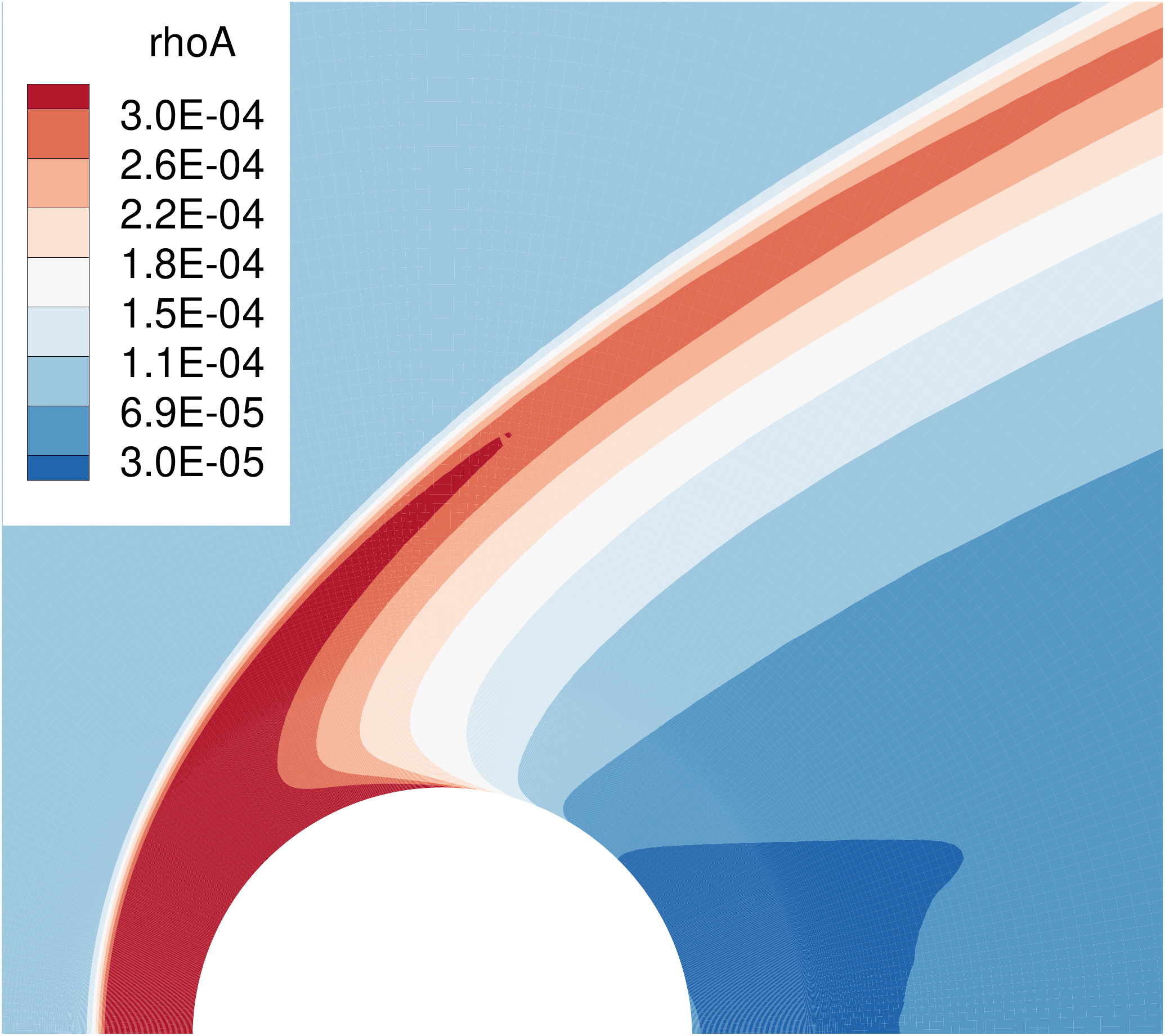}}
	\subfloat[]{\includegraphics[width=0.3\textwidth]
	{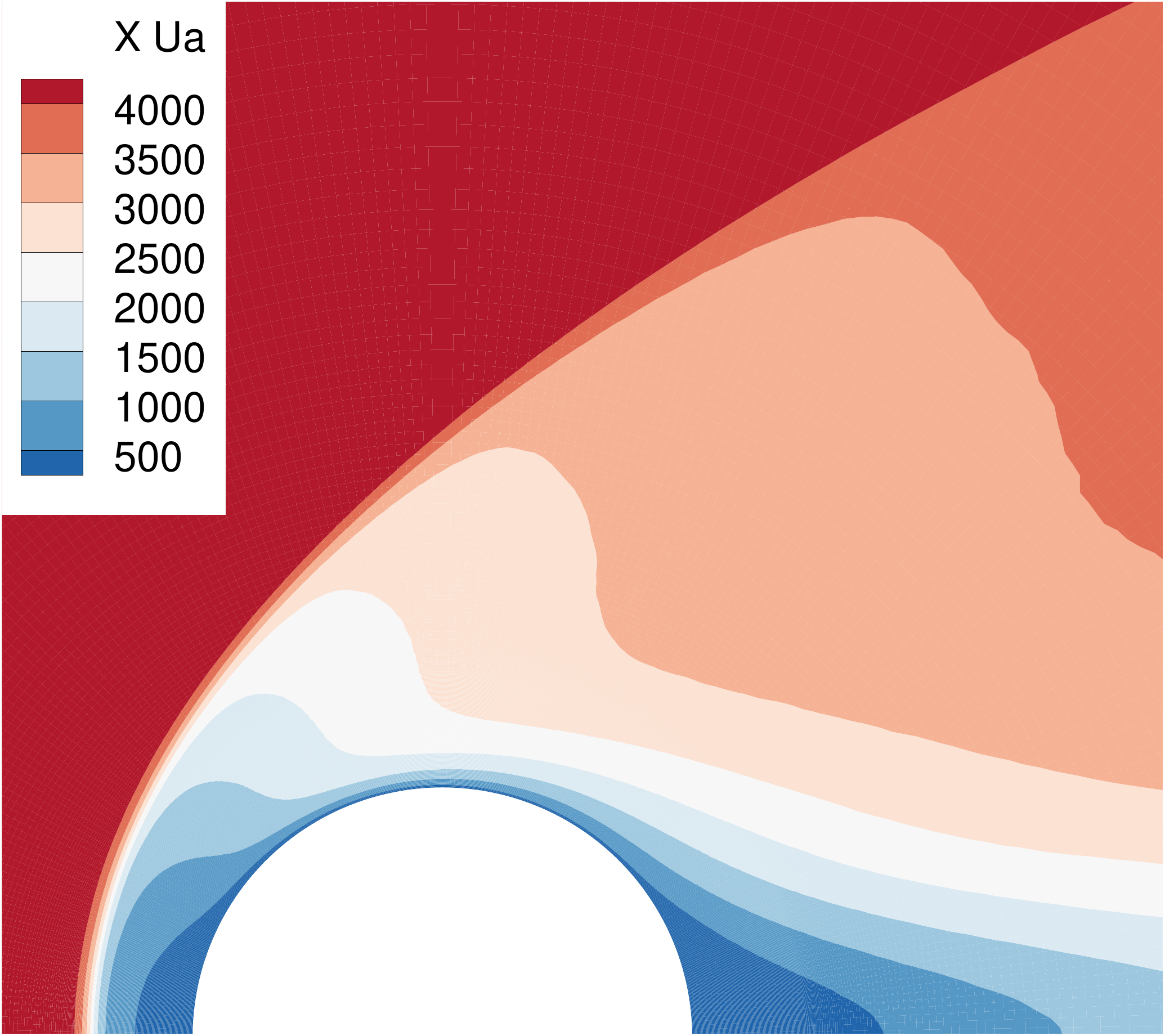}}
	\subfloat[]{\includegraphics[width=0.3\textwidth]
	{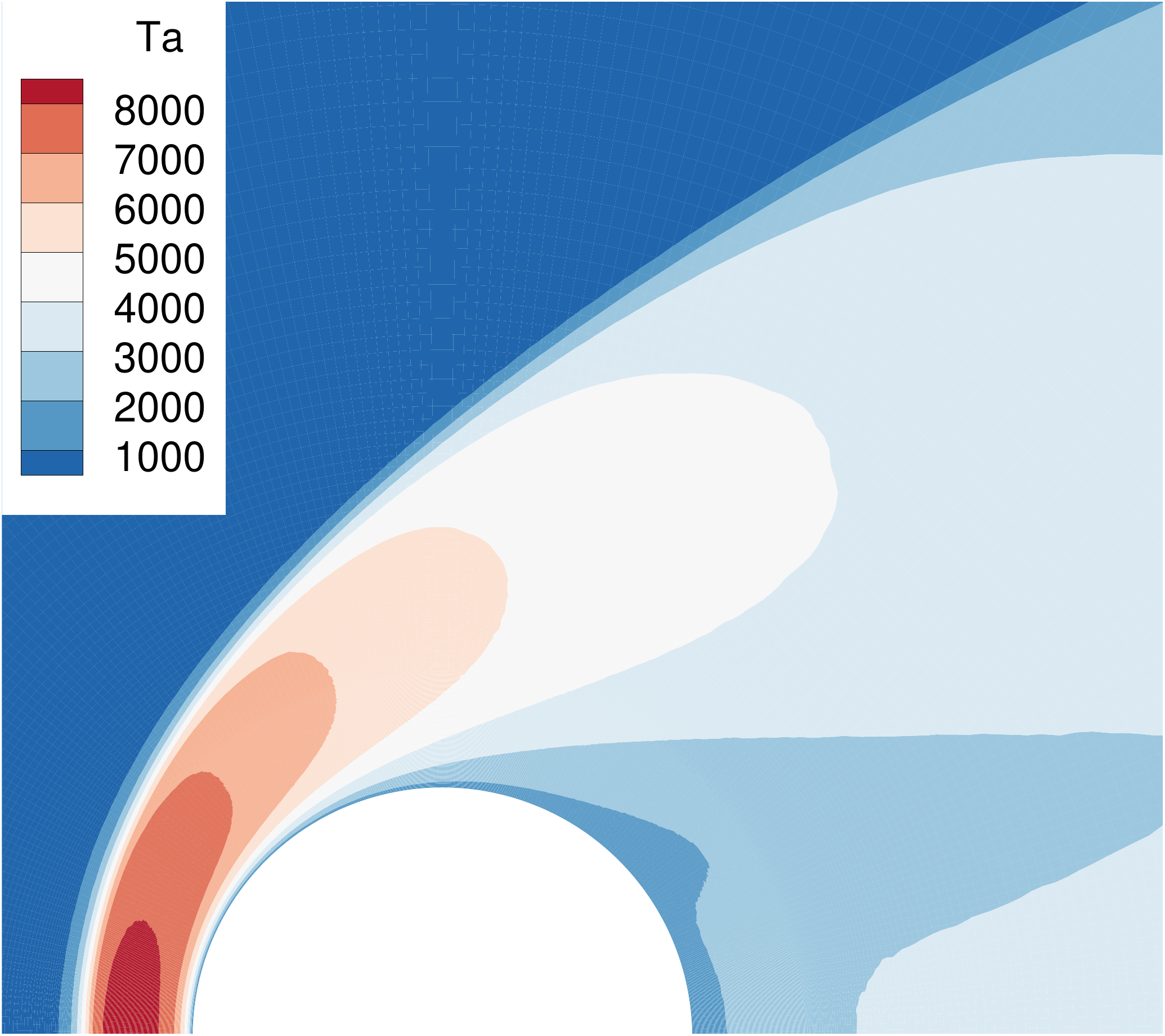}} \\
	\subfloat[]{\includegraphics[width=0.3\textwidth]
	{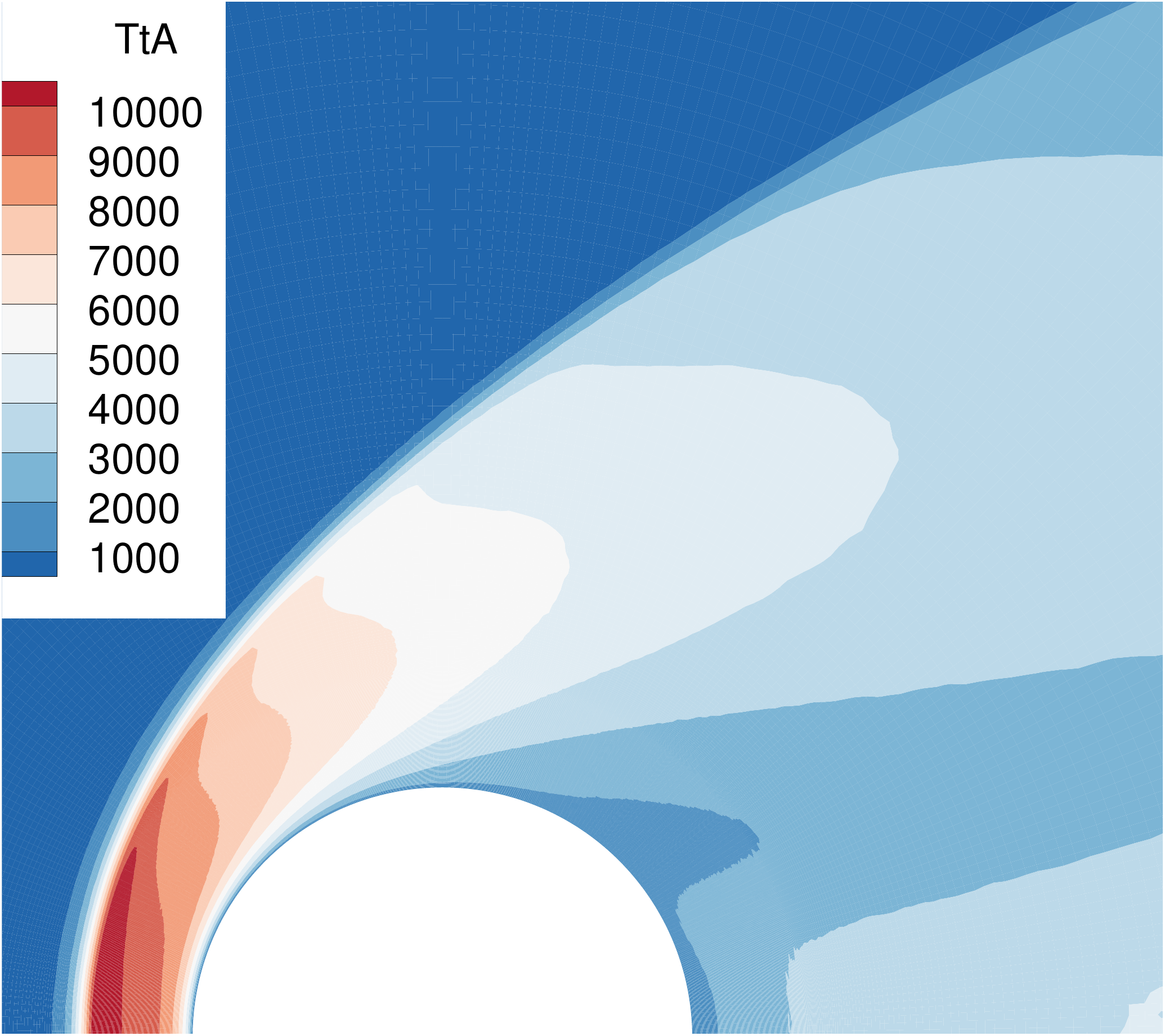}}
	\subfloat[]{\includegraphics[width=0.3\textwidth]
	{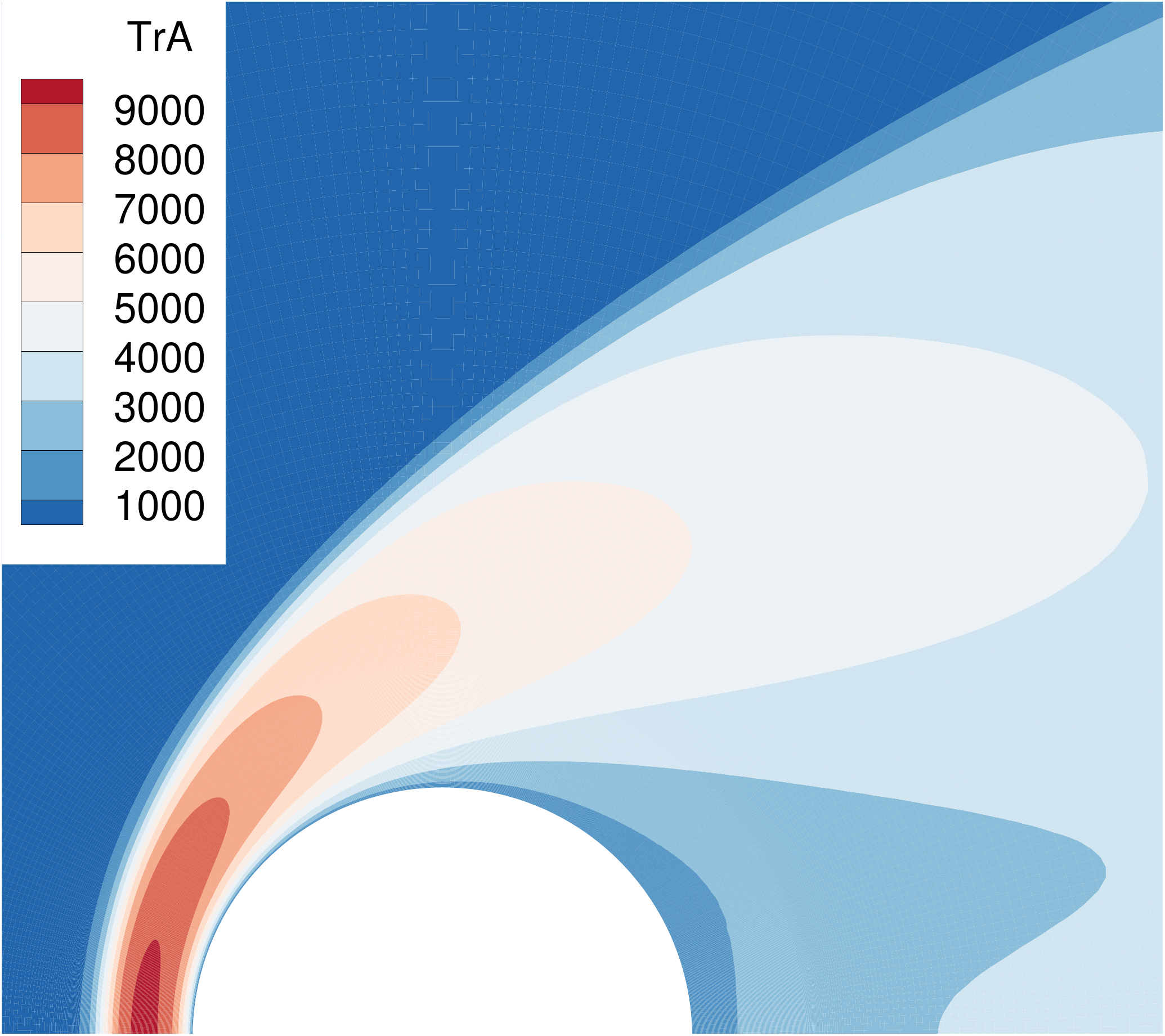}}
	\subfloat[]{\includegraphics[width=0.3\textwidth]
	{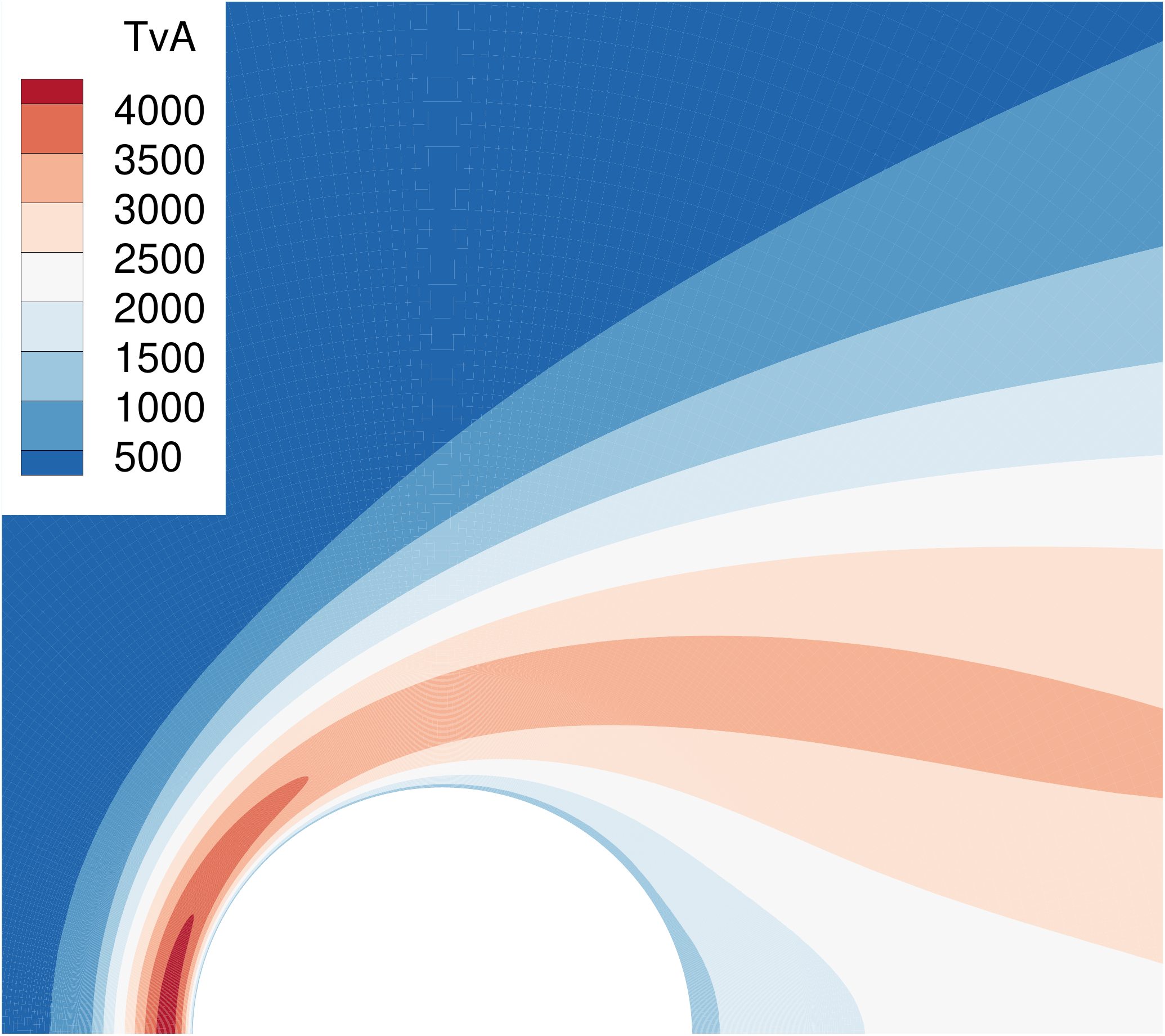}}
	\caption{Hypersonic flow at ${\rm Ma} = 15$ around a semi-circular
	cylinder for ${\rm Kn} = 0.01$.
	(a) Density, (b) $x$ direction velocity,
	(c) temperature, (d) translational temperature,
	(e) rotational temperature and (f) vibrational
	temperature contours.}
	\label{fig:cylinder}
\end{figure}

\begin{figure}[H]
	\centering
	\includegraphics[width=8cm]{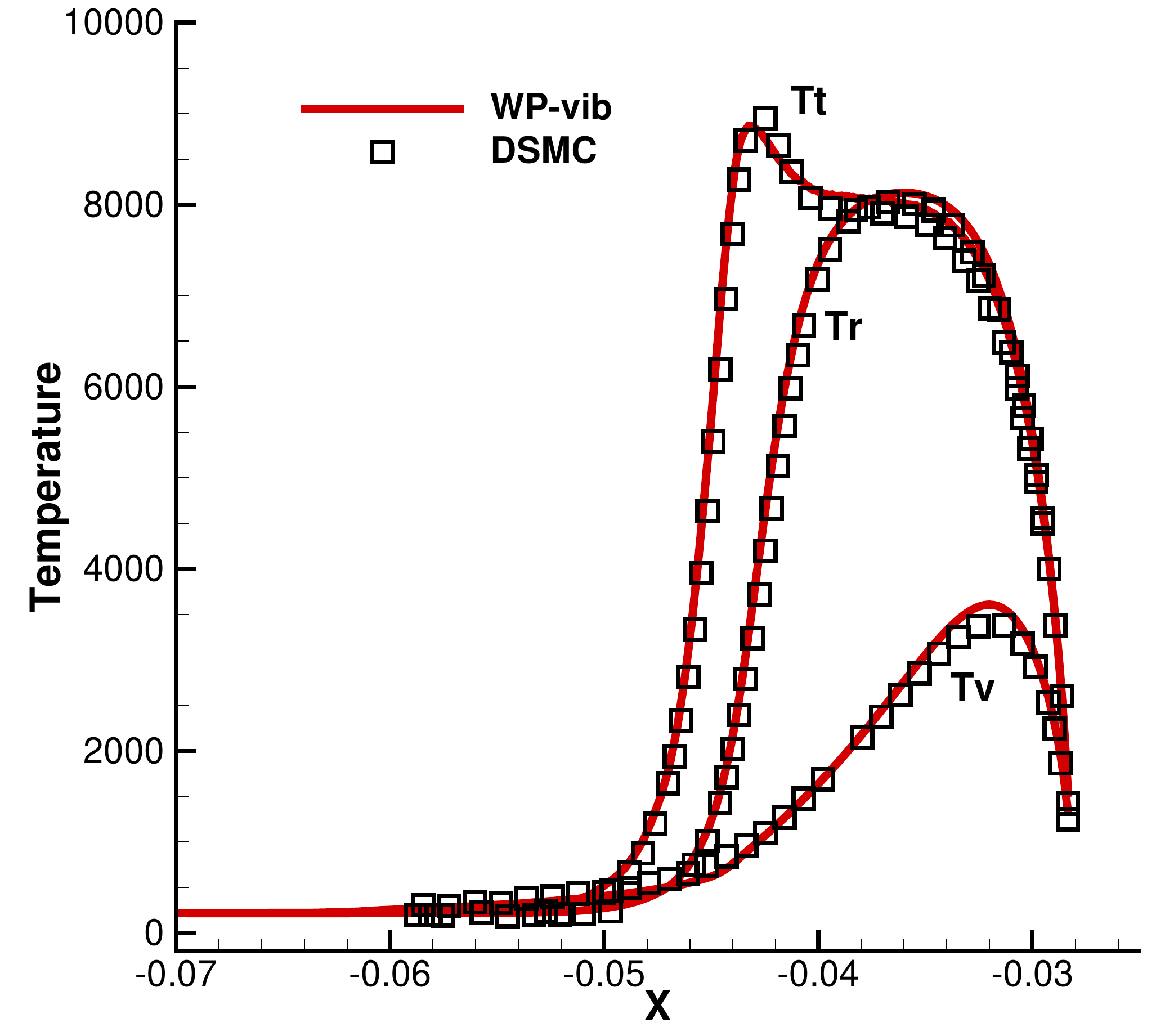}
	\caption{Temperature distributions along the $45^{ \rm \circ }$
	extraction line at ${\rm Ma} = 15$ and ${\rm Kn}=0.01$.}
	\label{fig:cylinder-T}
\end{figure}

High-speed flow passing over a semi-circular cylinder at a Mach number $15$ for ${\rm Kn} = 0.01$ is simulated \cite{deschenes2011extension}. 
The diameter of the cylinder $D = 0.08$ m. The Knudsen number is defined with respect to the diameter. The computational domain is discretized by $280 \times 200 \times 1$ quadrilateral cells. The initial reference number of particles $N_r$ is set as 2000.
The initial temperature of free stream gives $T_\infty = 217.5$ K, and the isothermal wall temperature is fixed at $T_w = 1000$ K. The rotational and vibrational collision numbers are evaluated by Eq.~\eqref{eq:Zr-tumuklu} and Eq.~\eqref{eq:Zv-tumuklu} with $Z_r^\infty = 12.5$ and $T^\ast = 91.5$ K. The CFL number is taken as 0.5.
Fig.~\ref{fig:cylinder} plots the contours of flow field computed by WP-vib, where an initial flow field provided by $10000$ steps of GKS calculation \cite{xu2001} is adopted, and $60000$ steps of averaging have been carried out starting from the $10000$th step. Fig.~\ref{fig:cylinder-T} shows the comparison between the WP-vib and DSMC method for the translational, rotational, and vibrational temperatures extracted along the $45^{\circ}$ line in the upstream. Accepted results have been obtained by the WP-vib.

\subsection{Flow around a sphere}
Supersonic flow at $\rm {Ma} = 4.25$ passing over a three-dimensional sphere in the transition regime at ${\rm Kn} = 0.031$ is computed for nitrogen gas. The reference length is chosen as the diameter of the sphere, i.e., $D = 2\times 10^{-3}$ m, for the definition of Knudsen number.

The initial condition for free stream is $T_{\infty} = 65$ K. Isothermal wall boundary condition at a constant temperature $T_w = 302$ K is used. Constant values of $Z_r = 3.5$ and $Z_v=10$ are adopted in this calculation. The surface mesh of the sphere is divided into $6$ blocks with $16 \times 16$ points in each block. The wall distance of the first layer of cells is $6.255\times 10^{-5}$ m.

\begin{figure}[H]
	\centering
	\subfloat[]{\includegraphics[width=0.3\textwidth]
		{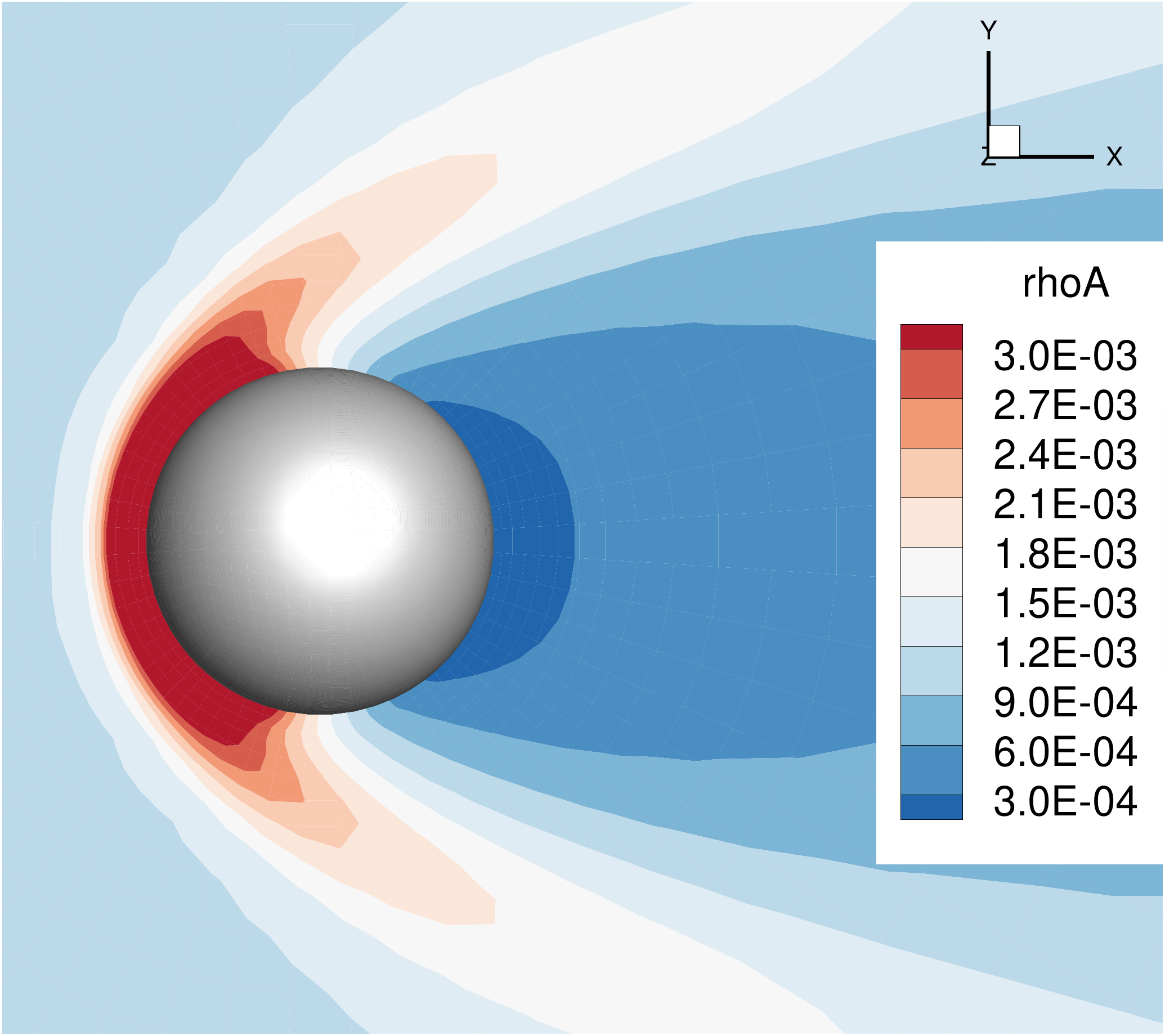}}
	\subfloat[]{\includegraphics[width=0.3\textwidth]
			    {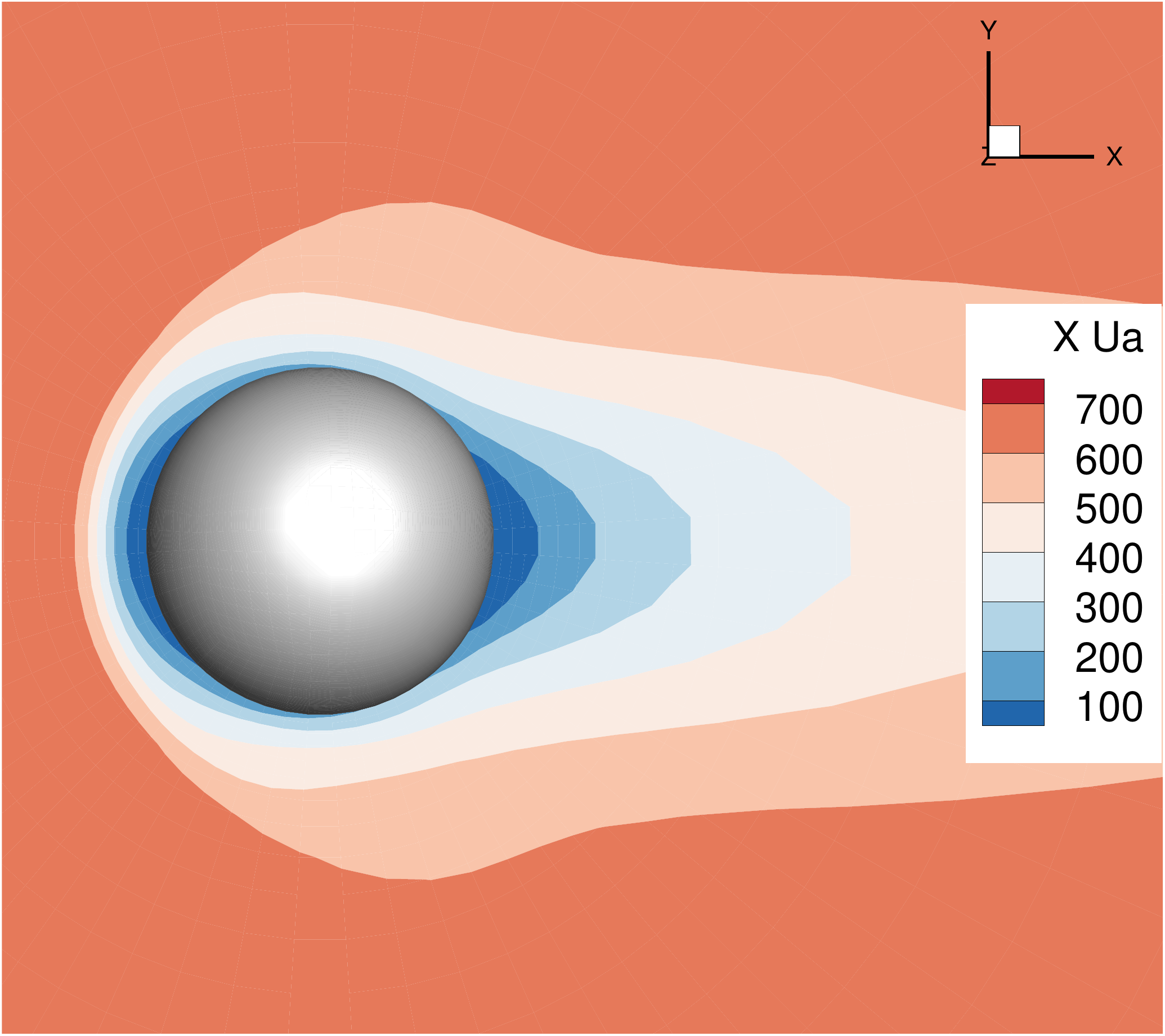} }
	\subfloat[]{\includegraphics[width=0.3\textwidth]
	{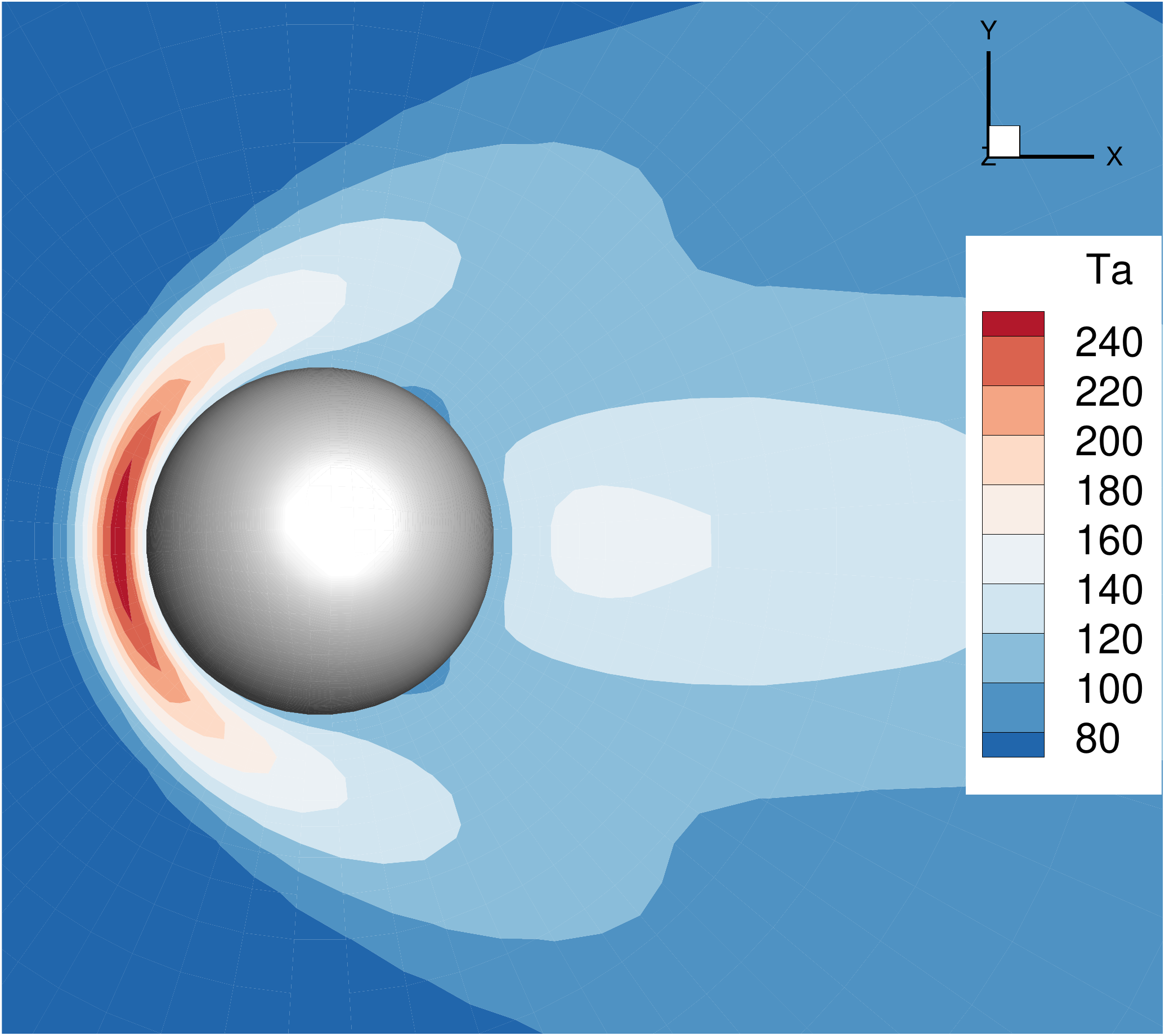}} \\
	\subfloat[]{\includegraphics[width=0.3\textwidth]
	{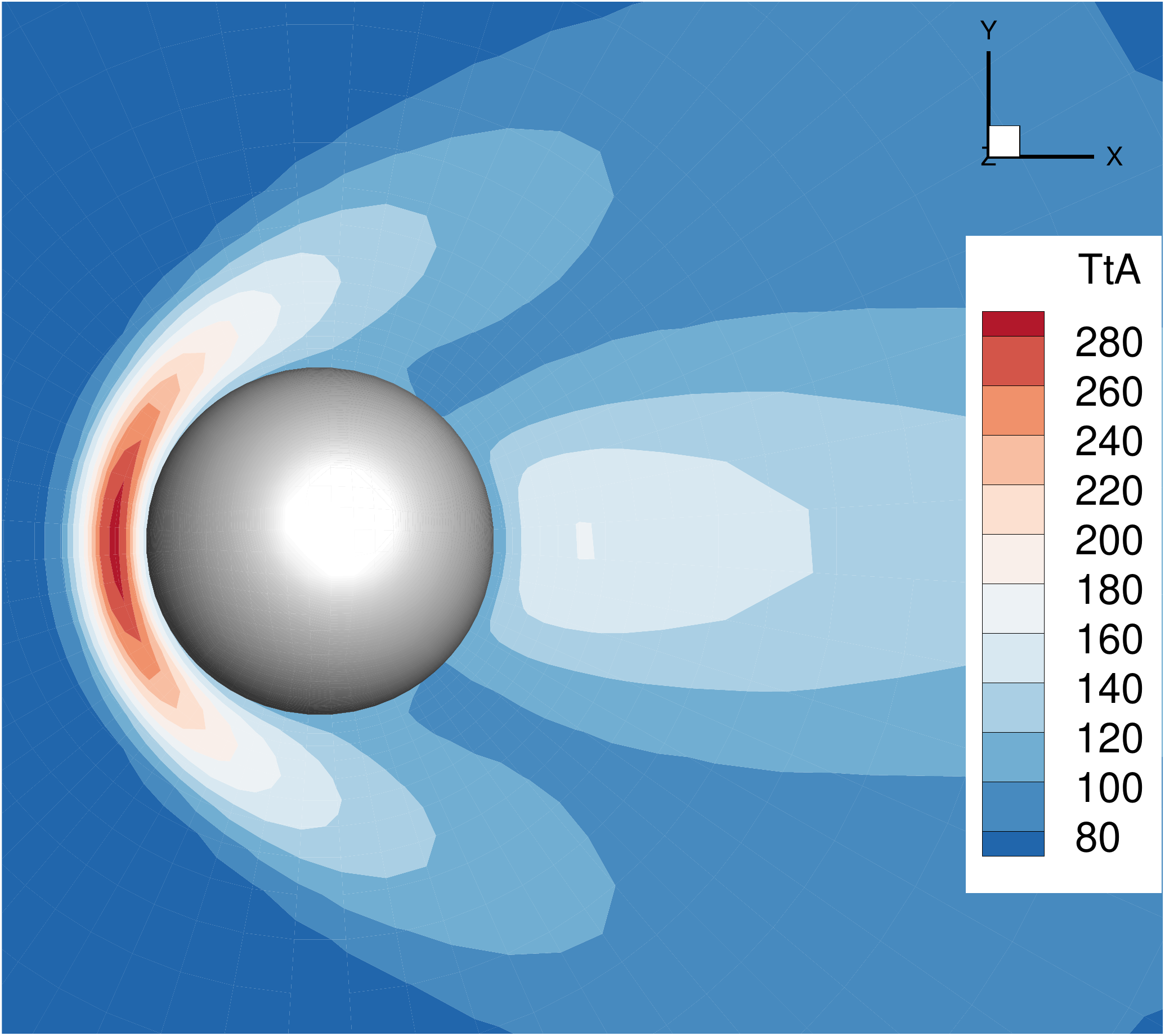}}
	\subfloat[]{\includegraphics[width=0.3\textwidth]
	{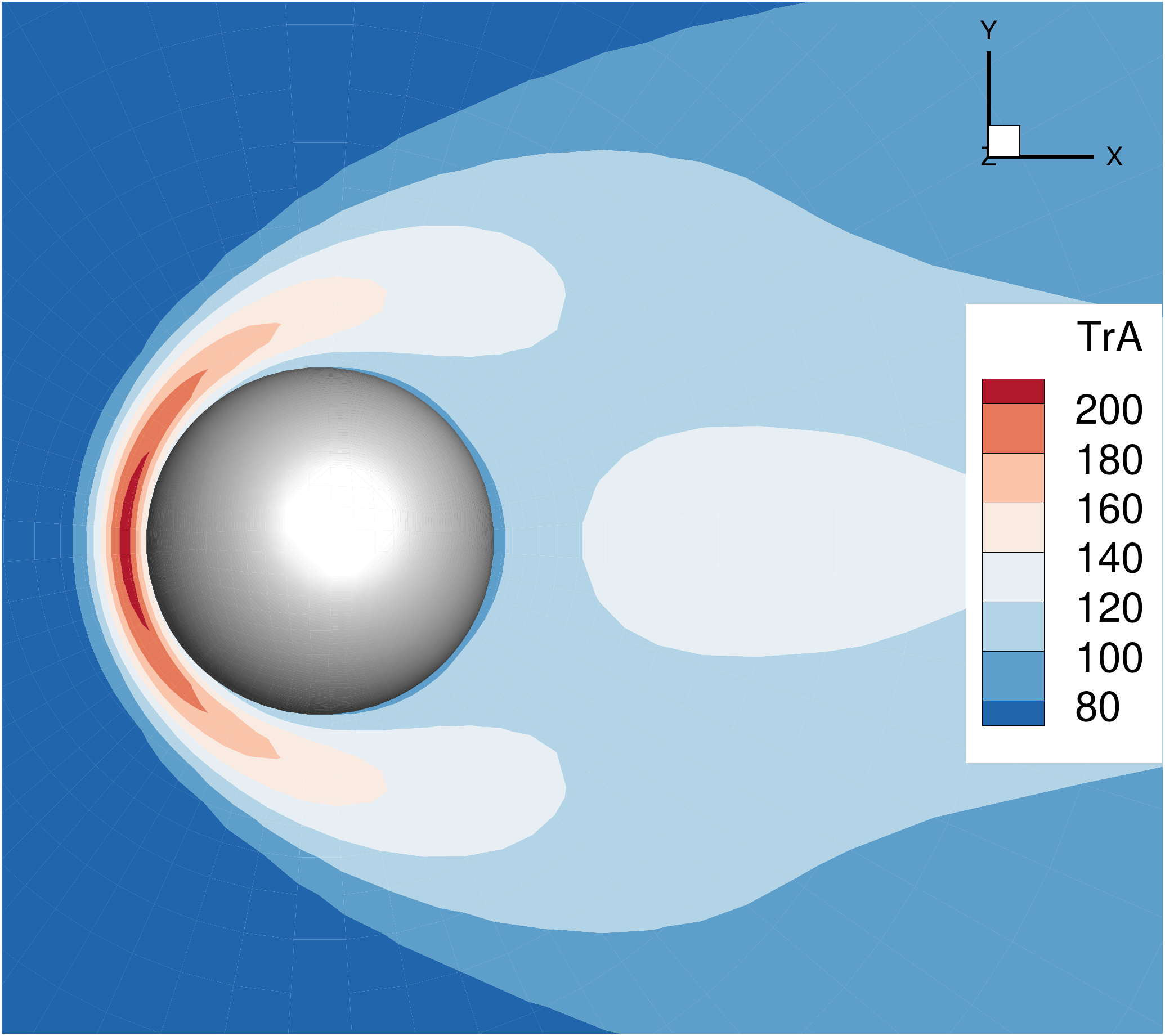}}
	\subfloat[]{\includegraphics[width=0.3\textwidth]
	{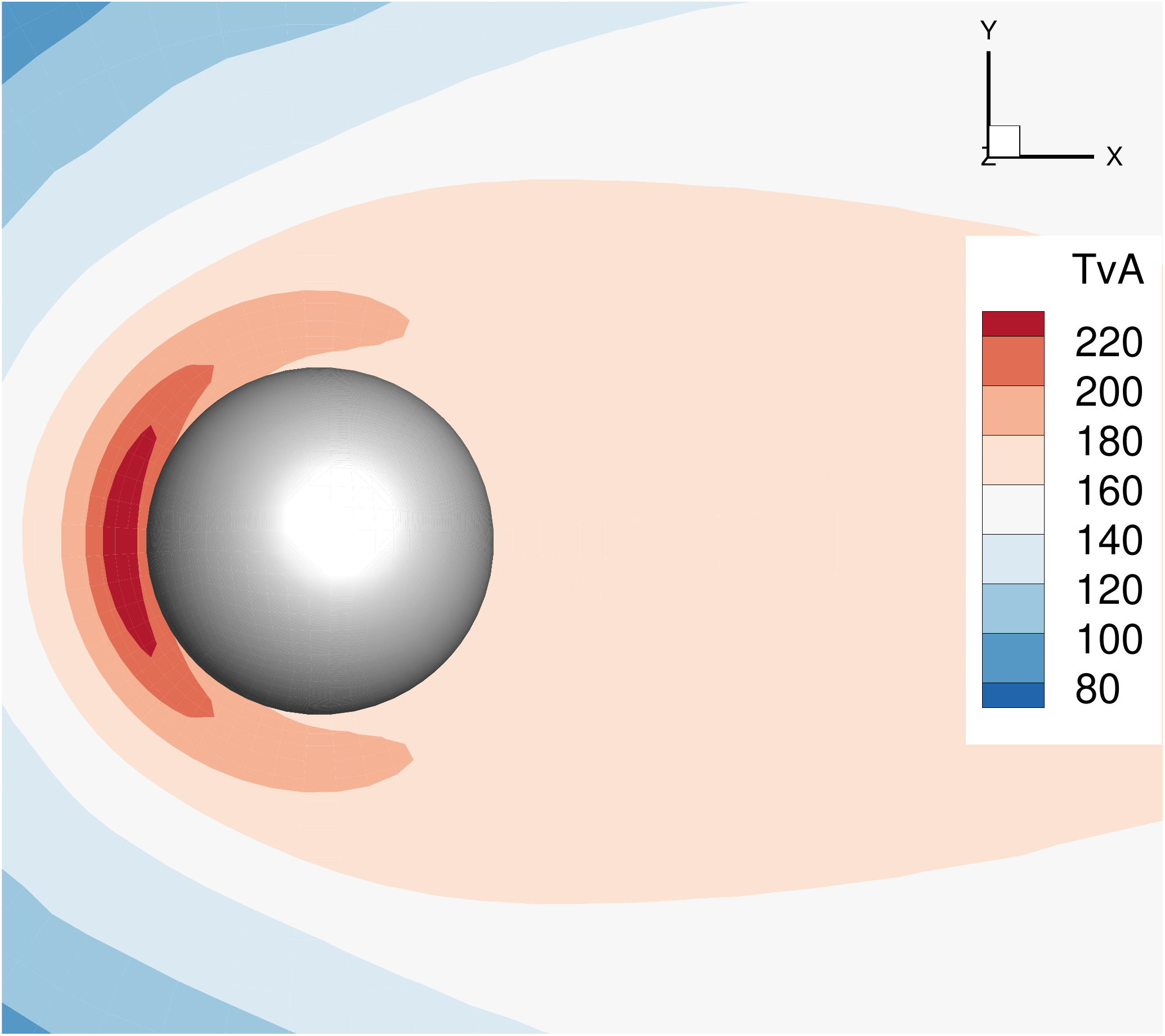}}
	\caption{Hypersonic flow at ${\rm Ma} = 4.25$ around a sphere for
	${\rm Kn} = 0.031$. (a) Density, (b) $x$ direction velocity,
	(c) temperature, (d) translational temperature,
	(e) rotational temperature and (f) vibrational
	temperature contours.}	
	\label{fig:sphere}
\end{figure}

\begin{figure}[H]
	\centering
	{\includegraphics[width=0.9\textwidth]
	{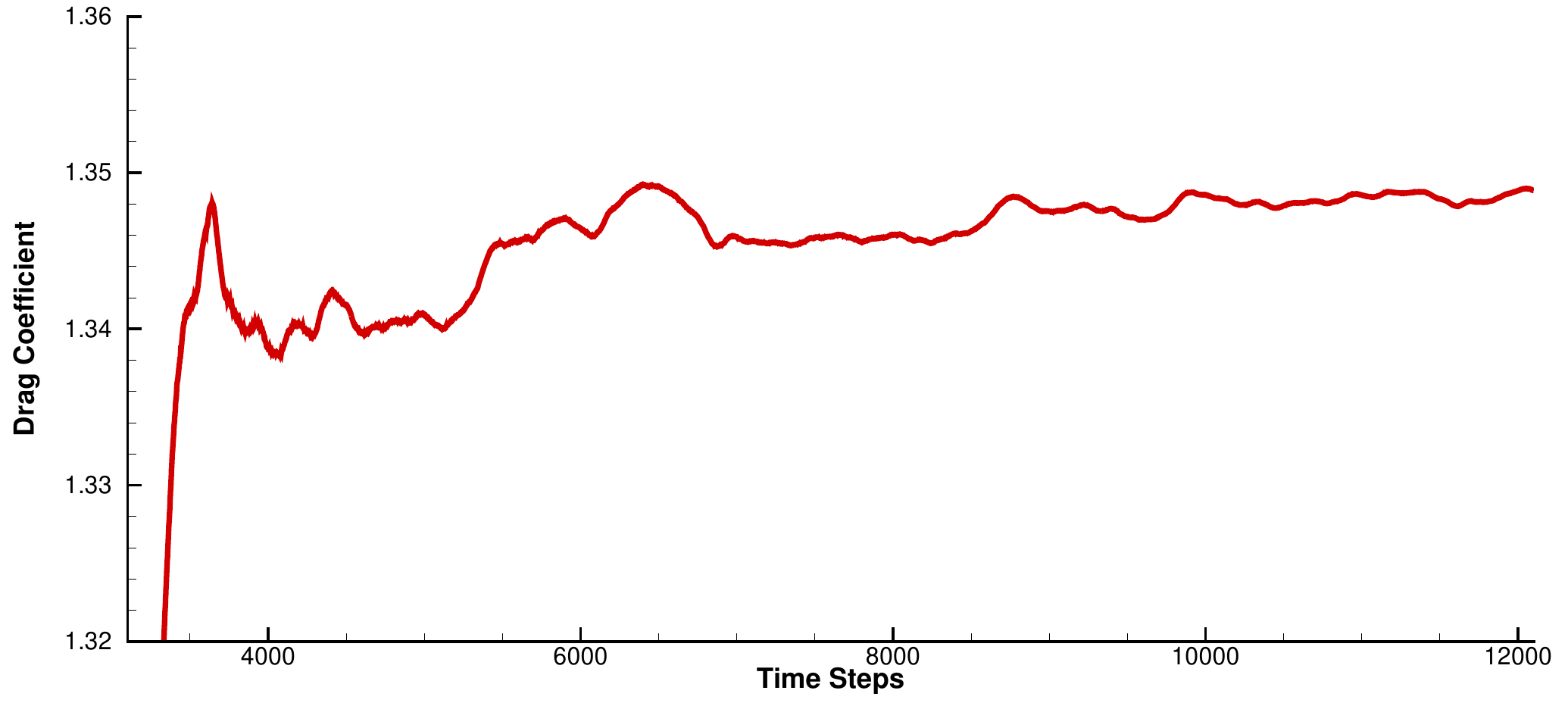}}
	\caption{Drag coefficient of hypersonic flow around a sphere at ${\rm Ma} = 4.25$, and $\rm{Kn=0.031}$.}.
	\label{fig:sphere-cd}
\end{figure}

In the calculation, the reference number of particles per cell is set at $N_r = 800$. An initial flow field provided by $500$ steps of GKS calculation \cite{xu2001} is adopted, and the time averaging for the steady solutions starts from the $2500$th step up to $9500$ steps. $\rm CFL=0.5$ is employed. The calculation takes 2 hours, running on Tianhe-2 with 2 nodes (48 cores, Intel Xeon E5-2692 v2, 2.2 GHz).
The distribution of density, velocity, temperatures are shown in Fig.~\ref{fig:sphere}. The drag coefficient computed by the WP-vib is compared in Tab.~\ref{tab:sphere-cd}, with those obtained
from experiment (Air) \cite{wendtJF}, the UGKWP method without vibrational model, and UGKS calculation \cite{jiang2019implicit}. Accurate results have been obtained with a relative error smaller than $0.1\%$. In Fig.~\ref{fig:sphere-cd}, the convergence history of the drag coefficent is plotted.
For this test case, the UGKS with $40\times40\times40$ discrete velocity points needs 222.5 hours on 6 nodes (48 cores).
The WP-vib shows great advantages in the aspects of computational efficiency and memory reduction.

\begin{table}[H]
	\centering
	\caption{Comparison of the drag coefficients.}
	\begin{tabular}{ccccc}
	\hline
	 & Experiment (Air) & WP-vib (Nitrogen) & UGKWP(Nitrogen) & UGKS   \\ \hline
	$C_d$ & 1.350  & 1.349  & 1.346 & 1.355 \\
	Error & -     & -0.03\% & -0.25\% & 0.39\%   \\ \hline
	\end{tabular}
	\label{tab:sphere-cd}
\end{table}

\subsection{Flow around a space vehicle}

Hypersonic flows at ${\rm Ma} = 6$ passing over a space vehicle at ${\rm Kn} = 10^{-3}$ and ${\rm Kn} = 10^{-5}$ are simulated for nitrogen gas.
According to the particle mean free path and the normal size of space vehicle (5m), the above Knudsen numbers correspond to the flight between 50km to 80km altitude.
These regimes can  be hardly recovered by the DSMC and Navier--Stokes solutions. At the hypersonic speed, all flow regimes can emerge at different part of flying vehicle.
 These cases can be used to test the efficiency and capability of the WP-vib for simulating three-dimensional hypersonic flow over complex geometry in the transition regime. 
 
\begin{figure}[H]
	\centering
	\subfloat[]{\includegraphics[height=3.5cm]{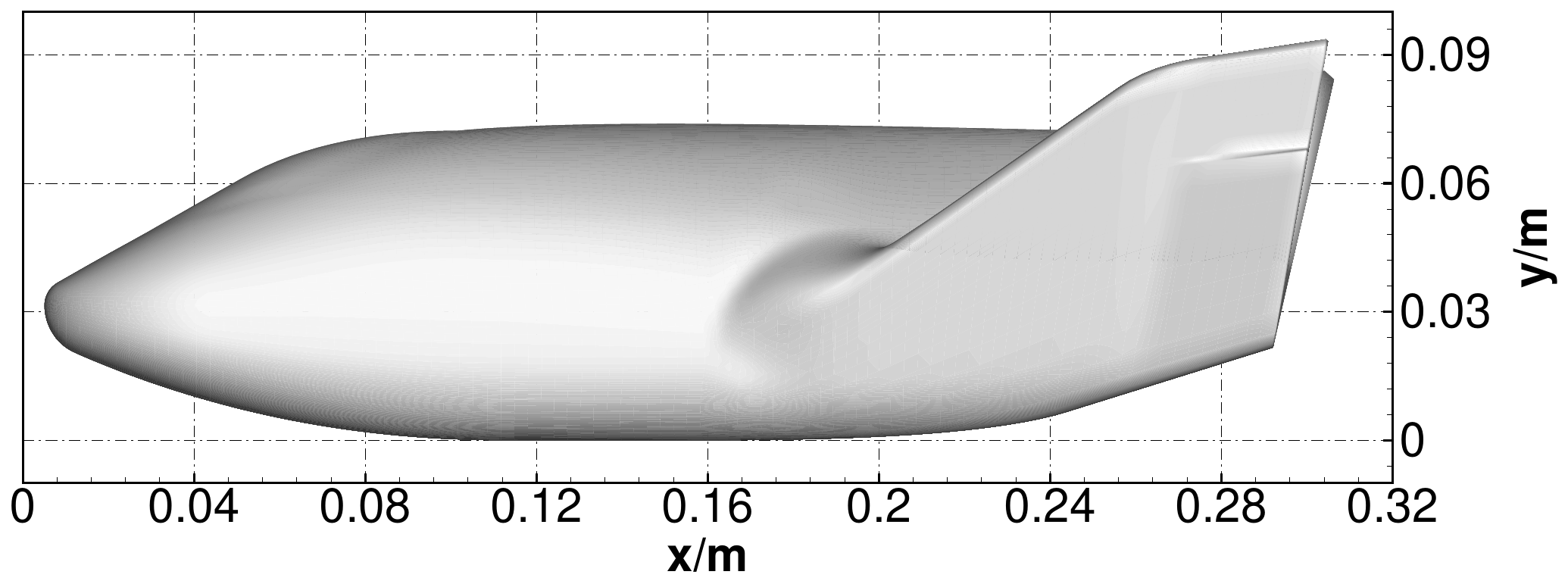}} \\
	\subfloat[]{\includegraphics[height=3.5cm]{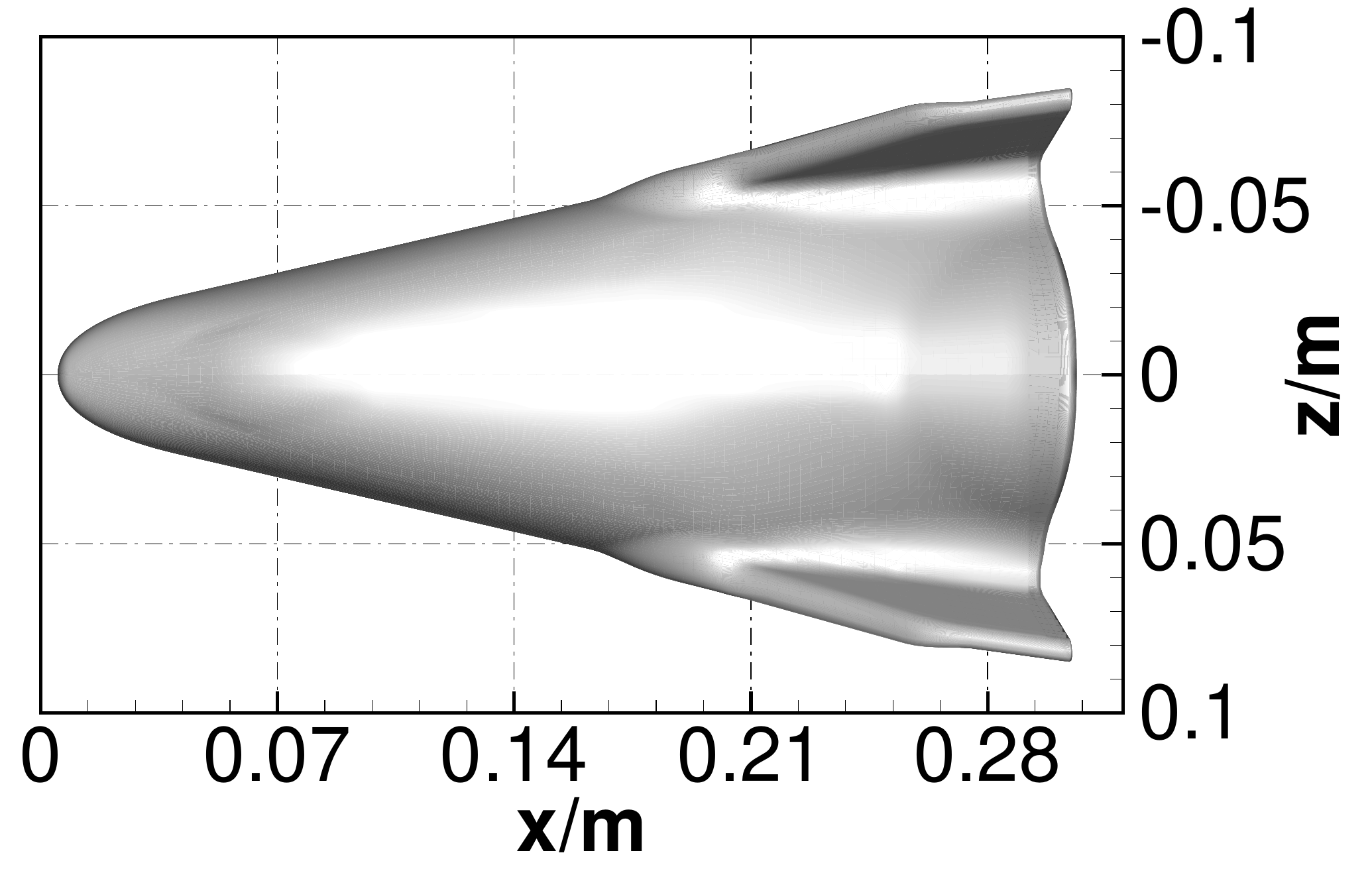}}
	\subfloat[]{\includegraphics[height=3.5cm]{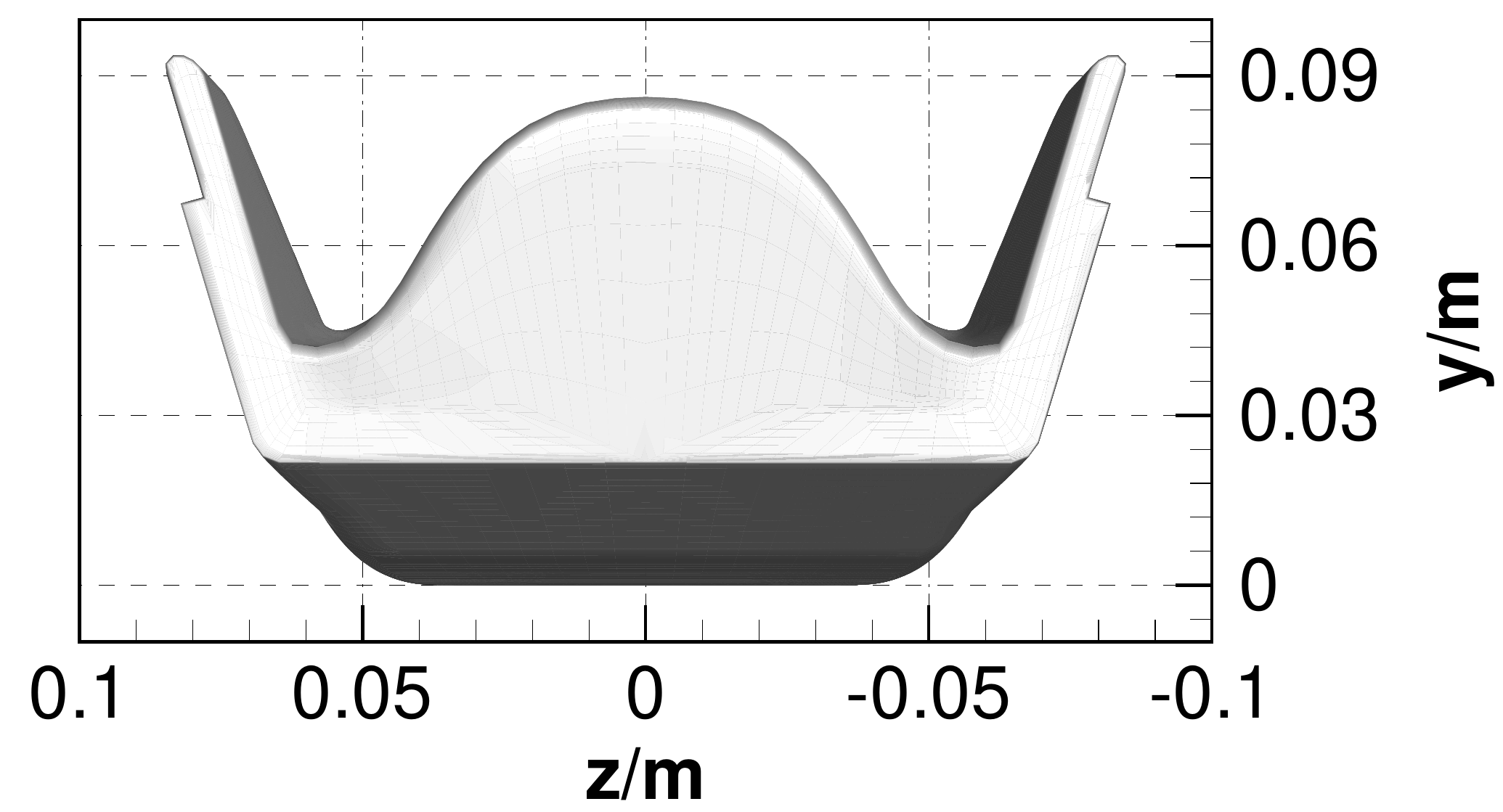}}
	\caption{Sketch of the space vehicle.}
	\label{fig:x38-geo}
\end{figure}

\begin{figure}[H]
	\centering
	\subfloat[]{\includegraphics[width=0.4\textwidth]
		{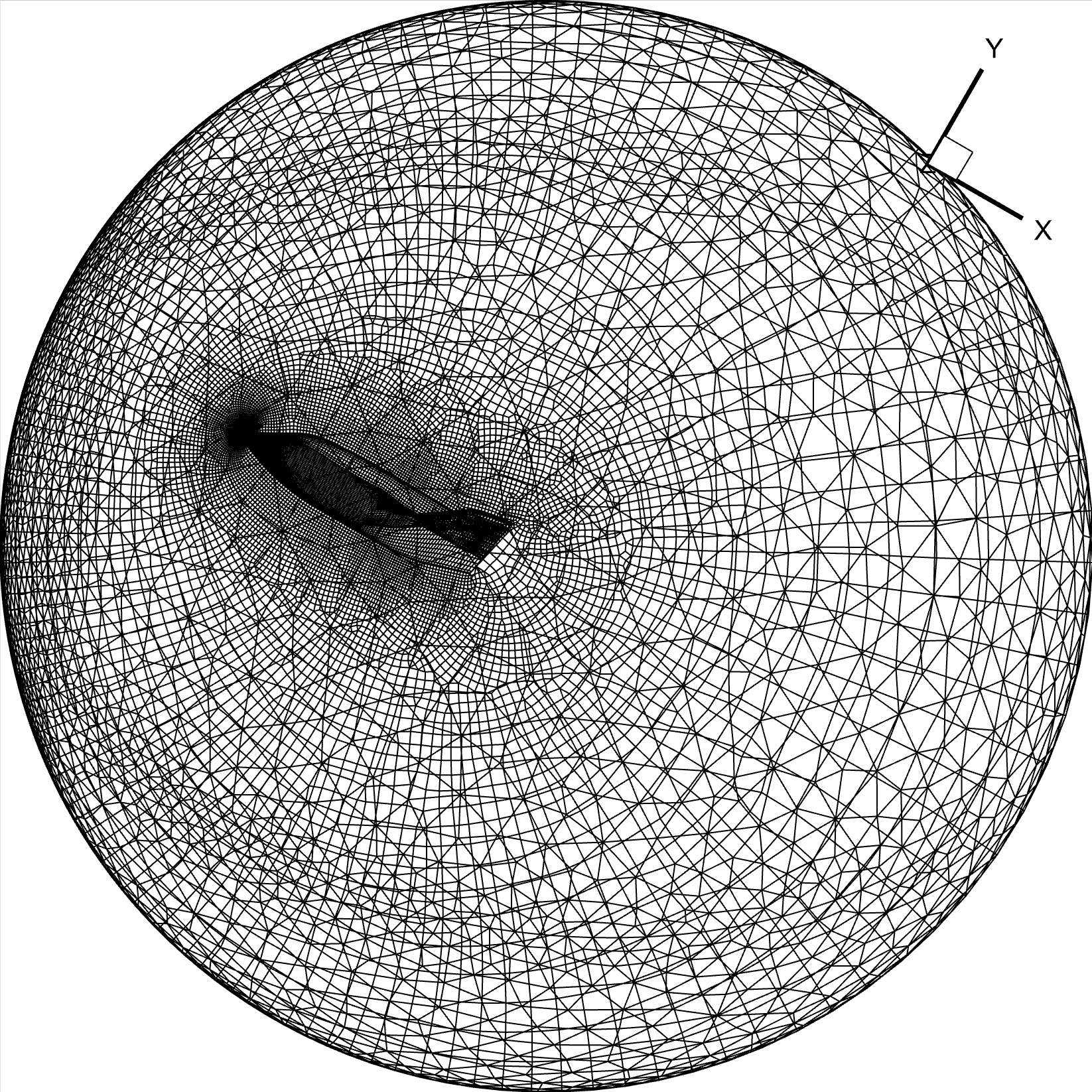}}
	\hspace{1mm}
	\subfloat[]{\includegraphics[width=0.4\textwidth]
			    {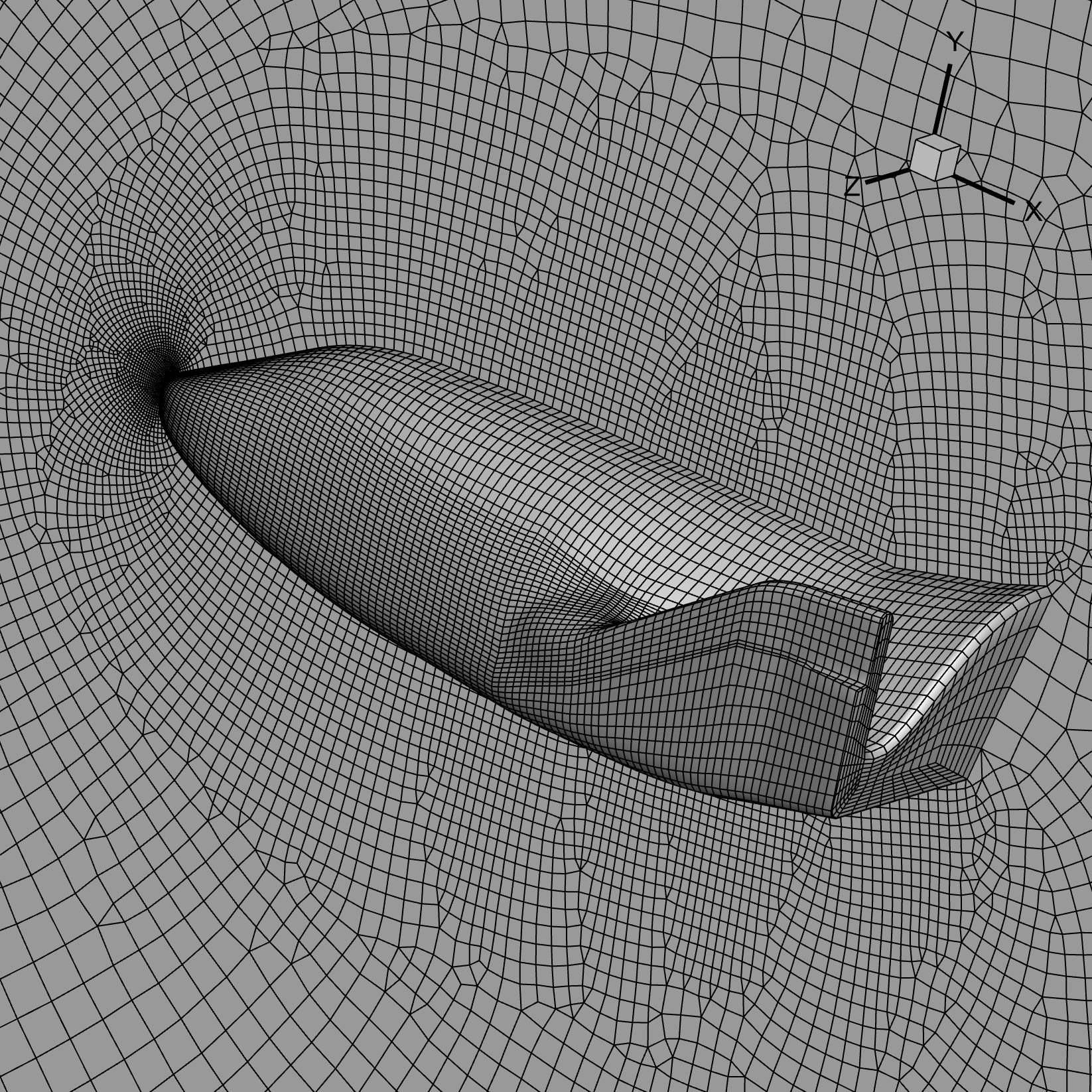} }
	\caption{Surface mesh of a space vehicle. (a) Global view and (b) local enlargement.}	
	\label{fig:x38-mesh}
\end{figure}

 The sketch of the vehicle is shown in Fig.~\ref{fig:x38-geo}. The reference length for the definition of the Knudsen number is $L_{ref} = 0.28$ m. Shown in Fig.~\ref{fig:x38-mesh}, the unstructured mesh of 560593 cells consists of 15277 pyramids and 545316 tetrahedra with the minimum cell height of $L_{ref}\times 10^{-3}$ near the front of the vehicle surface.
In order to have a clear understanding of the flow field, the local Knudsen number is defined as
\begin{equation*}
	{\rm Kn}_{GLL} = \frac{l_{mfp}}{\rho / \left| \nabla \rho \right|},
\end{equation*}
where $l_{mfp}$ is the local mean free path.
At the same time,
the local mesh Knudsen number for the determination of flow dynamics in each cell is defined by
\begin{equation*}
	{\rm Kn}_{mesh} = \frac{l_{mfp}}{\sqrt[3]{\Omega_i}},
\end{equation*}
where $\Omega_i$ is the cell volume.


The initial temperature of free stream is $T_{\infty} = 500 $ K, and the vehicle surface is treated as isothermal wall with a constant temperature $T_w = 500$ K. The angle of attack is $30^\circ$. Constant values of $Z_r = 5$ and $Z_v = 28$ are used in the cases. The reference number of particles per cell is set as $N_r = 400$. For both ${\rm Kn} = 10^{-3}$ and ${\rm Kn} = 10^{-5}$ tests, an initial flow field calculated by GKS at 15000 steps are adopted, and the time-averaging starts from 25000th steps up to 25000 steps. $\rm CFL = 0.5$ is employed.

\begin{figure}[H]
	\centering
	\subfloat[]{\includegraphics[width=0.45\textwidth]{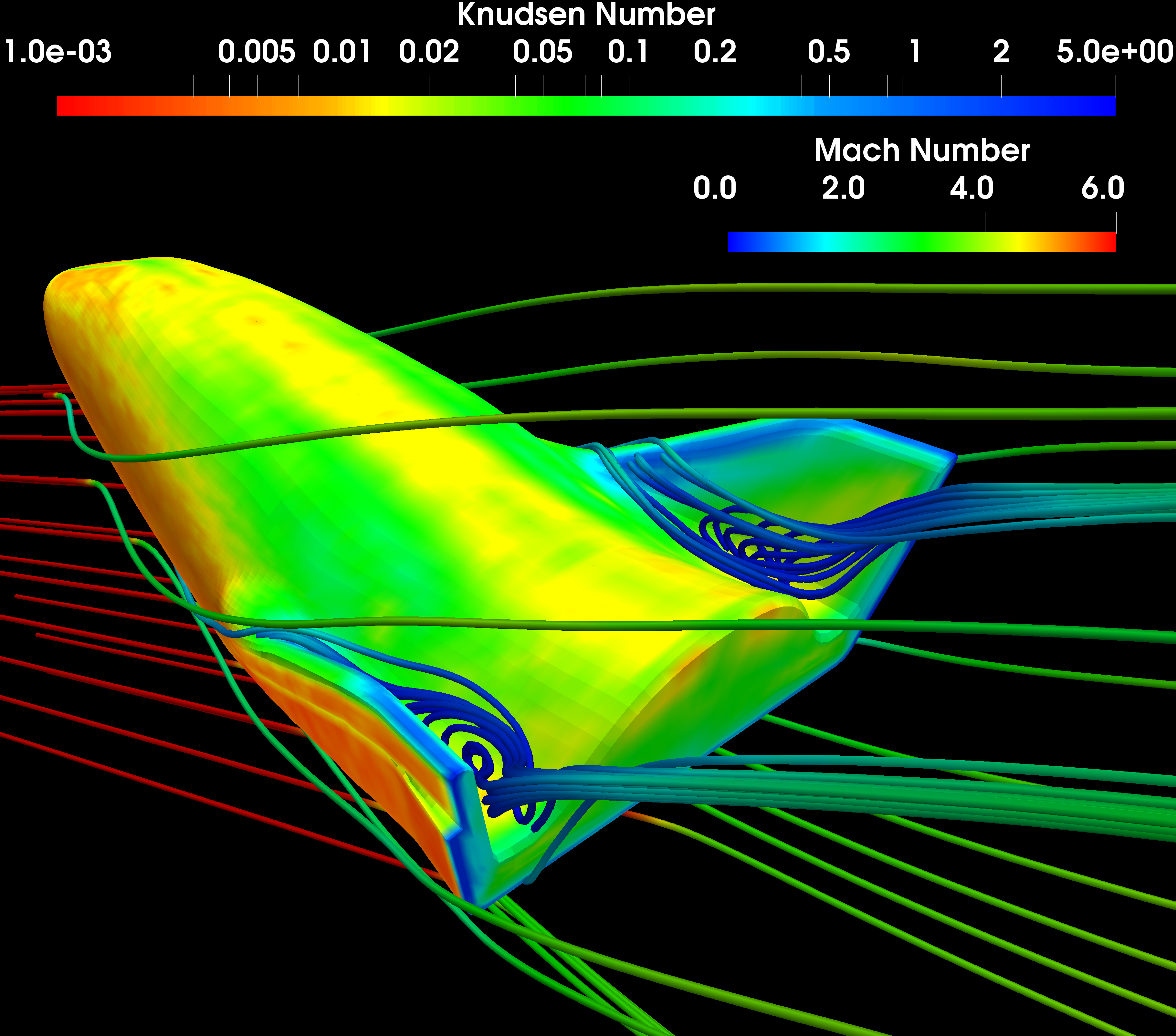}}
	\hspace{1mm}
	\subfloat[]{\includegraphics[width=0.45\textwidth]{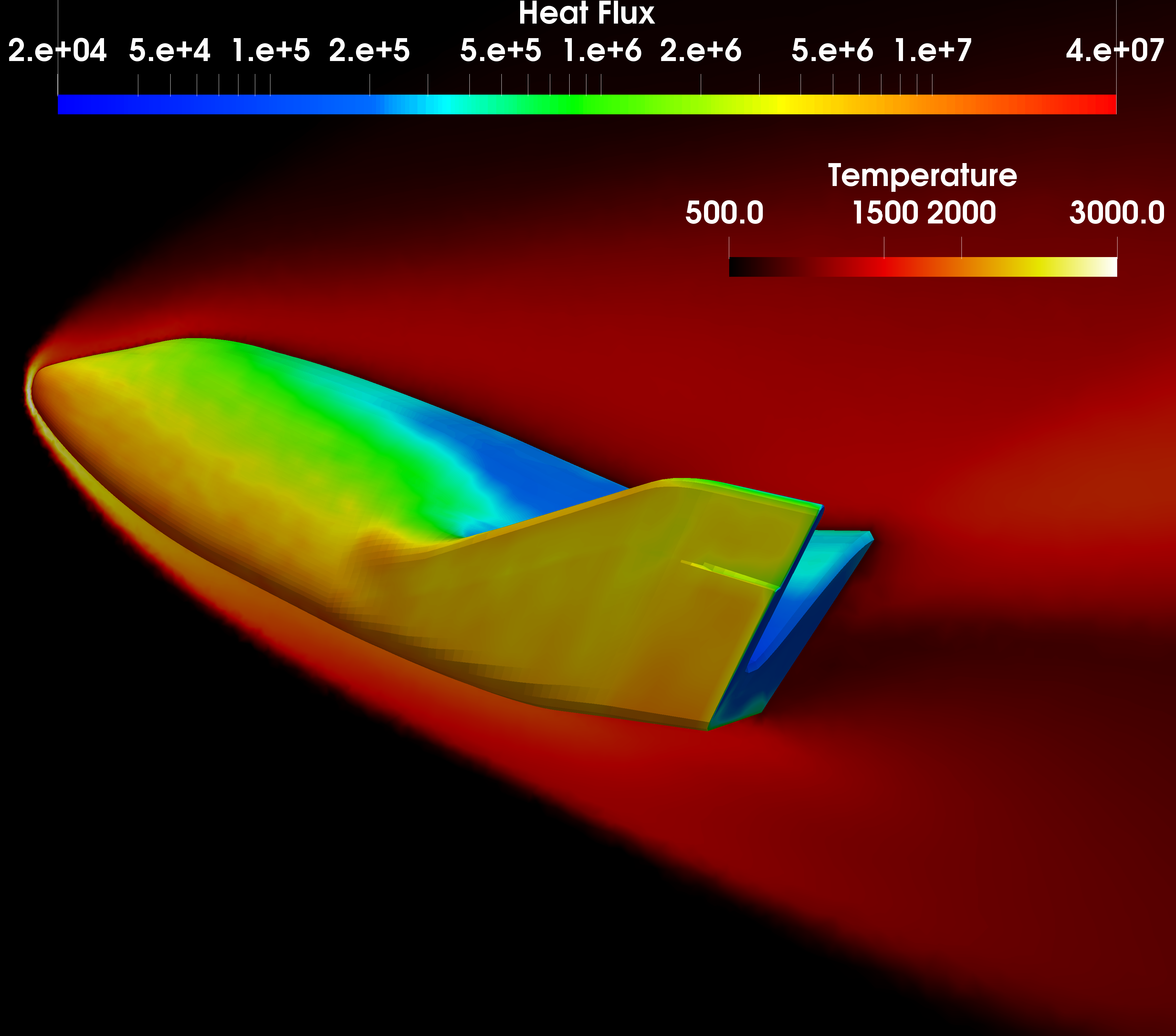}}\\
	\subfloat[]{\includegraphics[width=0.45\textwidth]{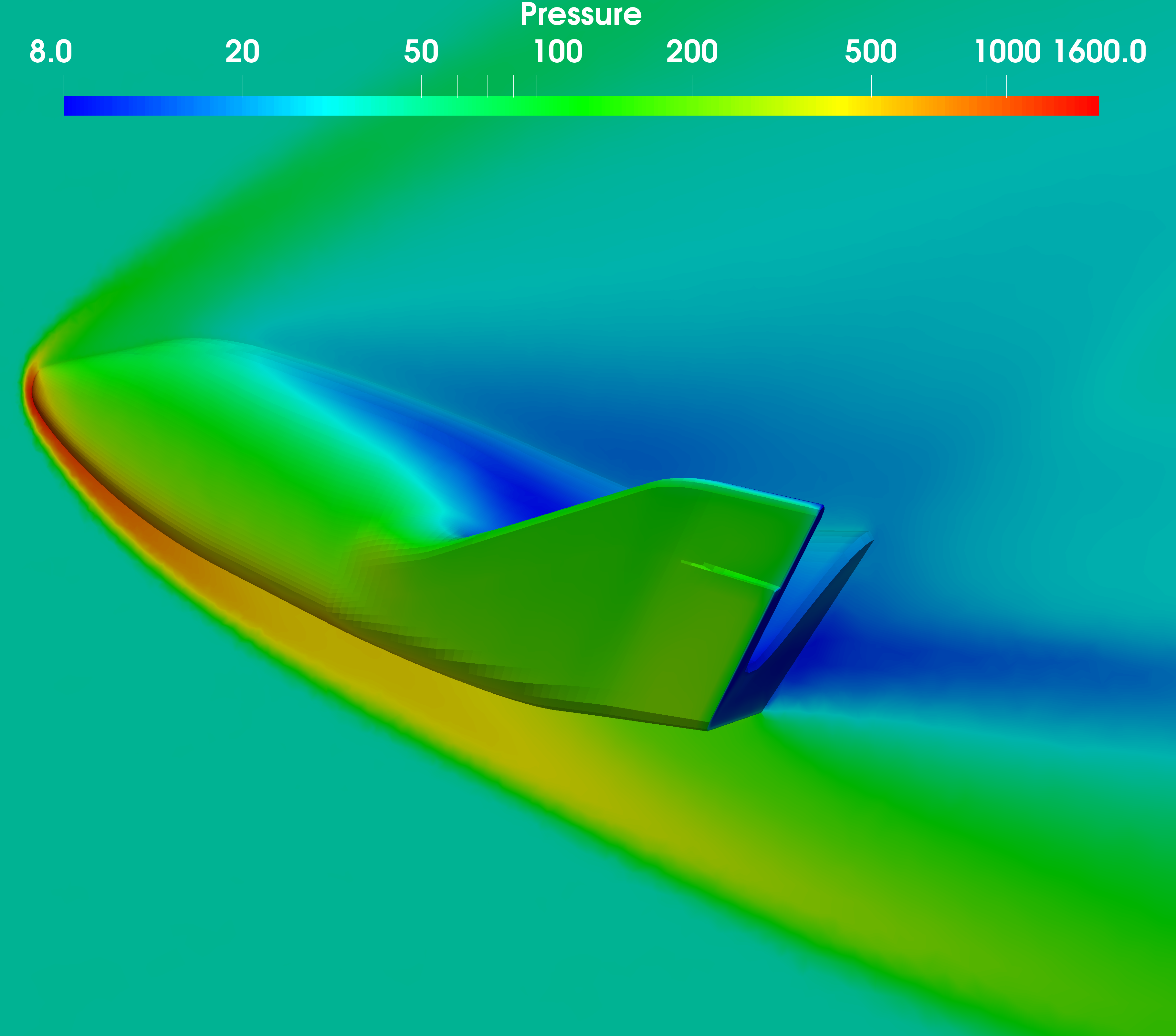}}
	\hspace{1mm}
	\subfloat[]{\includegraphics[width=0.45\textwidth]{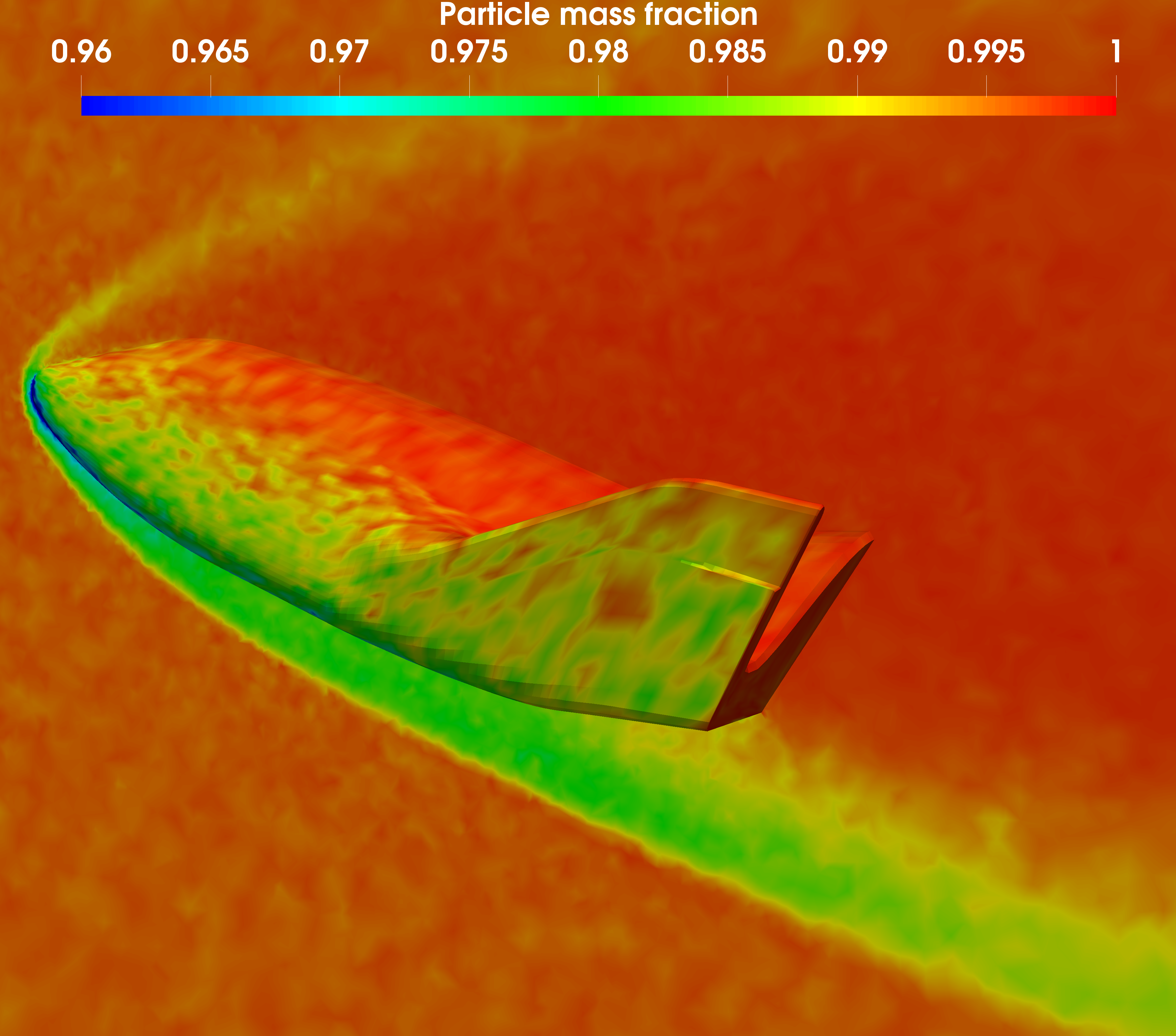}}
	\caption{Hypersonic flow at ${\rm Ma} = 6$ around a space vehicle for ${\rm Kn} = 10^{-3}$. (a) Local Knudsen number ${\rm Kn}_{GLL}$ on the surface and Mach number along the streamline, (b) heat flux (vehicle surface) and temperature (outside vehicle), (c) pressure, and (d) particle mass ratio distributions.}
	\label{fig:x38-Kn-Ma}
\end{figure}

For the case at ${\rm Kn} = 10^{-3}$, the distribution of local Knudsen number ${\rm Kn}_{GLL}$ around the surface of the vehicle, Mach number along the streamline, heat flux around the surface, temperature, pressure, and particle mass fraction defined by $(\rho - \rho^h)/\rho$ inside each cell are shown in Fig.~\ref{fig:x38-Kn-Ma}. It can be seen from Fig.~\ref{fig:x38-Kn-Ma}(a) the ${\rm Kn}_{GLL}$ has four orders of magnitude differences on the surface of vehicle. Fig.~\ref{fig:x38-Kn-Ma}(d) illustrate the particles are dominant in most parts of the computational domain.

\begin{figure}[H]
	\centering
	\subfloat[]{\includegraphics[width=0.5\textwidth]{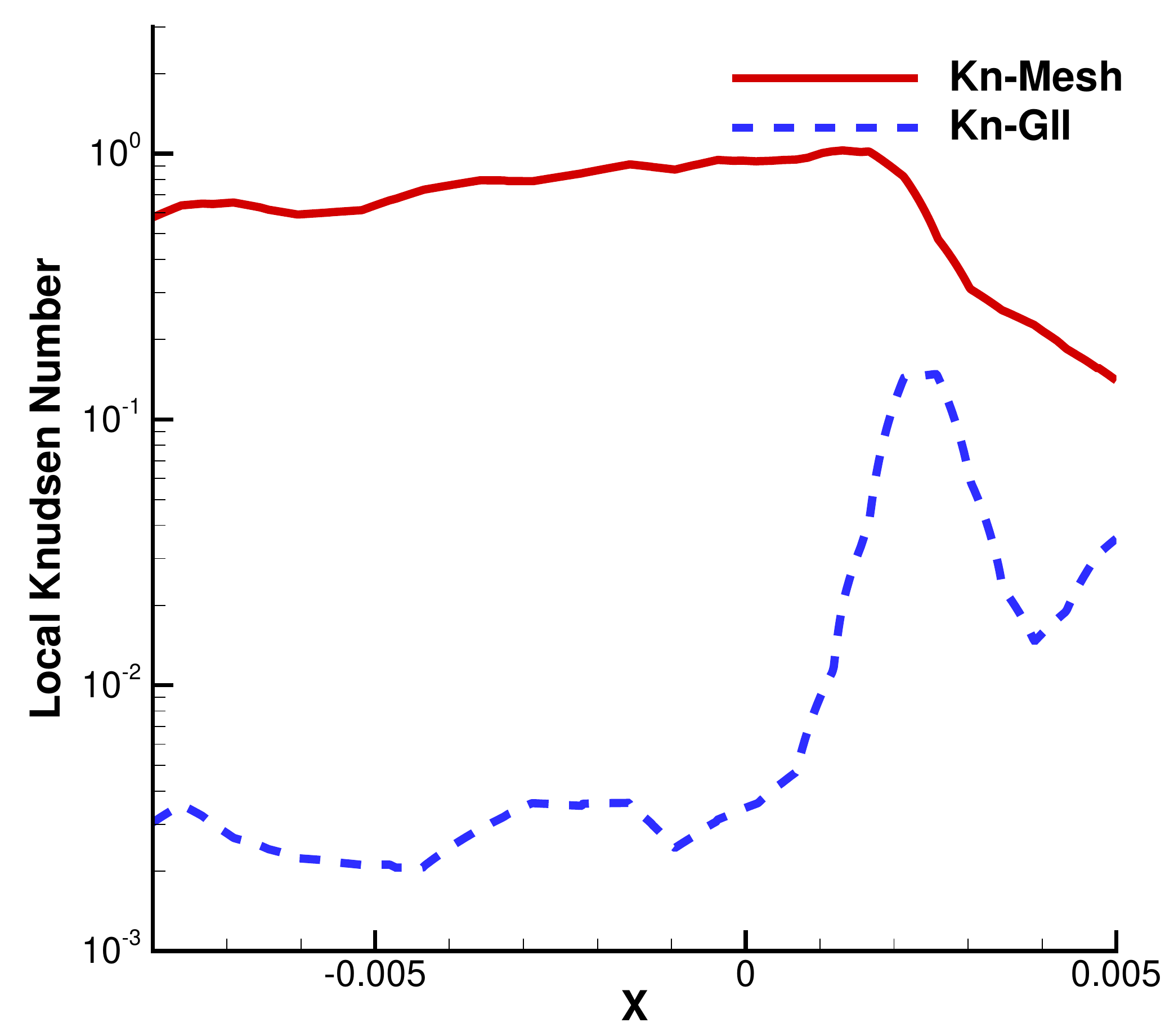}}
	\subfloat[]{\includegraphics[width=0.5\textwidth]{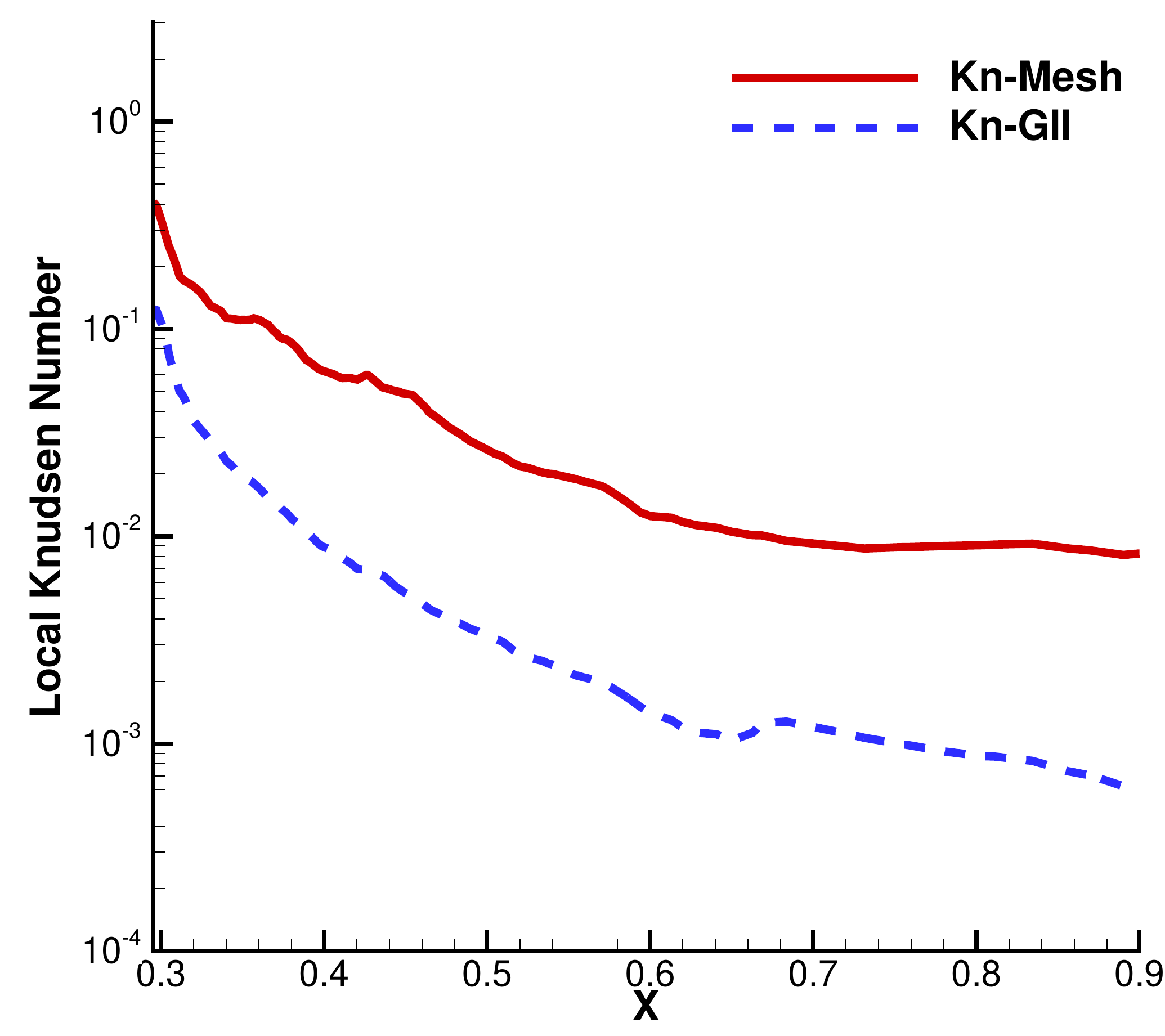}}
	\caption{Local Knudsen number distributions along the $y=0.03$ m line on the symmetry plane at ${\rm Ma} = 6$ and ${\rm Kn} = 10^{-3}$. (a) Windward and (b) leeward directions.}
	\label{fig:x38-Kn}
\end{figure}

\begin{figure}[H]
	\centering
	\includegraphics[width=8cm]{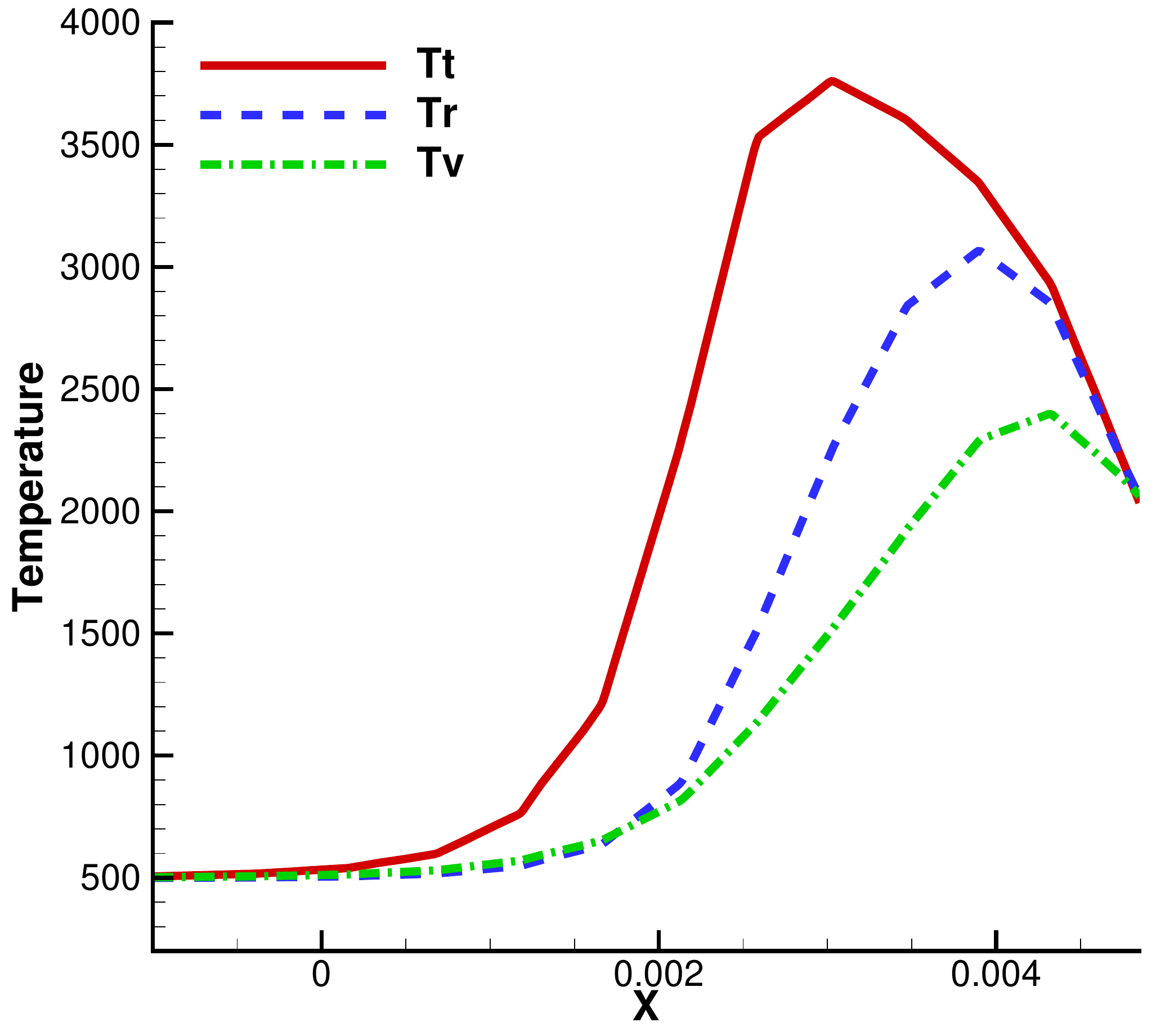}
	\caption{Temperature distributions along the $y=0.03$ m line on the symmetry plane at ${\rm Ma} = 6$ and ${\rm Kn} = 10^{-3}$.}
	\label{fig:x38-T}
\end{figure}

Figure~\ref{fig:x38-Kn} plots the distribution of the local Knudsen number and local mesh Knudsen number along the $y = 0.03$ m line on the symmetry plane in both windward and leeward (see Fig.~\ref{fig:x38-geo}). For DSMC method, the local mesh Knudsen number is restricted to be greater than 3, while the local mesh Knudsen number in the UGKWP method is less than 1 (see Fig.~\ref{fig:x38-Kn}). The UGKWP can save significant amount of computational resources in comparison with DSMC method.
Fig.~\ref{fig:x38-T} shows the distributions of translational, rotational and vibrational temperatures along the $y = 0.03$ m line on the symmetry plane in the front of vehicle, which displays the thermal non-equilibrium effect in the leading edge.

\begin{figure}[H]
	\centering
	\subfloat[]{\includegraphics[width=0.45\textwidth]{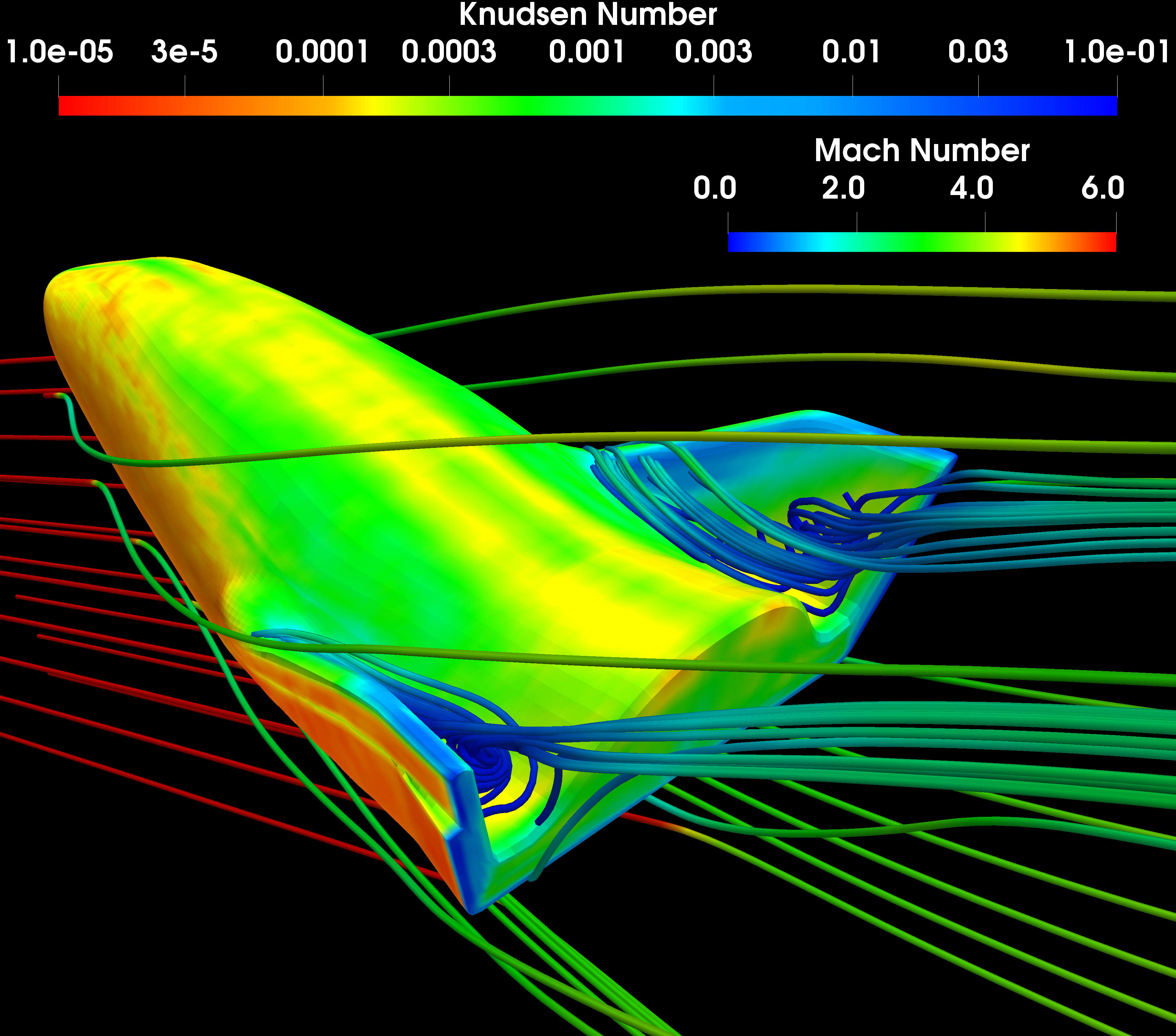}}
	\hspace{1mm}
	\subfloat[]{\includegraphics[width=0.45\textwidth]{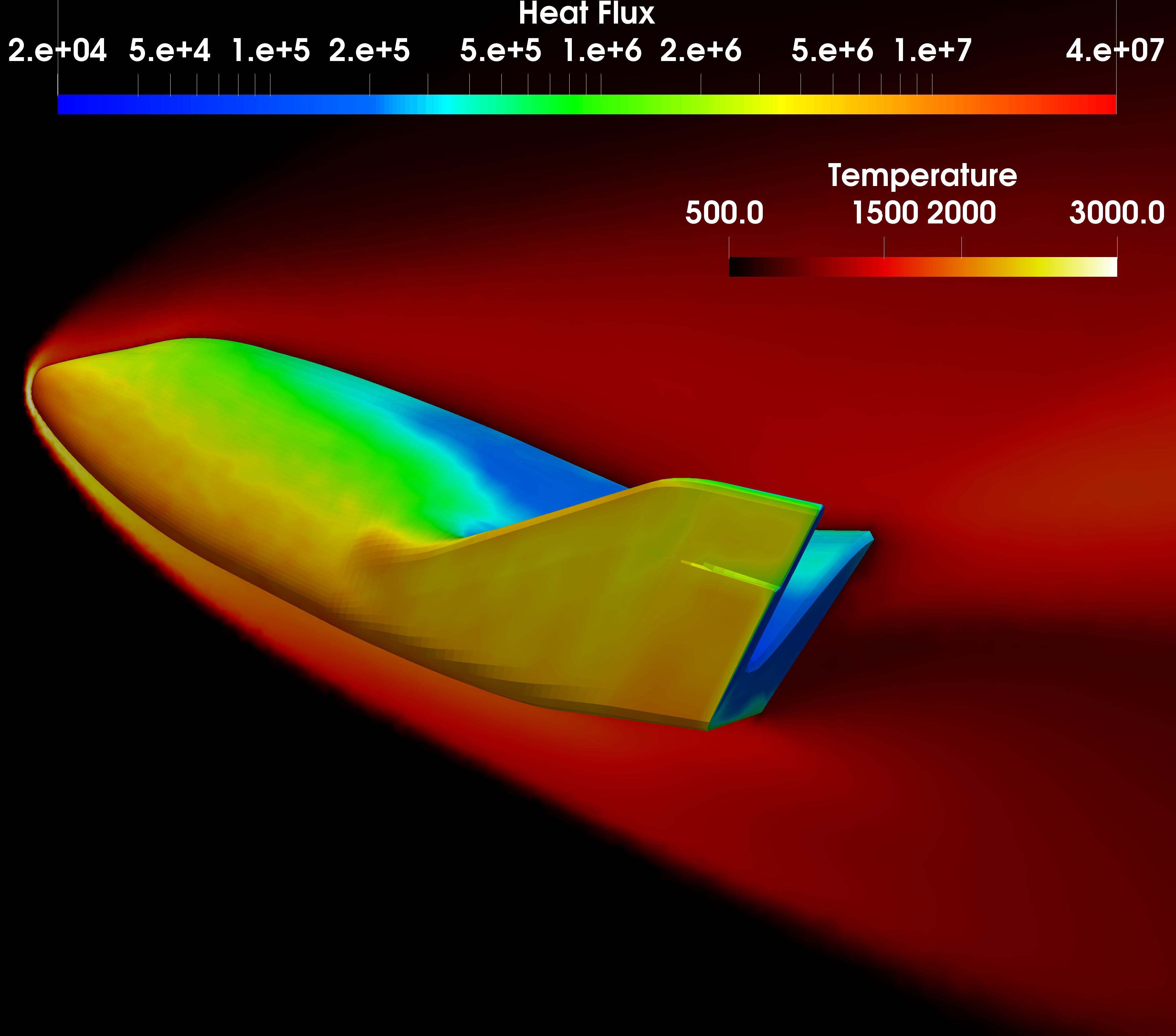}} \\
	\subfloat[]{\includegraphics[width=0.45\textwidth]{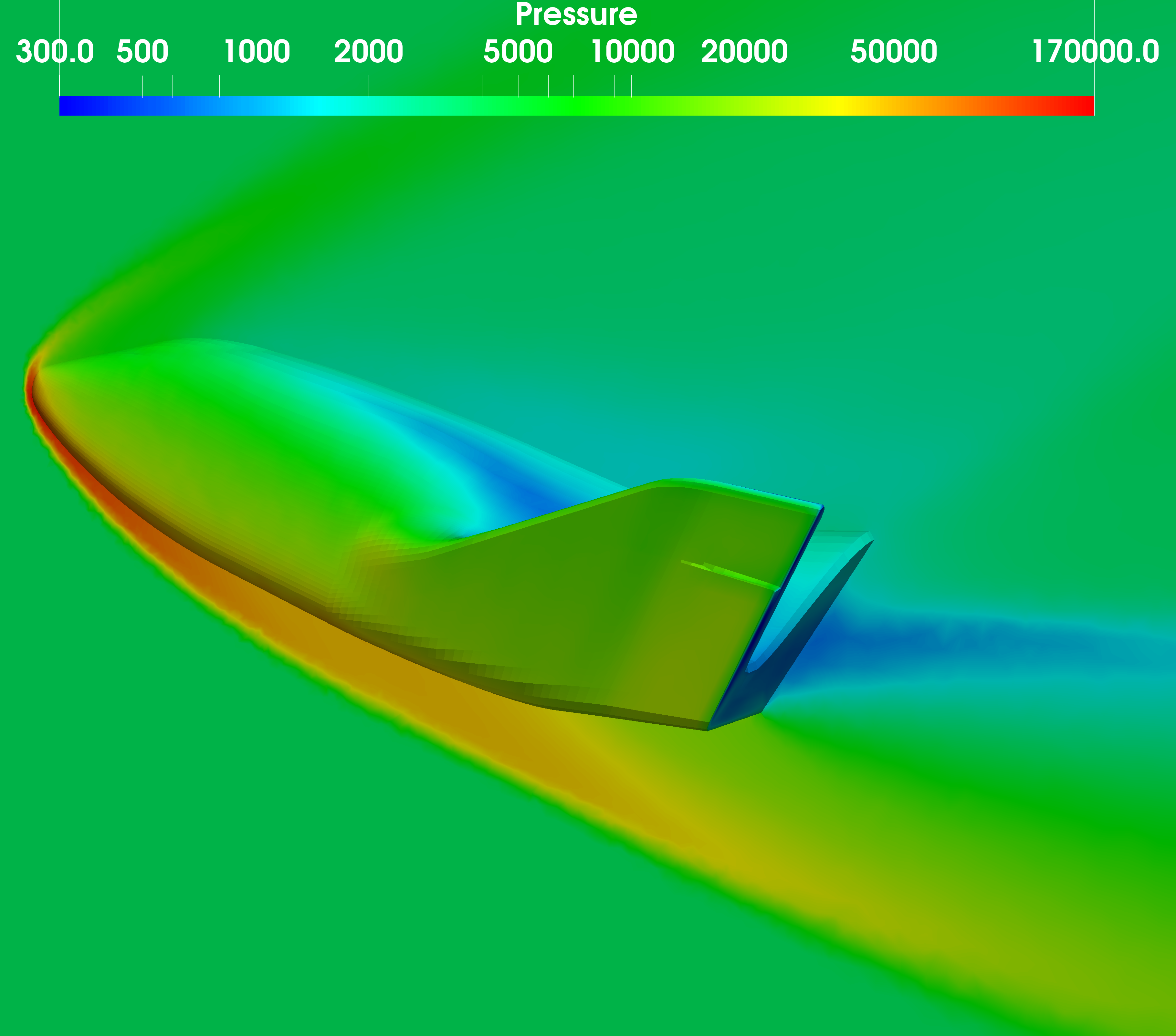}}
	\hspace{1mm}
	\subfloat[]{\includegraphics[width=0.45\textwidth]{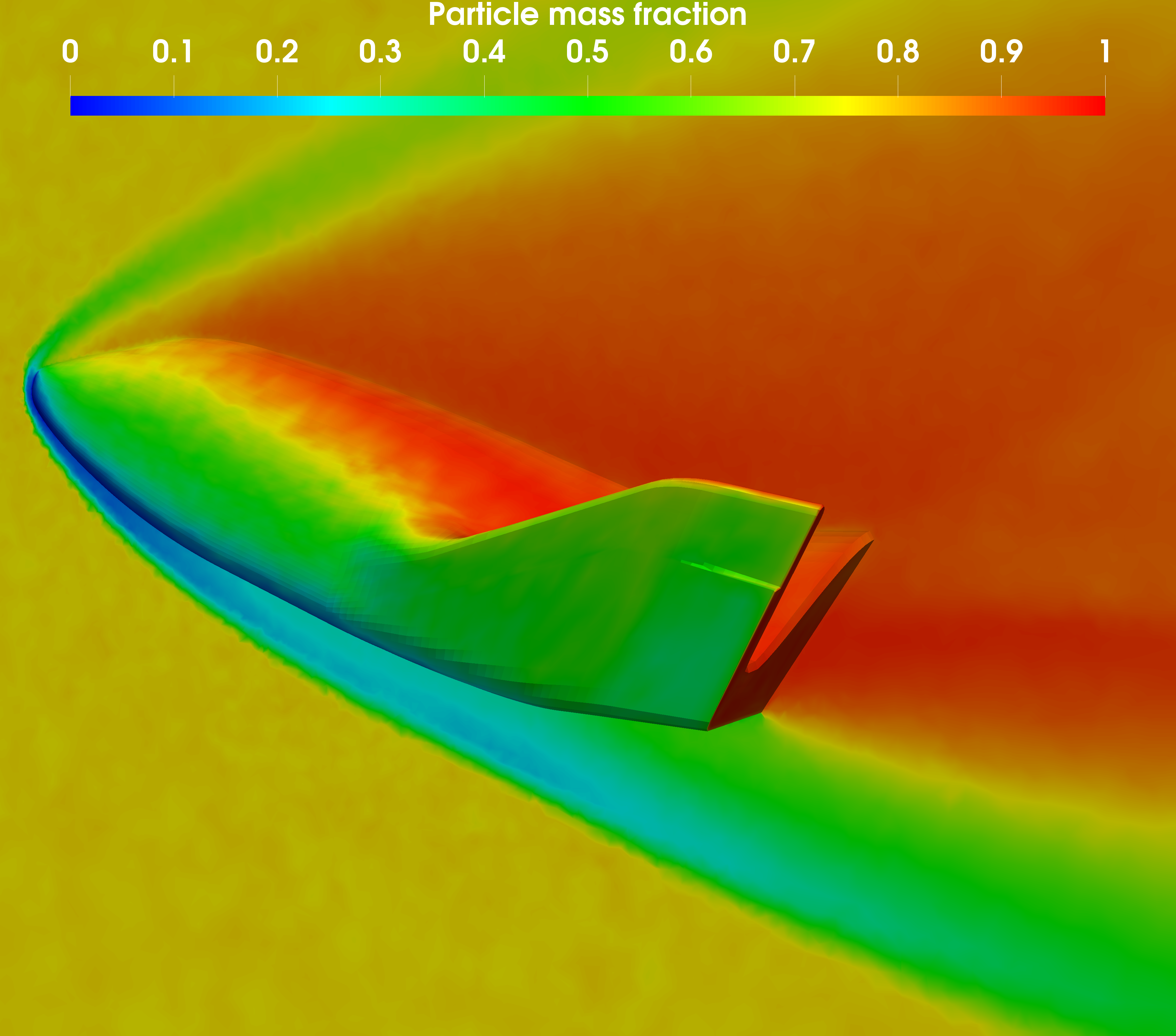}}
	\caption{Hypersonic flow at ${\rm Ma} = 6$ around a space vehicle for ${\rm Kn} = 10^{-5}$. (a) Local Knudsen number ${\rm Kn}_{GLL}$ on the surface and Mach number along the streamline, (b) heat flux (vehicle surface) and temperature (outside vehicle), (c) pressure, and (d) particle mass ratio distributions.}
	\label{fig:x38-kn1e-5-Kn-Ma}
\end{figure}

\begin{figure}[H]
	\centering
	\subfloat[]{\includegraphics[width=0.5\textwidth]{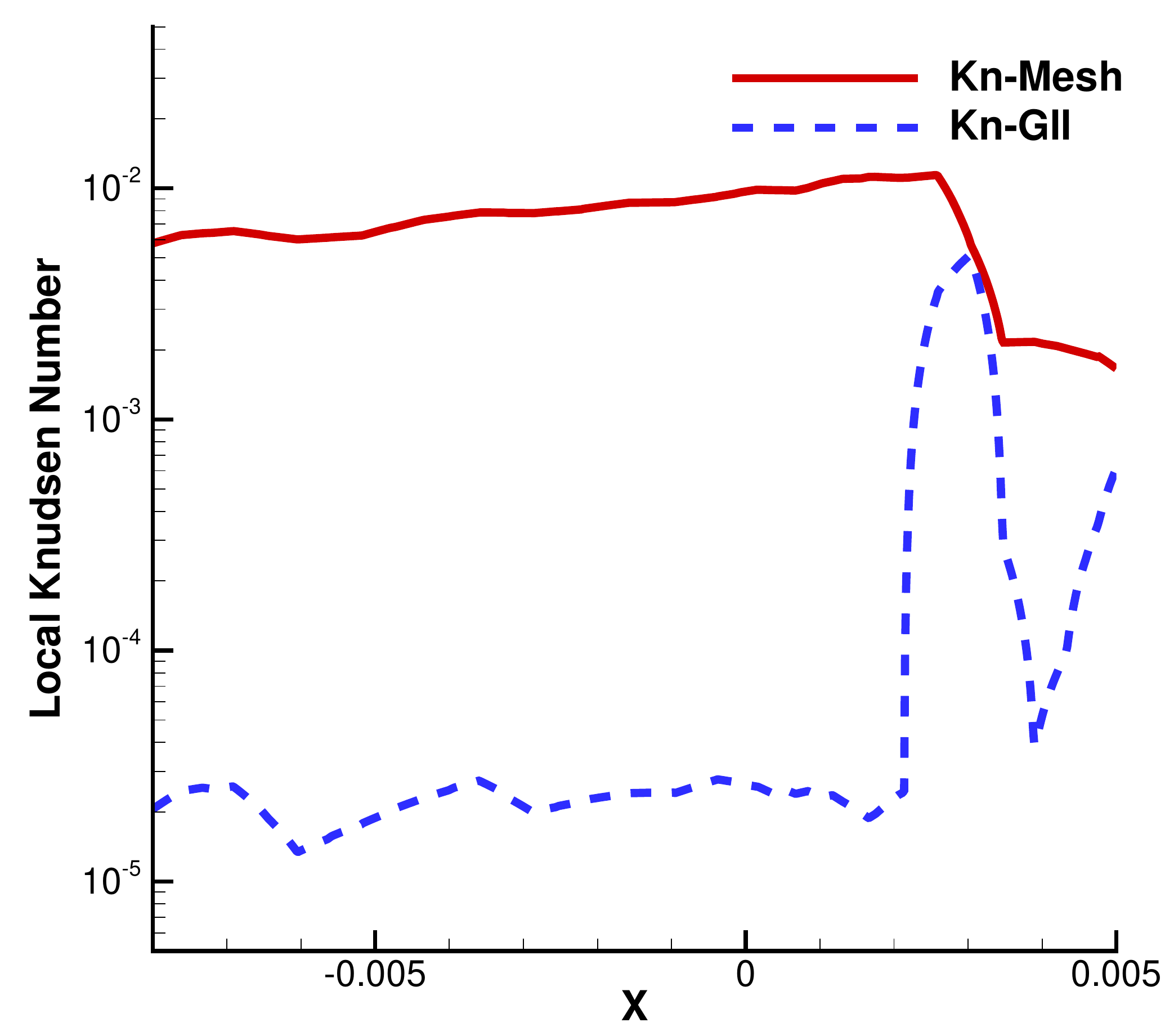}}
	\subfloat[]{\includegraphics[width=0.5\textwidth]{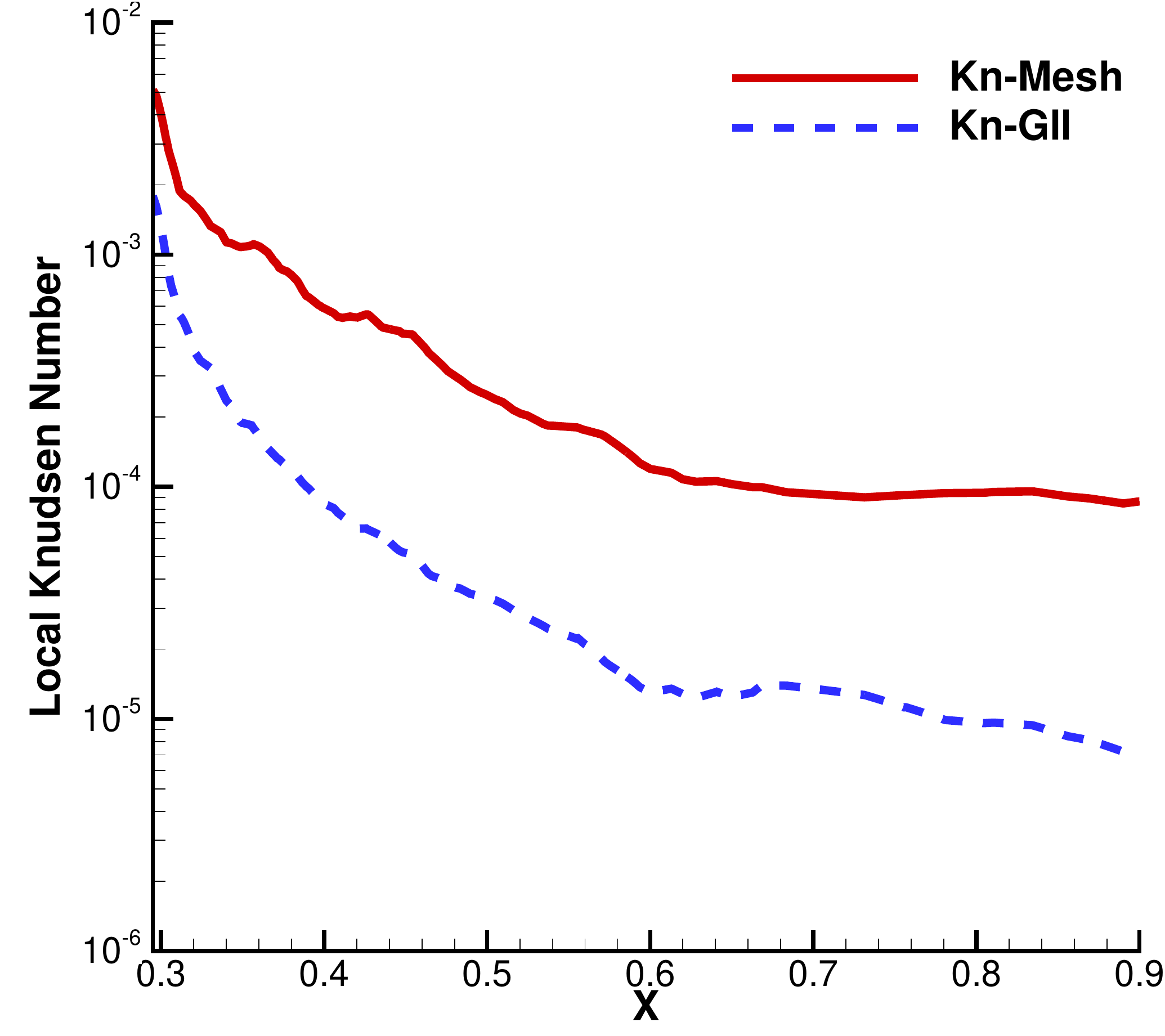}}
	\caption{Local Knudsen number distributions along the $y=0.03$ m line on the symmetry plane at ${\rm Ma} = 6$ and ${\rm Kn} = 10^{-5}$. (a) Windward and (b) leeward directions.}
	\label{fig:x38-kn1e-5-Kn}
\end{figure}

\begin{figure}[H]
	\centering
	\includegraphics[width=8cm]{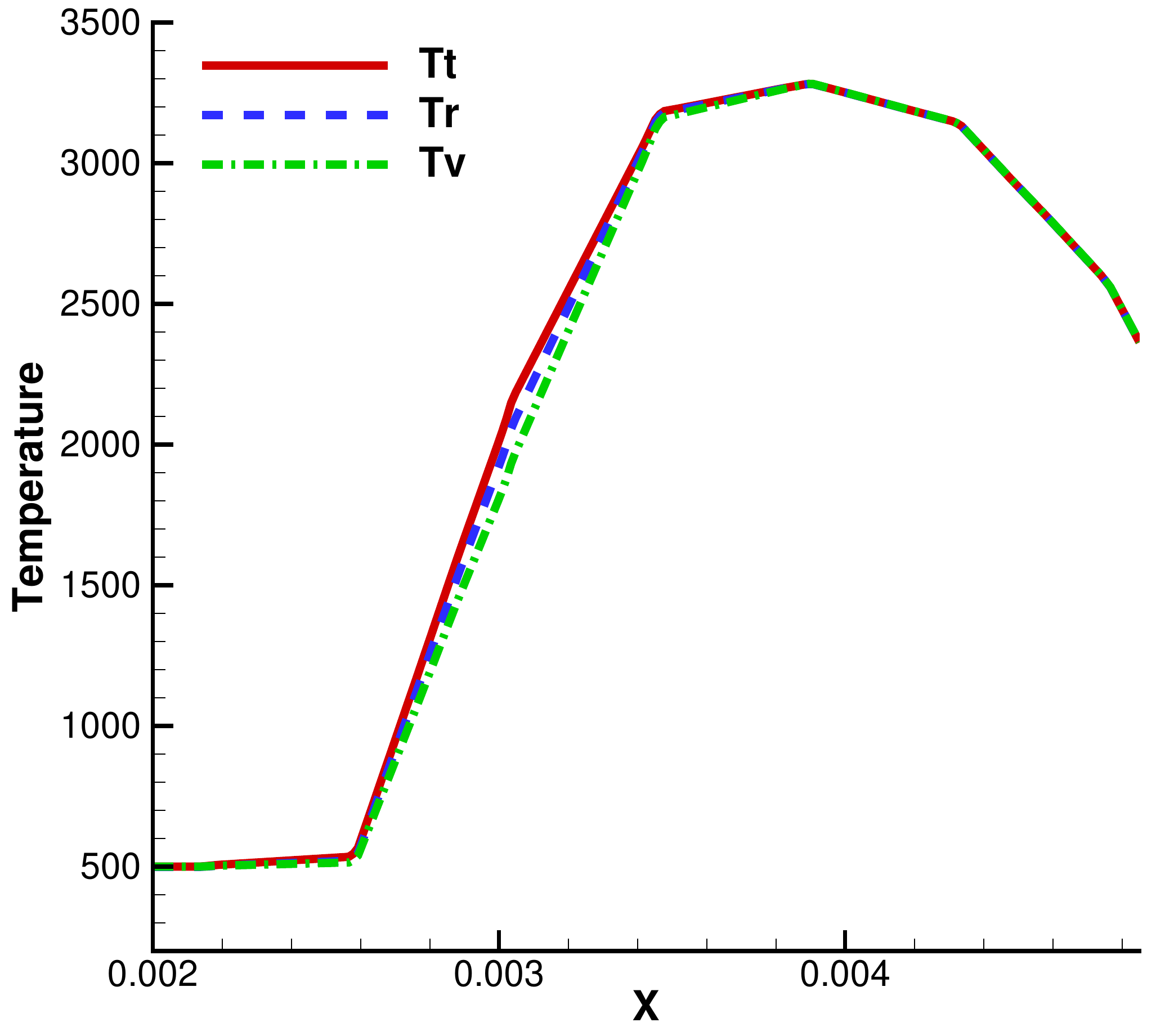}
	\caption{Temperature distributions along the $y=0.03$ m line on the symmetry plane at ${\rm Ma} = 6$ and ${\rm Kn} = 10^{-5}$.}
	\label{fig:x38-kn1e-5-T}
\end{figure}

For the case of ${\rm Kn} = 10^{-5}$, the distribution of local Knudsen number ${\rm Kn}_{GLL}$ around the surface of the vehicle, Mach number along the streamline, heat flux around the surface, temperature, pressure, and particle mass fraction distributions are shown in Fig.~\ref{fig:x38-kn1e-5-Kn-Ma}. From Fig.~\ref{fig:x38-kn1e-5-Kn-Ma}(a), in the computational domain there is still four orders of magnitude differences in local Knudsen number even at ${\rm Kn} = 10^{-5}$. Due to the wave-particle decomposition, the particle appears only at the region with relatively large cell's Knudsen number (see Fig.~\ref{fig:x38-kn1e-5-Kn-Ma}(d)). The analytical wave and stochastic particles are dynamically coupled in each cell, which can be hardly treated by a hybrid NS-DSMC method with sub-domains separated by a buffer zone.
The local Knudsen number and the local mesh Knudsen number along the $y=0.03$ m on the symmetry plane are plotted in Fig.~\ref{fig:x38-kn1e-5-Kn}. It shows a large variation of local Knudsen number as well.
The small mesh Knudsen number used in UGKWP indicates that the computational cost for the DSMC method will become unaffordable in this test. The translational, rotational, and vibrational temperatures are plotted in Fig.~\ref{fig:x38-kn1e-5-T}.
For both cases at ${\rm Kn} = 10^{-3}$ and ${\rm Kn} = 10^{-5}$, the simulations take 22.5 hours and 18.7 hours running on Tianhe-2 with 10 nodes (240 cores, Intel Xeon E5-2692 v2, 2.2 GHz), respectively. For UGKWP, there are no significant differences in terms of computational cost in transition regime.


\section{Conclusion}\label{sec:conclusion}

In this paper, a unified gas-kinetic wave-particle (UGKWP) method for diatomic gas with vibrational relaxation model is constructed.
Different from the discrete particle velocity space based unified gas-kinetic scheme (UGKS-DVM), the UGKWP method adopts a wave-particle decomposition for capturing the evolution of the gas-distribution function. More specifically, the wave is for the description of equilibrium part in the distribution function through macroscopic variables and the particle is for the non-equilibrium part.
The multiscale transport process in all Knudsen regimes is recovered through the automatic and dynamic distributions of the weights between the wave and particle decomposition.
In the continuum flow regime, the UGKWP gets back to the hydrodynamic Navier--Stokes solver, the so-called gas-kinetic scheme (GKS), without introducing any particles. In the highly rarefied regime, the UGKWP becomes a purely stochastic particle method.
As a result, the UGKWP can achieve a balance between the physical accuracy and numerical efficiency in the simulation of multiscale flow problem.
With inclusion of molecular vibrational degrees of freedom, in this paper the UGKWP method extends its applicable regime to high-speed and high-temperature flow with the excitation of vibrational mode for the diatomic gas.

The UGKWP method is validated through many test cases and the comparison with the results from DSMC and experiments measurements.
In comparison with the DVM-based UGKS, the UGKWP method shows great advantages in its high computational efficiency and memory reduction.
The UGKWP becomes a useful method in the simulation of three-dimensional high-speed high-temperature rarefied and continuum flow with affordable computational resources.

\section*{Author's contributions}

All authors contributed equally to this work.

\section*{Acknowledgments}
This work was supported by Hong Kong research grant council (16208021,16301222), and National Natural Science Foundation of China (12172316).

\section*{Data Availability}

The data that support the findings of this study are available from the corresponding author upon reasonable request.

\appendix

\section{Moments and derivative of the Maxwellian distribution function with vibrational mode}\label{sec:appendix-1}

In the unified gas-kinetic wave-particle method with the vibrational mode, the equilibrium flux $\vec{F}^{eq}_{ij}$ in Eq.~\eqref{eq:Feq} requires higher order moments of $ \vec{\xi}$ and $\varepsilon_v$. Here, we list the formula of the moments
\begin{equation*}
	\begin{aligned}
		&\int_{ -\infty }^{ \infty } \frac{\lambda_r}{\pi } \vec{\xi}^2
		e^{ - \lambda_r \vec{\xi}^2} {\rm{d}} {\vec{\xi}} = \frac{K_r}{2 \lambda_r} , \\
		&\int_{ -\infty }^{ \infty } \frac{\lambda_r}{\pi } \vec{\xi}^4
		e^{ - \lambda_r \vec{\xi}^2} {\rm{d}} {\vec{\xi}} = \frac{ K_r^2 + 2 K_r }{ 4 \lambda_r^2} ,\\
		&\int_0^{ \infty }           \frac{4 \lambda_v}{K_v(\lambda_v)} \varepsilon_v
		e^{ - \frac{ 4 \lambda_v }{ K_v(\lambda_v) }{ \varepsilon_v }} {\rm{d}} \varepsilon_v
		= \frac{ K_v(\lambda_v) }{4 \lambda_v}, \\
		&\int_0^{ \infty } \frac{4 \lambda_v}{K_v(\lambda_v)} \varepsilon_v^2
		e^{ - \frac{ 4 \lambda_v}{K_v(\lambda_v)}\varepsilon_v} {\rm{d}} \varepsilon_v
		= 2 \left( \frac{K_v(\lambda_v)}{ 4 \lambda_v} \right)^2.
	\end{aligned}
\end{equation*}

The distribution of the equilibrium state in space and time $( \vec{r}, t)$ can be expanded by the Taylor expansion
\begin{equation*}
	g^\ast(\vec{r}, t) = g_0 + \vec{r} \cdot \frac{\partial g}{\partial \vec{r}}
	                   + \frac{\partial g}{\partial t} t.
\end{equation*}
As an example, taking the $x-$ as the normal direction of the cell interface, the micro-slope $a$ can be defined by
\begin{equation*}
	a=\frac{1}{g}\bigg(\frac{\partial g}{\partial x}\bigg),
\end{equation*}
with the form
\begin{equation*}
	a =a_1
	+a_2 u
	+a_3 v
	+a_4 w
	+\frac12 a_5\vec{u}^2
	+\frac12 a_6\vec{\xi}^2
	+ a_7{\varepsilon_v}.
\end{equation*}
Applying the chain rule, the micro-slope $a$ can be determined by the derivative of macroscopic quantities evaluated at $({\vec r},t)$
\begin{equation*}
	\begin{aligned}
	&a_1 = \frac{1}{\rho }\frac{{\partial \rho }}{{\partial x}}
			 - {a_2}U - {a_3}V - {a_4}W
			 - \frac{1}{2}{a_5}\left( {{{\vec{U}}^2} + \frac{3}{{2{\lambda _t}}}} \right)
			 - \frac{1}{2}{a_6}\frac{{{K_r}}}{{2{\lambda _r}}}
			 - {a_7}\frac{K_v(\lambda_v)}{4\lambda_v}, \\
	&a_2 = \frac{\lambda_t}{\rho} R_1 - a_5 U,\\	
	&a_3 = \frac{\lambda_t}{\rho} R_2 - a_5 V,\\
	&a_4 = \frac{\lambda_t}{\rho} R_3 - a_5 W,\\
	&a_5 = \frac{4\lambda_t^2}{3 \rho} \left( B- U R_1 - V R_2 - W R_3 \right),\\
	&a_6 =  \frac{ 4 \lambda_r^2 } { K_r \rho }
	       \left( \frac{4}{ K_r } \frac{ \partial (\rho E_r)}{ \partial x }
		        - \frac{1}{ \lambda_r} \frac{ \partial \rho }{ \partial x} \right),\\
	&a_7 = \frac{ 4 e^{ 2 \Theta_v R \lambda_v } \lambda_v^2 }
	            { ( 4 \lambda_v R \Theta_v + K_v(\lambda_v)) \rho }
			\left( \frac{4}{K_v(\lambda_V)} \frac{ \partial (\rho E_v) }{ \partial x}
			     - \frac{1}{\lambda_v}      \frac{ \partial \rho }{ \partial x} \right),
	\end{aligned}
\end{equation*}
with the defined variables
	\begin{align*}
	\begin{aligned}
		B &= 2 \frac{\partial (\rho E - \rho E_{r} - \rho E_v )}{\partial x}
		   - (\vec{U}^2 + \frac{3}{2\lambda_t}) \frac{\partial \rho}{\partial x} ,\\
		R_1 &= 2 \frac{\partial \rho U}{\partial x}  - 2U \frac{\partial \rho}{\partial x} ,\\
		R_2 &= 2 \frac{\partial \rho V}{\partial x}  - 2V \frac{\partial \rho}{\partial x} ,\\
		R_3 &= 2 \frac{\partial \rho W}{\partial x}  - 2W \frac{\partial \rho}{\partial x}.\\
		\end{aligned}
	\end{align*}

\section{Sampling particles with vibrational mode}\label{sec:sampling}

In the collision processes, simulation particles $P_k(m_k, {\vec r}_k, {\vec u}_k, e_{r,k}, e_{v,k})$ will be resampled from a given Maxwellian distribution function to recover the gas distribution function on the microscopic level. With the primary variables ($\rho_s, {\vec{U}}_s, \lambda_{t,s}, \lambda_{r,s}, \lambda_{v,s}$), a Maxwellian distribution function is given by
\begin{equation*}
	g_{s} = \rho_s
	        \left( \frac{\lambda_{t,s}}{\pi} \right) ^{\frac{3}{2}}  e^{ -\lambda_{t,s} {\vec {c}}^2}
	        \left( \frac{\lambda_{r,s}}{\pi}               \right) e^{ -\lambda_{r,s} {\vec {\xi}}^2}
	        \frac{4 \lambda_{v,s}}{K_v(\lambda_{v,s})}
			e^{-\frac{ 4 \lambda_{v,s}}{K_v(\lambda_{v,s})} \varepsilon_{v}}.
\end{equation*}
The microscopic translational velocity ${\vec u}_k = (u_k, v_k, w_k)^T$ for each particle can be obtained from \cite{bird1994molecular}
\begin{equation*}
\begin{aligned}
	&u_k=U_s+ \sqrt{-\ln \left(r_1\right) / \lambda_{t,s}} \cos \left(2 \pi r_2\right),\\
	&v_k=V_s+ \sqrt{-\ln \left(r_1\right) / \lambda_{t,s}} \sin \left(2 \pi r_2\right),\\
	&w_k=W_s+ \sqrt{-\ln \left(r_3\right) / \lambda_{t,s}} \cos \left(2 \pi r_4\right),
\end{aligned}
\end{equation*}
where $U_s$, $V_s$, and $W_s$ are the components of ${\vec{U}_s}$. $r_1$, $r_2$, $r_3$, and $r_4$ are independent random numbers generated from the uniform distribution between the interval $(0,1)$. A symmetric sampling process is adopted to reduce the variance. Specifically, from a group of $r_1$, $r_2$, $r_3$, and $r_4$, a pair of simulation particles with microscopic velocity ${\vec {u}_k}$ and ${\vec {u}_k^\prime}$ are sampled, where the symmetric microscopic velocity is
\begin{equation*}
\begin{aligned}
	&u_k^\prime = U_s - \sqrt{-\ln \left(r_1\right) / \lambda_{t,s}} \cos \left(2 \pi r_2\right), \\
	&v_k^\prime = V_s - \sqrt{-\ln \left(r_1\right) / \lambda_{t,s}} \sin \left(2 \pi r_2\right), \\
	&w_k^\prime = W_s - \sqrt{-\ln \left(r_3\right) / \lambda_{t,s}} \cos \left(2 \pi r_4\right).
\end{aligned}
\end{equation*}

Given with a preset reference number $N_{r}$ for each cell, the number of particles to be sampled is determined by
\begin{equation*}
	N_s=
	\begin{cases}
		0, & \text { if } \Omega_s \rho_s \leq m_{\min }, \\
		2 \lceil \frac{\rho_s N_r}{2 (\rho - \rho^h + \rho^h e^{-\Delta t /\tau} )} \rceil, & \text { if } \Omega_s \rho_s>m_{\min },\end{cases}
\end{equation*}
where $\Omega_s$ is the cell volume and $m_{min}$ is the minimum mass to sample. In the sampling process, for the cases $N_s > 0$, the mass weight actually sampled for each simulation particle is
\begin{equation*}
	m_k = \frac{\Omega_s \rho_s}{N_s},
\end{equation*}
which guarantees the mass density $\rho_s$ in the volume $\Omega_s$ after the sampling process.

The rotational energy $e_{r,k}$ and vibrational energy $e_{v,k} $ for simulated particles are calculated by
\begin{equation*}
	e_{r,k} = \frac{K_r}{4 \lambda_{r,s}}
	\quad
	{\rm and}
	\quad
	e_{v,k} = \frac{K_v(\lambda_{v,s})}{4 \lambda_{v,s}}.
\end{equation*}
The position $\vec{r}_k$ is derived from the uniform distribution on $\Omega_s$. Thus far, we get all information of a simulated particle with a given Maxwellian distribution function.

\begin{figure}[H]
	\centering
	\includegraphics[width=0.9\textwidth]{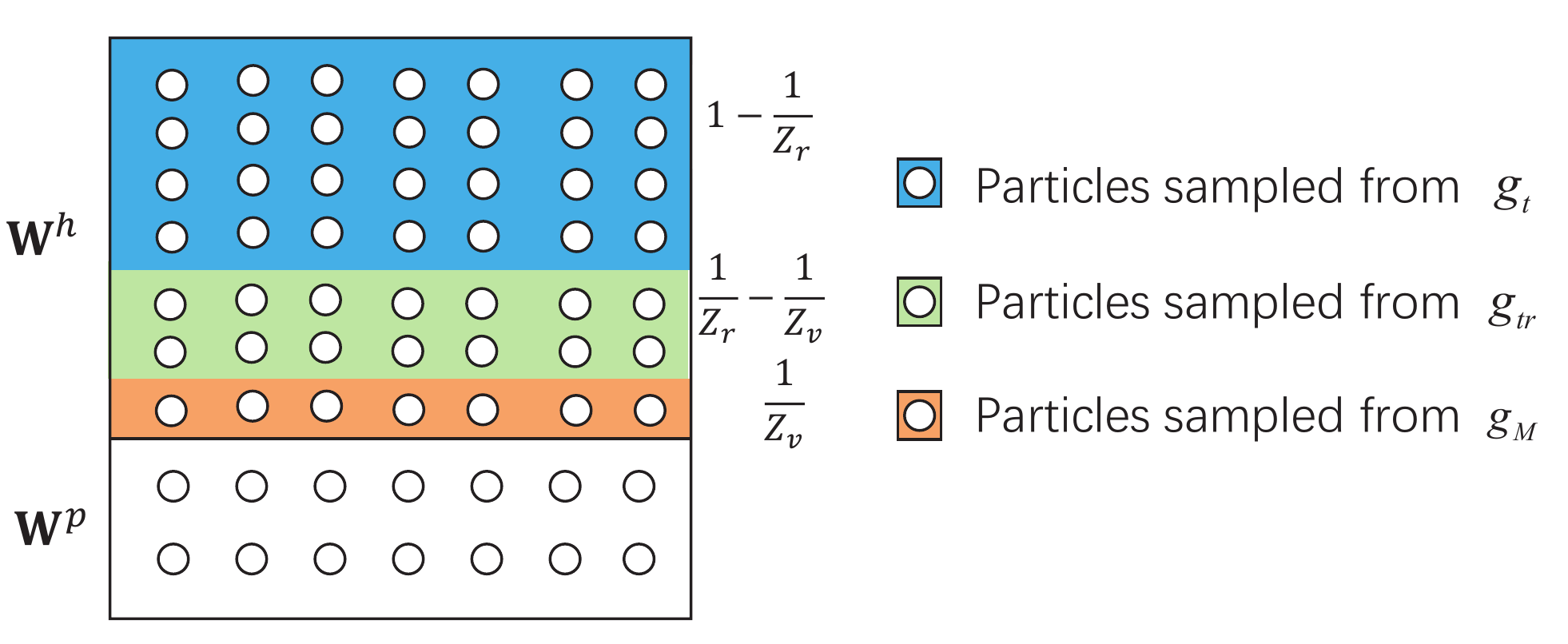}
	\caption{Sampling particles for UGKP method with the vibrational model.}
	\label{fig:sample-Wh}
\end{figure}

In the current study, the vibrational model has the distribution function
\begin{equation*}
	g^\ast = \left( 1 - \frac{1}{Z_r} \right)g_t
	       + \left( \frac{1}{Z_r} - \frac{1}{Z_v} \right)g_{tr}
		   + \left( \frac{1}{Z_v} \right)g_M ,
\end{equation*}
which contains three Maxwellian distribution functions with different weights. Therefore, three types of simulated particles  $P^t_k$, $P^{tr}_k$, and $P^{M}_k$ corresponding to $g_t$, $g_{tr}$, and $g_{M}$ respectively (see Fig.~\ref{fig:sample-Wh}) should be sampled to recover the distribution function
\begin{equation*}
	\begin{aligned}
		& P^t_k \sim \left( \rho^{h, t}_i, {\vec U_i}, \lambda_t, \lambda_r, \lambda_v \right), \\
		& P^{tr}_k \sim \left( \rho^{h, tr}_i, {\vec U_i}, \lambda_{tr}, \lambda_{tr}, \lambda_v \right), \\
		& P^{M}_k \sim \left( \rho^{h, M}_i, {\vec U_i}, \lambda_M, \lambda_M, \lambda_M \right),
	\end{aligned}
\end{equation*}
with
\begin{equation*}
	\rho_i^{h,t} = \left( 1 - \frac{1}{Z_r} \right)
	\rho_i^h,
	\quad
	\rho_i^{h,tr} = \left( \frac{1}{Z_r} - \frac{1}{Z_v} \right)
    \rho_i^h,
	\quad
    {\rm{and}}
	\quad
	\rho_i^{h,M} = \left( \frac{1}{Z_v} \right)
	\rho_i^h.
\end{equation*}

\section{Upstream and downstream condition of a shock structure with vibrational mode}
\label{sec:app-shock}

Since the vibrational degrees of freedom depend on the temperature, the specific heat ratio is not a constant in the computational domain. For normal shock structure, the Rankine--Hugoniet relation under the constant specific heat ratio $\gamma = 7/5$ is no longer valid. Instead, the relation between upstream and downstream states should be obtained by imposing conservation laws with a non-constant specific heat ratio
\begin{equation}\label{eq:shock-T}
	\frac{\lambda_2}{\lambda_1}
	=
	\frac{\left( {\rm Ma}_2^2 \gamma_2 \right) / 2 + \gamma_2 / \left( \gamma_2 - 1 \right)}
	     {\left( {\rm Ma}_1^2 \gamma_1 \right) / 2 + \gamma_1 / \left( \gamma_1 - 1 \right)},
\end{equation}
\begin{equation} \label{eq:shock-u}
	\frac{u_2}{u_1}
	=
	\sqrt{ \frac{ \left[ 1/2 + {\rm Ma}_2^2 / \left(\gamma_1 - 1\right)\right] }
	            { \left[ 1/2 + {\rm Ma}_1^2 / \left(\gamma_2 - 1\right)\right] } },
\end{equation}
\begin{equation}\label{eq:shock-p}
	\frac{p_2}{p_1}
	=
	\frac{ 1 + \gamma_1 {\rm Ma}_1^2}{ 1 + \gamma_2 {\rm Ma}_2^2 },
\end{equation}
\begin{equation}\label{eq:shock-Ma}
	\frac{ \left( 1 + \gamma_1 {\rm Ma}_1^2 \right)^2 }
	     { \gamma_1 {\rm Ma}_1^2 \left[ \gamma_1 / \left( \gamma_1 - 1 \right)
		                       + \left( \gamma_1 {\rm Ma}_1^2 \right) / 2 \right] }
	=
	\frac{ \left( 1 + \gamma_2 {\rm Ma}_2^2 \right)^2 }
	     { \gamma_2 {\rm Ma}_2^2 \left[ \gamma_2 / \left( \gamma_2 - 1\right)
		                       + \left( \gamma_2 {\rm Ma}_2^2 \right) / 2 \right] },
\end{equation}
where the subscripts “1” and “2” denote the state at upstream and downstream, respectively. The relation between specific heat ratio and the internal degrees of freedom is
\begin{equation} \label{eq:gamma}
	\gamma = \frac{ 7 + K_v }{ 5 + K_v}.
\end{equation}
Substituting Eq.~\eqref{eq:Kv} into Eq.~\eqref{eq:gamma}, the expression for specific heat ratio with respect to temperature $\lambda$ can be obtained
\begin{equation}\label{eq:gamma-T}
	\gamma = \frac{ 7 \left( e^{ 2 R \lambda \Theta_{v} } - 1 \right) + 4 R \lambda \Theta_v }
			      { 5 \left( e^{ 2 R \lambda \Theta_{v} } - 1 \right) + 4 R \lambda \Theta_v }.
\end{equation}
Due to the complexity of Eq.~\eqref{eq:gamma-T}, explicit determination of the downstream is difficult,
therefore, implicit iteration of Eqs~\eqref{eq:shock-Ma}, \eqref{eq:shock-T} and \eqref{eq:gamma-T} is carried out to get the downstream temperature and Mach number. Then, the velocity and pressure in the downstream are determined by Eqs~\eqref{eq:shock-u} and \eqref{eq:shock-p}.


%% References


% \end{CJK*}
\newpage
\bibliographystyle{elsarticle-num}
% \section*{\refname}
\bibliography{wpvib.bib}






% \end{thebibliography}

\end{document}